%% file: main.tex
\documentclass[10pt,journal,compsoc]{IEEEtran}
\PassOptionsToPackage{hyphens}{url}
\usepackage[utf8]{inputenc}
\usepackage{graphicx}
\usepackage{subcaption}
\usepackage{amsmath}
\usepackage{amssymb}
\usepackage{algorithm}
\usepackage{algpseudocode}
\usepackage{listings}

\usepackage{siunitx}
\usepackage{amsthm}
\usepackage{xcolor}
\usepackage{booktabs}
\usepackage{tabularx}
\usepackage{multirow}
\usepackage{soul}
\usepackage{comment}
\usepackage[pagebackref=true,breaklinks=true,letterpaper=true,colorlinks,bookmarks=false]{hyperref}
\usepackage{url}
\usepackage{bm}
\usepackage{tikz} 
\usepackage{wrapfig}
\usetikzlibrary{shapes,snakes}

\DeclareMathOperator*{\argmin}{arg\,min}

\usepackage{xifthen}
\usepackage{mathtools}

\usepackage{xparse}

\NewDocumentCommand{\vect}{ O{} O{} m }{\bm{#3}\ifthenelse{\isempty{#1}}{}{^{(#1)}}\ifthenelse{\isempty{#2}}{}{_{#2}}}

\NewDocumentCommand{\mat}{ O{} O{} m }{\bm{#3}\ifthenelse{\isempty{#1}}{}{^{(#1)}}\ifthenelse{\isempty{#2}}{}{_{#2}}}
\NewDocumentCommand{\mmat}{ O{} O{} m }{{\mbox{\bfseries\itshape #3}}\ifthenelse{\isempty{#1}}{}{^{(#1)}}\ifthenelse{\isempty{#2}}{}{_{#2}}}

\NewDocumentCommand{\ten}{ O{} O{} m }{\bm{\mathcal{#3}}\ifthenelse{\isempty{#1}}{}{^{(#1)}}\ifthenelse{\isempty{#2}}{}{_{#2}}}

\newcommand{\kl}{k_l}
\newcommand{\ku}{k_u}

\renewcommand{\O}[1]{\mathcal{O}(#1)}
\newcommand{\OO}[1]{\mathcal{O}\left(#1\right)}

\usepackage[dont-mess-around]{fnpct}
\usepackage{etoolbox}

\begin{document}
\sloppy

\title{Distributed non-negative RESCAL  with Automatic Model Selection for Exascale  Data}

\author{Manish ~Bhattarai,
        Namita ~Kharat,
        Erik ~Skau,
        Benjamin ~Nebgen,
        Hristo ~Djidjev,
        Sanjay ~Rajopadhye,
        James ~P. ~Smith,
        and ~Boian ~Alexandrov
\IEEEcompsocitemizethanks{\IEEEcompsocthanksitem Manish Bhattarai, Namita ~Kharat, Benjamin ~Nebgen, James ~P. ~Smith and Boian ~Alexandrov are with the Theoretical Division, Los Alamos National Laboratory, Los Alamos,
NM, 87544.
\IEEEcompsocthanksitem Erik-Skau is with Computer, Computational, and Statistical Science Division, Los Alamos National Laboratory, Los Alamos,
NM, 87544. 
\IEEEcompsocthanksitem Hristo~Djidjev is with the Institute of Information and Communication Technologies, Bulgarian Academy of Sciences, Sofia, Bulgaria and the Computer, Computational, and Statistical Science Division, Los Alamos National Laboratory, Los Alamos,
NM, 87544. 
\IEEEcompsocthanksitem Sanjay ~Rajopadhye and Namita ~Kharat are with Department of Computer Science, CSU, Fort Collins, CO 80523.}
\thanks{Manuscript received December 1, 2021; }}

\IEEEtitleabstractindextext{%

\begin{abstract}
With the boom in the development of computer hardware and software, social media, IoT platforms, and communications, there has been an exponential growth in the volume of data produced around the world.  Among these data, relational datasets are growing in popularity as they provide unique insights regarding the evolution of communities and their interactions. Relational datasets are naturally non-negative, sparse, and extra-large. Relational data usually contain triples, (subject, relation, object), and are represented as graphs/multigraphs, called knowledge graphs, which need to be embedded into a low-dimensional dense vector space. Among various embedding models, RESCAL allows learning of relational data to extract the posterior distributions over the latent variables and to make predictions of missing relations. However, RESCAL is computationally demanding and requires a fast and distributed implementation to analyze extra-large real-world datasets. Here we introduce a distributed non-negative RESCAL algorithm for heterogeneous CPU/GPU architectures with automatic selection of the number of latent communities (model selection), called pyDRESCALk. We demonstrate the correctness of pyDRESCALk with real-world and large synthetic tensors, and the efficacy showing near-linear scaling that concurs with the theoretical complexities. Finally, pyDRESCALk determines the number of latent communities in an 11-terabyte dense and 9-exabyte sparse synthetic tensor.

\end{abstract}

}

\maketitle

\IEEEraisesectionheading{\section{Introduction}\label{sec:introduction}}
\IEEEPARstart{E}{xtremely} large high-dimensional datasets are generated daily as biproducts of daily work in fields such as business, commerce, surveillance activities, social media networks, computer macro-simulations, and large-scale experiments to name a few
\cite{data_volume}. The size of global data grows exponentially and is expected to reach about 175 zettabytes (more than one trillion gigabytes) by 2025 \cite{IDC}. Part of this data is relational in nature, that is, data formatted in linked tables (i.e., related) \cite{dvzeroski2009relational}.
Relational data usually contains triples, (\emph{subject; relation; object}), and is represented as a graph (or multigraph), called knowledge graph \cite{ehrlinger2016towards}, where the nodes correspond to the entities of the analyzed system, while the edges store the relations between these entities. Examples of worldly known knowledge graph projects are YAGO \cite{suchanek2007yago}, DBpedia \cite{auer2007dbpedia}, and the Google Knowledge Graph (containing $\sim$ 70 billion relational facts) \cite{singhal2012introducing}. The development of knowledge graphs  led to increasing activities in statistical relational learning that integrates semantics and probabilistic graphical models in semantic networks \cite{gao2021combination}. 

To be of a practical use, the large (and usually very sparse) knowledge graphs need to be embedded into a low-dimensional dense vector space. The embedding means representing the knowledge graph entities in a vector space with a scoring function, defined to measure the likelihood of each triplet, \emph{(subject, relation, object)}. Beginning from a random  representation, the embedding uses some optimization algorithm to maximize the global likelihood of all available triplets \cite{dai2020survey}.
The knowledge graph embedding models provide unique information about how various entities are linked, empowering the search-engine to offer relevant results and information retrieval as well as enabling prediction of previously unknown relations from available sparse, noisy, and incomplete data \cite{wang2017knowledge}. Also, knowledge graphs are the basis of semantic webs used to create "webs of data" that are machine readable\footnote{\url{https://www.w3.org/2013/data/}}.  An embedding model is defined by its scoring function which measures the plausibility of the relations between the entities, that is, the plausibility of the triple, ($e_i$; $e_j$; $r_k$). By maximizing the total plausibility of the observed relations, the scoring function determines the concrete embedding model. The broadly used scoring functions can be separated into two main classes: distance- vs. similarity-based scoring functions, which define (a) translational distance interaction and (b) semantic matching interaction models, respectively \cite{choudhary2021survey}.

Tensor factorization and tensor networks \cite{cichocki2017tensor} are well-known factorization techniques used for knowledge graph embedding with semantic matching interactions \cite{ali2021bringing}. 
In a tensor representation, each triplet of the knowledge graph is represented by the value, $X_{ijk}$ of a three dimensional tensor, $\ten{X} \in \mathbb{R}^{n \times n \times m}_+ $  (with $n$-entities
and $m$-relations). This value could be binary, $\{0,1\}$, and in this case, $X_{ijk} =1$ means that the triplet $(i,j,k)$ exists, or can be an arbitrary non-negative value that measures the strength of the $k$-relation between the $i$- and $j$-entities.

RESCAL \cite{nickel2011three} was the first embedding model based on tensor factorization, and was used for embeddings of the knowledge graph YAGO \cite{nickel2012factorizing}. 
Although other tensor-based embeddings were proposed \cite{nickel2016holographic,kazemi2018simple,balavzevic2019tucker}, RESCAL and its modifications continue to serve as an excellent embedding model \cite{nickel2013tensor,yang2014embedding,trouillon2016complex,han2021time,ma2021temporal}. 
The non-negative RESCAL embedding approximates a  tensor, $\ten{X} \in \mathbb{R}^{n \times n \times m}_+$, by a bilinear product of a low-rank non-negative matrix $\mat{A} \in \mathbb{R}^{n \times k }_+$ and a small core tensor $\ten{R} \in \mathbb{R}^{k \times k \times m}_+$. Thus,  $$\ten{X}_{i,t,j} \approx \sum_{s=1}^{k}\mat{A}_{i,s}\ten{R}_{s,t,s}\mat{A}_{s,j},$$ with the constraints; $\mat{A},\ten{R} \geq 0$. 

\begin{figure}[htb]
    \includegraphics[width=\columnwidth]{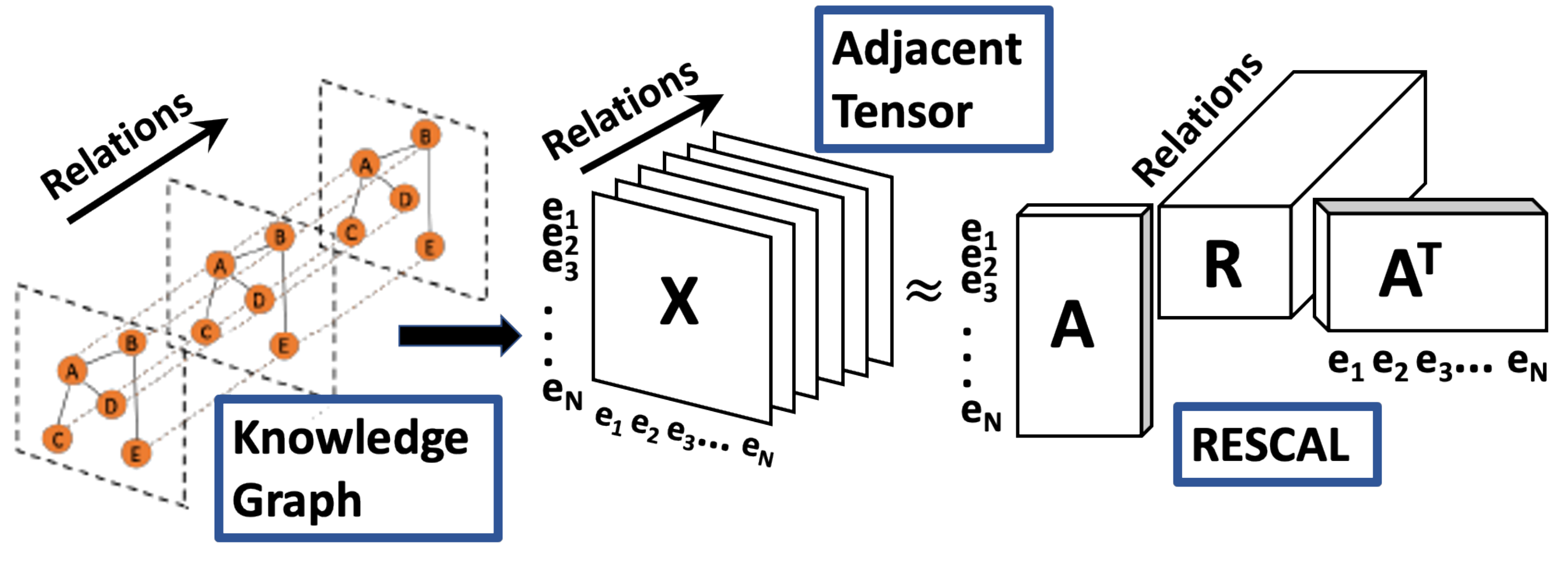}
\caption{RESCAL tensor embedding of a knowledge graph.}
	\label{fig:RESCAL_KG} 
	\vspace{-0.5em}
\end{figure}

The columns of $\mat{A}$ represent the latent communities and the  slices of  $\ten{R}$ represent the relations between the latent communities, as shown in Figure ~\ref{fig:RESCAL_KG}. Furthermore, Figure~\ref{fig:RESCAL_KG} provides  a graphic representation of a knowledge graph expressed as an entity-relation/adjacent tensor $\ten{X}$ which is then decomposed with RESCAL to extract a matrix $\mat{A}$ and a tensor $\ten{R}$.

The non-negative RESCAL model is well-known for its ability to unravel major interactions in dynamic asymmetric pairwise relationship tensors, while constraining the extracted components to be non-negative, which leads to parts-based components \cite{lee1999learning} and explainable latent communities \cite{krompass2013non}. 
Most of the refinements and modifications of RESCAL are targeting better speed and scaling to make it suitable for analyzing extra-large and sparse real-world datasets. However, due to the quadratic run-time and memory complexity with respect to the embedding dimensions, existing RESCAL implementations fail to scale to very large tensors. Hence, existing RESCAL implementations cannot currently be used to analyze the knowledge graphs arising in most of the applications of practical importance such as social network analysis and business interactions.

To address the bottlenecks associated with decomposing such large tensors, we introduce a new efficient distributed algorithm for non-negative RESCAL factorization with a superior scaling and speed, capable of working on modern heterogeneous CPU/GPU architectures to factorize extra-large dense and sparse tensors. We integrated our algorithm  with a model selection method based on the stability of the extracted latent communities (the columns of matrix $A$) \cite{truong2020determination}, and call it pyDRESCALk. We evaluate pyDRESCALk on several extra-large synthetic tensors as well as on real-world data, and show that in all cases the predicted latent communities are highly correlated with the predetermined ground-truth solutions and pyDRESCALk determines accurately the latent dimension. Our scalability results show that pyDRESCALk scales nearly linearly when applied to large, dense and/or sparse  datasets for both CPU/GPU architectures. Finally, we demonstrate how pyDRESCALk determines the number of latent communities and extracts them from $10$~Terabytes(TBs) dense as well as from a $9$~Exabytes(EBs) sparse ($10^{-5}$ density) synthetic datasets. To the best of our knowledge, pyDRESCALk is the first RESCAL implementation that can work on such a large scale.

The main contributions of this paper are as follows:

\begin{itemize}
   \item We developed pyDRESCALk, which is the first distributed RESCAL implementation with the  ability to estimate latent features (determine the model).\smallskip 
   \item pyDRESCALk works for extra-large non-negative tensors, with dense as well as sparse structure.\smallskip
    \item pyDRESCALk is the first distributed RESCAL framework for relational datasets on distributed GPU/CPU architectures.\smallskip
    \item We demonstrate that pyDRESCALk is able to decompose $10$TB dense and $9$EB sparse data-tensors.\smallskip
     \item We  released the pyDRESCALk library\footnote{\url{https://github.com/lanl/pyDRESCALk}}\cite{pyDRESCALk} for the reproducibility of the presented results and availability to researchers. 
\end{itemize}

The remainder of the paper is organized as follows: Section \ref{sec:background} gives a summary of related work, non-negative RESCAL, and the model selection algorithm. In Section~\ref{sec:preliminary}, the preliminary notations and concepts are presented whereas Section~\ref{sec:dnmfk} provides insight into our distributed decomposition and clustering algorithms for estimation of latent components. Section~\ref{sec:complexity} presents the complexity analysis of the distributed framework; Section~\ref{sec:experiments} demonstrates the efficacy of the pyDRESCALk via correctness and scalability. Finally~\ref{sec:conclusions} concludes the paper and suggests future directions.  
\section{Background}\label{sec:background}

\subsection{non-negative RESCAL}
non-negative RESCAL \cite{krompass2013non} simultaneously decomposes the $t$-th slice of the adjacent tensor of a knowledge graph, $\ten[][]{X}$: $\ten[][t]{X}$ for $1 \leq t \leq m$, into a non-negative matrix product
$$\ten[][t]{X} = \mat{A} \ten[][t]{R} \mat{A}^{\top}\;.$$
where $\mat{A} \in \mathbb{R}^{n \times k}_+$ and each $\ten[][t]{R} \in \mathbb{R}^{k \times k}_+$ for $1 \leq t \leq m$ is a matrix, that is as the $t$-th slice of the core tensor $\ten[][]{R} \in \mathbb{R}^{k \times k \times m}_+$. To remove a scaling ambiguity, RESCAL typically constrains the columns of $\mat{A}$ to be  normalized, i.e., $|| \mat[][i]{A} || = 1$ for $1 \leq i \leq k$. In this decomposition, $k$ is the number of the latent communities of the knowledge graph, and the columns of $\mat{A}$ provide the community membership weights for each entity in each community. Each slice $\ten[][t]{R}$ encodes the relations between these communities needed to coalesce the entity relations in $\ten[][t]{X}$. 

In practice, exact equality in the decomposition is unattainable, so the decomposition is determined by solving the constrained minimization problem
\begin{align}
    \label{eqn:minproblem}
    \begin{split}
    \argmin_{ \mat{A},  \ten[][t]{R}} & \quad \sum_{t=1}^m ||   \ten{X}_t -  \mat{A}   \ten{R}_t  \mat{A}^{\top} ||_F^2\\
    \text{subject to} &\quad \mat{A} \geq 0;|| \mat[][i]{A} || = 1; \ten[][t]{R} \geq 0\;.
    \end{split}
\end{align}

\subsection{Multiplicative Update Algorithm } 
The non-negative minimization problem in Equation~\ref{eqn:minproblem} is typically solved with an alternating optimization procedure, where $\mat{A}$ and $\ten{R}$ updates are individually optimized in an iterative fashion. To solve the $\mat{A}$ and $\ten{R}$ subproblems, we use the multiplicative update schemes employed in \cite{krompass2013non}. The non-negative constraint is ensured with the iterative update rules,
\begin{equation}
\label{eqn:updates}
  \begin{gathered}
    (\ten{R}_t)_{ij} \leftarrow{} (\ten{R}_t)_{ij} \dfrac{ [\mat{A}^\top \ten{X}_t \mat{A}]_{ij}}{[\mat{A}^\top \mat{A} \ten{R}_t \mat{A}^\top \mat{A}]_{ij} + \epsilon} \text{ \ \ for } t = 1,...,m \\
    \mat{A}_{ij} \leftarrow{}  \mat{A}_{ij}\dfrac{\left[\sum_{t=1}^T  \ten{X}_t \mat{A} \ten{R}_t^\top +  \ten{X}_t^\top  \mat{A}   \ten{R}_t\right]_{ij} }{\left[ \mat{A}\left( \ten{R}_t \mat{A}^\top  \mat{A} \ten{R}_t^\top +  \ten{R}_t^\top  \mat{A}^\top  \mat{A}   \ten{R}_t \right)\right]_{ij} + \epsilon} \;,
  \end{gathered}
\end{equation}
as long as both $\mat{A}$ and $\ten{R}$ initialization are non-negative, and $\epsilon \sim 10^{-16}$ is added to avoid divisions by zero. Normalization of $\mat{A}$ is done at the end of the optimization with the appropriate inverse scaling applied to $\ten{R}$.

\begin{figure*}[htb]
  \centering
  \includegraphics[width = 1.0 \textwidth]{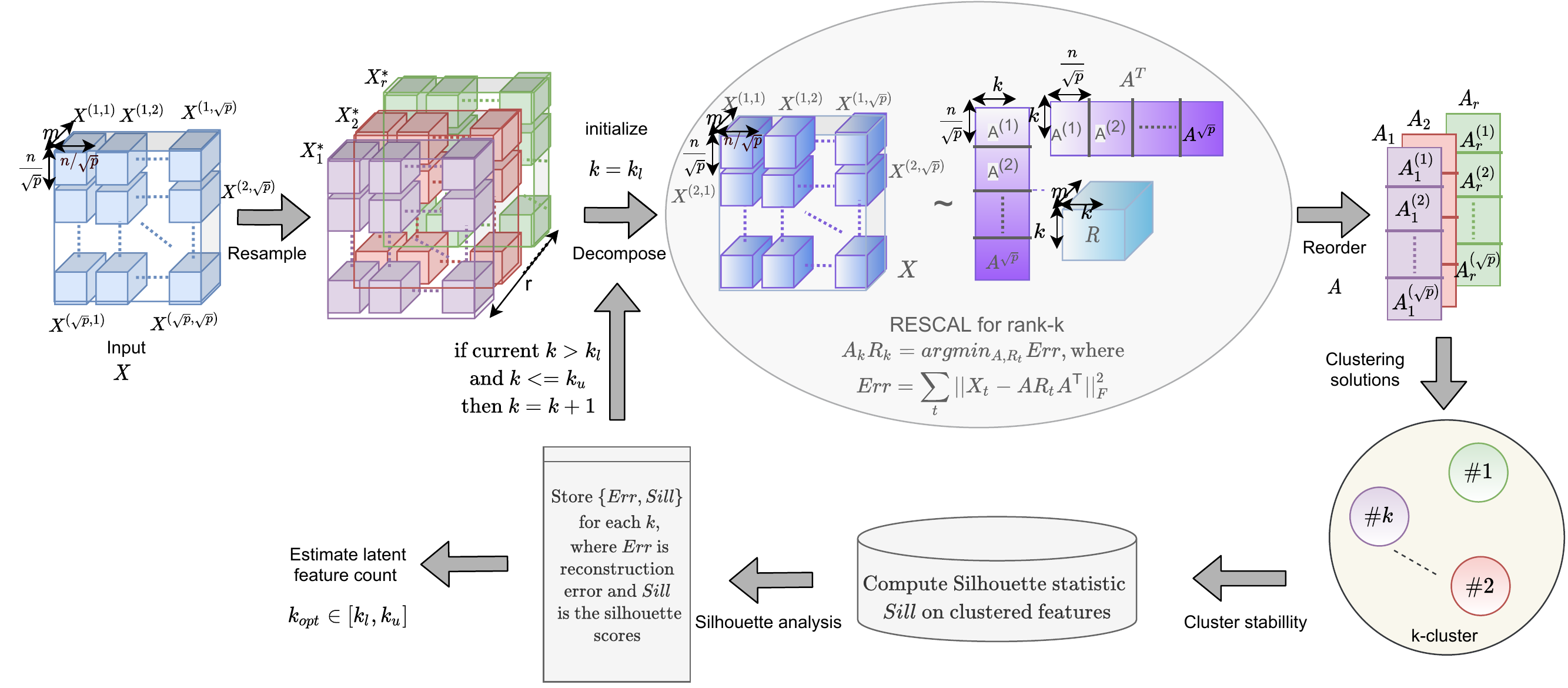}
  \caption{Diagram of steps to determine latent dimension for non-negative RESCAL.} 
  \label{fig:rescal-k}
\end{figure*}

\begin{table}[htp]
	\centering
	\caption{Notations} 
	\label{tab:notations}
	\begin{tabular}{|c|c|l|}
		\toprule
		\textbf{Notation} & \textbf{Dimensions} & \textbf{Description} \\
		\midrule
		$\ten{X}$ & $n \times n \times m$ & Input tensor \\
		$\ten{X}^{q}$ & $n \times n \times m$ & Perturbed tensor \#q \\
		$\mat{A}$ & $n \times k$ & Outer RESCAL factor \\
		$\ten{R}$ & $k \times k \times m$ & Inner RESCAL factor \\
		$k$   & scalar & Low rank \\
		$\kl$   & scalar & Lower bound of low rank \\
		$\ku$   & scalar & Upper bound of low rank \\
		$p$ & scalar & Count of parallel processes \\
		$r$ & scalar & Number of perturbations\\
		$\ten{A}$ & $n \times k \times M$ & A tensor of outer RESCAL factors $(\mat{A})$\\
		$\ten{R^*}$ & $k \times k \times m \times M$ & A tensor of inner RESCAL factors $(\ten{R})$\\
		$\ten[][:,:,q]{A}$ & $n \times k \times 1$ & The $q^{th}$ slice of tensor $\ten{A}$. \\
		$\ten[][:,:,:,q]{R^*}$ & $k \times k \times m \times 1$ & The $q^{th}$ slice of tensor $\ten{R^*}$. \\
		$\ten[i]{A}$ & $\frac{n}{p} \times k \times M$ & sub-tensor of $\ten{A}$ on the $i^{th}$ processor. \\
		$\ten[i][q]{A}$ & $\frac{n}{p} \times k$ & The $q^{th}$ sub slice of $\ten{A}$ on $i^{th}$ processor. \\
		$\vect{s}$ & $k\times 1$ & Average silhouettes of each cluster \\
		$\vect{e}$ & $k \times 1$ & Average reconstruction error \\
		\bottomrule
	\end{tabular}
\end{table}
\subsection{RESCAL with Automatic Model Selection} \label{sec:model_selection}
\label{sec:rescalk}
RESCAL tensor factorization requires prior knowledge of the latent dimensionality, $k$ (the number of latent communities), which usually is not available. Model selection, that is, the determination of $k$, is a process for estimating the number of parameters in the model without prior information, which is a difficult and well-known problem. Different heuristics, based on various criteria, have been proposed to solve this problem. Some of them employ information criteria, such as, Akaike’s information criterion (AIC) \cite{akaike1974new}, Bayesian information criterion (BIC) \cite{schwarz1978estimating}, minimum description length (MDL) \cite{rissanen1978modeling}. Others explore Baysian methods, such as, Automatic Relevance Determination (ARD) \cite{mackay1994automatic,bishop1999bayesian, morup2009tuning}. A different approach is to estimate $k$ based on the stability of the solutions of factorization for different values of $k$ \cite{brunet2004metagenes, alexandrov2013deciphering,sanchez2021automatic,desantis2020factorization}, which showed a superior performance, when applied to a large number of synthetic datasets with a predetermined number of latent features \cite{nebgen2021neural}. 

The stability approach was recently utilized for RASCAL \cite{truong2020determination}. Here we integrated our algorithm with stability approach model selection, and for completeness, provide the main steps of this approach below:

\begin{enumerate}
    \item \textit{Resampling:} Based on the adjacent tensor of the knowledge graph, $\ten{X}$, we create an ensemble of $r$ random tensors, $[\ten{X}^q]_{q=1,...,r}$, with means equal to the original adjacent tensor $\ten{X}$. Each one of these random tensors $\ten{X}^q$ is generated by perturbing the elements of $\ten{X}$ by a small uniform noise $\delta$, such that: $X^q_{ijk} = X_{ijk} +\delta$, for each $q=1,...,r$. For each $k$ explored, RESCAL minimizations result in $r$ solutions; one for each member of this random ensamble of tensors, $\ten{X}^q$.
    \item \textit{Custom clustering of the RESCAL solutions:} For each explored latent dimension, $k\in [k_{min},k_{max}]$, the minimizations of the $r$ random tensors, $[\ten{X}^q]_{q=1,...,r}$, results in $r$ pairs $\{\mat{A}[k]^q; \ten{R}[k]^q\}_{q=1,...,r}$. Further, we clusters the set of the $r*k$ latent communities, the columns of $\mat{A}[k]^q$. The clustering we use is similar to k-means, but it holds in each one of the clusters exactly one column from each of the $r$ RESCAL solutions $\mat{A}[k]^q$. This constraint is needed since each RESCAL minimization gives exactly one solution $\mat{A}[k]^q$ with the same number of columns, $k$. In the clustering, the similarity between the columns is measured by cosine similarity.
    \item \textit{Robust $\mat{A}^k$ and $\ten{R}^k$ for each $k$:} The medians of the clusters, $\mat{\widetilde{A}}^k$, are the robust solution for each explored $k$. The corresponding coefficients $\ten{R}_{regress}^k$ are calculated by regression of $\ten{X}_t$ on $\mat{\widetilde{A}}^k$, slice by slice.
    \item \textit{Silhouette statistics:} We explore the stability of the obtained clusters, for each $k$, by calculating their Silhouettes \cite{ROUSSEEUW198753}. Silhouette statistics quantifies the cohesion and separability of the clusters. The Silhouettes values range between $[-1, 1]$, where $-1$ means unstable cluster, while $+1$ means perfect stability.
    \item \textit{Reconstruction error:} Another metric we use is the relative reconstruction error, $e = ||\ten{X} - \ten{X}^{rec}||/||\ten{X}||$, where $\ten{X}^{rec} = \widetilde{\mat{A}}^k\ten{R}^k_{regress}\widetilde{\mat{A}}^{k^T}$, which measures the accuracy of the reproduction of initial data by the robust solution with $k$ latent features. 
    \item \textit{RESCAL final solution:} The number of latent features, $k_{opt}$, is determined as the maximum number of stable clusters corresponding to a good accuracy of the reconstruction. The corresponding $\widetilde{\mat{A}}^{k_{opt}}$ and $\ten{R}^{k_{opt}}_{regress}$ are the final robust RESCAL solutions.
\end{enumerate}

\subsection{Related Work}
In the literature, a significant amount of work has been dedicated to distributed tensor decompositions such as distributed PARAFAC, distributed Tucker, and distributed tensor train (TT) for both sparse/dense data and CPU/GPU hardware, which are scalable for large datasets~\cite{de2014distributed,sidiropoulos2014parallel,wang2015fast,traore2019singleshot,chakaravarthy2017optimizing,chakaravarthy2018optimizing,choi2018high,chakaravarthy2019optimizing,wang2019distributed,bhattarai2020distributed}. In other words, several optimal distributed design solutions have already been proposed for decomposition of very large multi-dimensional datasets. For decomposition of knowledge graph/relational datasets, relational models such as the bilinear RESCAL tensor decomposition \cite{nickel2012factorizing} have successfully extracted the underlying entity-relation features. This highlights the need fora high-performance RESCAL model to decompose real-world, large knowledge graphs. However, there are only a few previous works on distributed/parallel RESCAL, and they are only able to demonstrate scalability on relatively small datasets \cite{nickel2012factorizing,al2021parallel}.The largest datasets used by \cite{al2021parallel} and \cite{nickel2012factorizing} are of size $135\times 135\times 49$(\num{8e6} non-zero elements) and sparse $3000417\times 3000417\times 38$(\num{4e7} non-zero elements) respectively. Compared to these, we were able to factorize dense tensor of size $396800\times 396800\times 20 $(\num{3e13} non-zero elements) and sparse tensor of size $373555200\times 373555200\times 20$(\num{3e14} non-zero elements) respectively.   In previous parallel implementations, the authors perform the factor updates $\mat{A}$ and $\ten{R}$ by splitting the data $\ten{X}$ along the third axis (i.e., relations) and updating the corresponding slices as per Equation~\ref{eqn:updates}. The resulting residuals are then aggregated through map-reduce. This implementation is only efficient if $m\gg n$ as the local computation on each process is small. However, for real-world datasets where the number of the entities is much larger than the number of relations (i.e., $n\gg m$),  local computation would be a bottleneck and such implementations fail to perform decomposition on such large datasets. To address these limitations, rather than slicing along the third dimension, we instead slice along the first two dimensions via a square 2D virtual grid to achieve better scalability. Another major challenge in the application of RESCAL to large, real-world datasets is determination of the correct $k$. In the previous work, the authors do not address the problem of model selection. The framework presented in this paper is the first to automatically estimate the number of latent features($k$) for very large datasets. 

An overview of the various knowledge graph models for non-distributed applications can be found in the PyKEEN documentation\footnote{\url{https://pykeen.readthedocs.io/en/stable/index.html}}. The existing implementation of the  distributed knowledge graph tensor embedding models such as TransE\cite{bordes2013translating}, RESCAL\cite{nickel2011three}, DistMult\cite{yang2014embedding}, and ComplEx\cite{trouillon2016complex} are based on PyTorch\cite{pbg,han2018openke,lerer2019pytorch}. As the distributed knowledge graph models incorporate underlying PyTorch big-graph\cite{lerer2019pytorch}, they are limited by the PyTorch distributed APIs and are not optimized to address the communication-computation bottleneck for distributed implementation across multiple nodes. To address this bottleneck, our distributed pyDRESCALk framework is designed with a 2D virtual grid-based MPI topology such that the algorithmic efficiency is maximized by minimizing the global communication cost. In addition, our framework accurately estimates the underlying latent components and provides realistic insight into the latent communities and their interactions. Moreover, our scaling tests, performed on the Grizzly and Kodiak supercomputers at Los Alamos National Laboratory, show the algorithm's scalability, and efficient node utilization on  large-scale data.

\section{Foundations}
\label{sec:preliminary}
In this section, we introduce mathematical notations for the pyDRESCALk algorithms, MPI collectives, and the multiplicative update algorithms. 
\subsection{Preliminaries}

Table~\ref{tab:notations} summarizes the notations used in this paper. A tensor is represented with a bold italic uppercase script letter $\ten{X}$, a matrix is represented with bold uppercase letter $\mat{X}$ and a vector is represented with lowercase math letter $\vect{x}$.  $\ten{X}^q$ is a perturbation of input tensor $\ten{X}$. $\ten{A}$ and $\ten{R^*}$ denote tensors with each slice (unit partition of tensor along an axis) representing a matrix and tensor, respectively. $\ten[][:,:,i]{A}$ represents the latent space of {\em k} for the $i^{th}$ perturbation. The median low rank factors are $\widetilde{\mat{A}}$ and $\widetilde{\ten{R}}$. $\ten[i][:,:,q]{A}$ is the sub slice ($\mat[i]{A}$) of $q^{th}$ perturbation on $i^{th}$ processor. Finally, $\vect{s}$ represents the Silhouette statistics of each data point in all the clusters. The vector $\vect{e}$ stands for the reconstruction error for a given perturbation at a given {\em k}.

\subsection{MPI Terminology}
In our distributed implementation, we use MPI specific point-to-point collective communication operations, specifically, {\bf all\_gather}, {\bf all\_reduce}, and {\bf broadcast}. For example, if we consider $\vect{a}$ to be a vector with $n$ elements distributed across $p$ processes, each process will have roughly $n/p$ elements. On application of the {\bf all\_gather} collective, all processes gather all local data vectors so that each process holds a copy of $\vect{a}$ of length $n$. Whereas on the application of {\bf all\_reduce} collective, an element-wise sum of the data across the processes is performed, resulting in data the same length across all processes. On the other hand, the application of {\bf broadcast} collective results in broadcasting the data from the source MPI process to the destination MPI processes. The details of these MPI collectives can be found in~\cite{chan2007collective}.

\subsection{Generic RESCALk}
\begin{algorithm}[!ht] 
	\caption{$\operatorname{Rescalk}(\ten{X}, \kl, \ku, r)$ -- Generic {\em RESCAL with model selection}}\label{alg:rescalk}
	\begin{algorithmic}[1]
		\Require $\ten{X} \in \mathbb{R}^{m\times n\times n}_{+}$, $\kl$, $\ku$, $r$
		\For{ $k$ \textbf{in} $k_l$ to $k_u$}\label{line:nmfk_fork}
		\For{$q$ \textbf{in} $1$ to $r$} \label{line:nmfk_forpert}
		\State $\ten[][]{X}^q$ = \textit{Perturb}($\ten{X}$) \Comment{Resampling to create ensembles}\label{line:rescalk_pert}
		\State $\ten[][:,:,q]{A}, \ten[][:,:,:,q]{R^*} = \operatorname{RESCAL}(\ten[][]{X}^q, k)$ \label{line:rescalk_rescal}
		\EndFor \label{line:rescalk_endforpert}
		\State $\ten[][:,:,k]{A'} = \operatorname{customCluster}(\ten{A})$\Comment{Custom clustering of factor solutions}\label{line:nmfk_cluster} 
		\State $\widetilde{\mat{A}}$ = $\operatorname{median}(\ten{A}')$ \label{line:rescalk_medians}
		\State $\vect[][k]{s} = \operatorname{clusterStability}(\ten{A'})$\Comment{Quality of clusters}\label{line:nmfk_sil}
		\State $\ten[][:,:,:,k]{R'} =
		\operatorname{regression}(\ten{X},\ten{A'})$\Comment{Regression(Performing RESCAL updates for $\ten{R}$}\label{line:nmfk_reg}
		\State $\vect[][k]{e} =
		\operatorname{reconstructError}(\ten{X},\ten[][:,:,k]{A'}, \ten[][:,:,:,k]{R'})$\Comment{Reconstruction Error Calculation}\label{line:nmfk_err}
		\EndFor
		\Ensure $\vect{s} \in \mathbb{R}^{(k_u - k_l) \times 1}_{+}$, $\vect{e} \in \mathbb{R}^{ (k_u - k_l) \times 1}_{+}$, $\ten{A'} \in \mathbb{R}^{n\times k\times r}_{+}$, $\ten{R'} \in \mathbb{R}^{k\times k\times m\times r}_{+}$
	\end{algorithmic}
\end{algorithm}

The details of the latent feature estimation, listed in the previous section, are given in Algorithm~\ref{alg:rescalk}, which describes the generic (sequential) RESCALk i.e. RESCAL implementation with model selection. For each $k$ in the interval [$k_l$, $k_u$], the RESCAL decomposition is performed for $r$ independent minimizations/runs on each resampled data tensor. First, a perturbation of the original input $\ten{X}$ (line~\ref{line:rescalk_pert}) using a random sample from a distribution (for example, \emph{uniform}) is performed. Then a RESCAL decomposition of this tensor is computed with a random or NNDSVD (non-negative double SVD) \cite{atif2019improved} initialization as shown in (line~\ref{line:rescalk_rescal}). Next, for each of the $r$ perturbations of the tensor $\ten{X}$, we get the corresponding $r$ different factorizations for each $k$.  These ensembles of factors are stored as tensors $\ten{A}$ = ($\ten[][:,:,1]{A}$, $\ten[][:,:,2]{A}$, $\dots$, $\ten[][:,:,r]{A}$) and $\ten{R^*}$ = ($\ten[][:,:,:,1]{R^*}$, $\ten[][:,:,:,2]{R^*}$, $\dots$, $\ten[][:,:,:,r]{R^*}$). Each basis factor matrix, $\ten[][:,:,i]{A}, i=1,\dots,r$, contains {\em k} columns that are each treated as a vector in $n$-dimensional space. Provided that this perturbation (line~\ref{line:rescalk_pert}) is within a reasonable error bound, each vector of the matrix $\mat{A}$ will be close to a corresponding vector in the solution space of the unperturbed matrix. Therefore, we can define $k$ such columns $C_1,\dots,C_k$.

Further, to cluster the $r$ RESCAL solutions, we use custom k-means clustering~\cite{hartigan1979algorithm}, to reorder the columns of each solution matrix $\ten[][:,:,q]{A}$ so that all first columns of each matrix belongs to one cluster, all the second columns belong to another, and so on. The optimal reordering is the permutation that maximizes the cosine similarities between the columns of $\ten[][:,:,q]{A}$ and the columns of the current centroid. Our clustering is detailed later in Section~\ref{sec:distclust}. 

In order to determine the number of latent communities $k_{opt}$, we compute $\vect[][k]{s}$ and $\vect[][k]{e}$ at each $k$. Here, $\vect[][k]{s}$ (line~\ref{line:nmfk_sil}) quantifies the quality of the clusters while $\vect[][k]{e}$ (line~\ref{line:nmfk_err}) quantifies the quality of reconstruction, which is the relative error defined as $\frac{\Arrowvert\ten{X}-\mat{A}\ten{R}\mat{A}^T\Arrowvert_{F}}{\Arrowvert\ten{X}\Arrowvert_{F}}$. Using $\vect{s}_k$ and $\vect{e}_k$, the optimal number of latent features, $k_{opt}$ is estimated  based on the combination of low reconstruction error and high silhouette score for  the input tensor, $\ten{X}$. In the case $k>k_{opt}$, the $k$ clusters are not separated well and the average cluster quality $\vect{s}$ will be low. On the other hand, for $k<k_{opt}$, the reconstruction error ($\vect{e}$) will be too high since the product of $\widetilde{\mat{A}}$ and $\widetilde{\ten{R}}$  will not closely approximate $\ten{X}$. Therefore, $k=k_{opt}$ when the cluster quality $\vect{s}_k$ is high and relative error $\vect{e}_k$ is low. Distributed implementations of the procedures used in lines~\ref{line:rescalk_pert}, \ref{line:rescalk_rescal}, \ref{line:nmfk_cluster}--\ref{line:nmfk_sil} are described later in section~\ref{sec:distcluststability}. 

\subsection{Optimization algorithm }
Multiplicative update (MU) is an alternating optimization algorithm that solves the sub-problems in equation~ \ref{eqn:minproblem} using  multiplicative updates.  The multiplicative update algorithm iterates through updating the factors $\mat{A}$ and $\ten{R}$ until convergence is achieved. The convergence is also dependent upon the initialization of the factors. We found that utilizing a custom NNDSVD-based initialization leads to a faster convergence compared to random initialization. The update rules are given in equation~\ref{eqn:updates}.

\begin{figure*}[htp]
	\centering
	\includegraphics[width=1\textwidth]{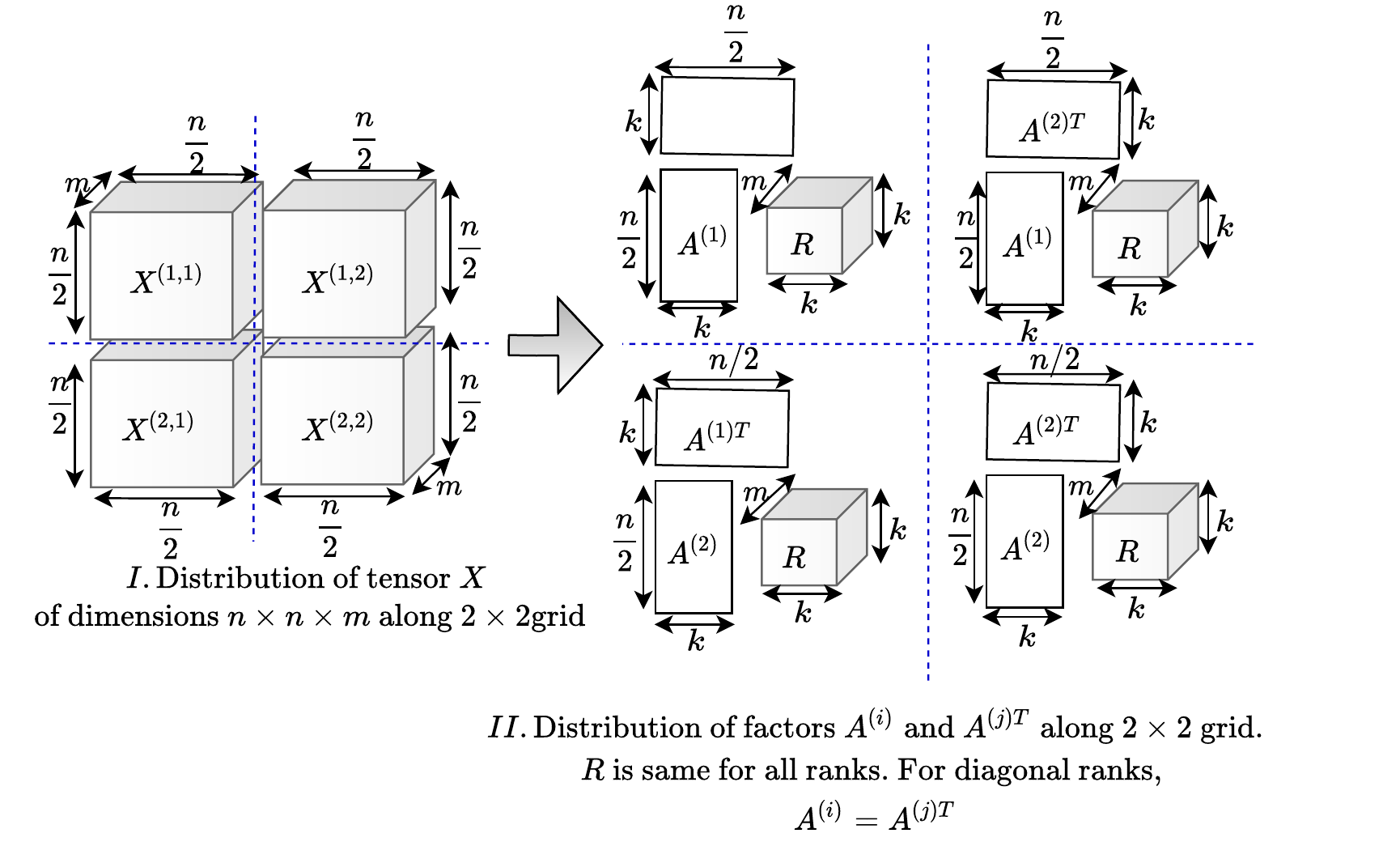}
	\caption{The input matrix $\mat{A}$ distributed in a virtual 2D grid, the factor matrices $\mat{W}$ and $\mat{H}$ in a 1D grid of processors.} \label{fig:dist-nmf} 
\end{figure*}

\section{pyDRESCALk}
\label{sec:dnmfk}
Here, we present distributed implementations for the individual components of generic {\em RESCAL} with automatic model selection, including: computing a distributed perturbation (line~\ref{line:rescalk_pert}), distributed implementation of RESCAL (line~\ref{line:rescalk_rescal}), distributed custom clustering (line~\ref{line:nmfk_cluster}), and distributed cluster stability (line~\ref{line:nmfk_sil}) .

\subsection{Distributed RESCAL}
\label{sec:drescal}
\begin{algorithm}[htp]
    \caption{$\mat{{AB}}$ = $\operatorname{distMM}(\mat{A^i},\mat{B^j},comm)$ -- Distributed matrix multiplication of $\mat{A}$ and $\mat{B}$ along row/column communicator}\label{alg:distMMt}
	\begin{algorithmic}[1]
    \Require  $\mat{A}\in \mathbb{R}_{+}^{x\times r}$, $\mat{B}\in \mathbb{R}_{+}^{r\times y}$ and \textbf{comm}$\in$ \{rowComm,colComm\} where rowComm and colComm are row and column subcommunicators respectively. 
    
    \Require $\mat{A}$ and $\mat{B}$ distributed along a 1D grid processor such that $\mat{A}^i\in \mathbb{R}_{+}^{\frac{x}{p_1}\times r}$ and $\mat{B}^j\in \mathbb{R}_{+}^{r \times \frac{y}{p_2}}$ where $p_1$ and $p_2$ are the number of processors  along the distribution of $\mat{A}$ and $\mat{B}$ respectively.
    \State $\mat[i,j]{{U}}$ = $\mat{A^i}\cdot\mat{B^j}$ \label{alg:distmm_mul} 
    \If{\textit{comm}=rowComm}
    \State $\mmat{AB}$ = $\sum_{i} \mat[i,j]{U}$ \Comment{$all\_reduce$ across row procs } \label{alg:distmm_row}
    \ElsIf{\textit{comm}=colComm}
    \State $\mmat{AB}$ = $\sum_{j} \mat[i,j]{U}$ \Comment{$all\_reduce$ across column procs }\label{alg:distmm_col}
    \EndIf
    \State \Return $\mmat{AB}$
    
    \end{algorithmic} 
\end{algorithm}

\begin{algorithm}[htp]\footnotesize
    \caption{$\mat{A}$,$\ten{R}$ = $\operatorname{RESCAL}(\ten{X}, k)$ -- Distributed RESCAL algorithm}\label{alg:distrescal}
  \begin{algorithmic}[1]
  \Require $\ten{X} \in \mathbb{R}_{+}^{n \times n \times m}$ , $k$ is the rank of approximation and $max\_iters$ is the number of iterations
  
    \Require $\ten{X}$ distributed across $\sqrt p \times \sqrt p$ grid of processors such that $\ten[i,j]{X} \in \mathbb{R}_{+}^{\frac{n}{\sqrt{p}} \times \frac{n}{\sqrt{p}} \times m}$, $\mat[i]{A} \in \mathbb{R}_{+}^{\frac{n}{\sqrt{p}}  \times k}$ and  $\ten{R} \in \mathbb{R}_{+}^{k\times k\times m}$
    \State Initialize $\mat[i]{A}$,$\mat[j]{A}$ and  $\ten{R}$ = $\operatorname{rand}(\frac{n}{\sqrt p}, k)$, $\operatorname{rand}(k, \frac{n}{\sqrt p})$  and  $\operatorname{rand}(m,k,k)$ 
      \label{alg:init}
    \For{$l$ in $[1,max\_iters]$} \label{alg:outer_itr}
       \State $\mmat{ATA}$ = $\operatorname{distMM}(\mat[j]{A},\mat[i]{A},rowComm)$\label{alg:ATA} 
         \For{$t$ in $m$} \label{alg:inner_itr}
            \Statex \textbf{/* Update  $\ten{R}$ given $\mat{A}$ */} 
             \State $\mmat{XA}$ = $\operatorname{distMM}(\ten[i,j]{X}[t],\mat[j]{A},colComm)$ \label{alg:XA}
             \State $\mmat{ATXA}$ = $\operatorname{distMM}((\mat[i]{A})^T,\mmat{XA},rowComm)$ \label{alg:ATXA}
            \State $\mmat{RATA}$ = $\ten{R}[t]\cdot\mmat{ATA}$ \label{alg:RATA}
            \State $DenoR$ = $\mmat{ATA}\cdot\mmat{RATA}$ + $\epsilon$ \label{alg:Rt}
            \State $\ten{R}[t]$ $\times$ = $\mmat{ATXA}$/$DenoR$ \label{alg:r_update}
         \Statex \textbf{/* Update $\mat{A}$ given $\ten{R}$ */}
            \State $\mmat{XART}$ = $\mmat{XA}\cdot\ten{R}[t]^T$ \label{alg:XART}
             \State $\mmat{AR}$ = $\mat[i][]{A} \cdot \ten{R}[t]$ \label{alg:AR}
          \State $\mmat{XTAR}$ = $\operatorname{distMM}(\ten[i,j]{X}[t]^T,\mmat{AR},rowComm)$\label{alg:XTAR}
          \State $\operatorname{broadcast}(XTAR)$ along columns from diagonal ranks \label{alg:broadXTAR}
          \State $NumA$ += $\mmat{XART}$+ $\mmat{XTAR}$ \label{alg:NumA}
          \State  $\mmat{ATAR}$ = $\mmat{ATA}\cdot\ten{R}[t]$  \label{alg:ATAR}
          \State  $\mmat{ART}$ = $\mat[i]{A}\cdot\ten{R}[t]^T$  \label{alg:ART}
          \State  $\mmat{ARTATAR}$ = $\mmat{ART}\cdot\mmat{ATAR}$ 
          \State  $\mmat{ATART}$ = $\mmat{ATA}\cdot\ten{R}[t]^T$ \label{alg:ATART}
          \State  $\mmat{ARATART}$ = $\mmat{AR}\cdot\mmat{ATART}$ 
          \State $DenoA$ += $\mmat{ARTATAR}+\mmat{ARATART}+\epsilon$ \label{alg:DenoA}
     \EndFor
     \State $\mat[i]{A}$ $\times$ = $NumA$/$DenoA$\label{alg:a_update}
     \State $\operatorname{broadcast}(\mat[i][l]{A})$ from diagonal ranks along rows to obtain $\mat[j][l]{A}$.  \label{alg:broadA}
 
    \EndFor
    
    \Ensure $\mat[i]{A} \in \mathbb{R}_{+}^{ \frac{n}{\sqrt p} \times r}$,$\ten{R} \in \mathbb{R}_{+}^{ m \times r \times r}$ , $\mat[j]{A} \in \mathbb{R}_{+}^{ r \times \frac{n}{\sqrt p}}$
     and $\mat{A},\ten{R} \approx \operatorname*{argmin}_{\widetilde{\mat{A}} \geq 0,\widetilde{\ten{R}} \geq 0} ||\ten{X}-\widetilde{\mat{A}} \widetilde{\ten{R}}\widetilde{\mat{A}}^T||_{F}^{2}$
    \end{algorithmic} 
\end{algorithm}

Our proposed distributed RESCAL algorithm is presented in Algorithm~\ref{alg:distrescal}. The data partitioning is carried out in a 2D square virtual grid as shown in Figure~\ref{fig:dist-nmf} such that each processor has it’s own local version of data $\ten[i,j][]{X}$ and factors $\mat[i][]{A}$ and $\mat[j][]{A}$. As illustrated, only the factor $\mat{A}$ is partitioned into a 1D grid whereas an exact copy of $\ten{R}$ is maintained across all the processors. $\ten{X}$ is distributed along a processor of grid $\sqrt p \times \sqrt p$ totaling to p processors. The resultant data $\ten{X}$, factors $\mat{A}$ and $\ten{R}$ are partitioned into $\frac{n}{\sqrt p} \times \frac{n}{\sqrt p} \times m$, $\frac{n}{\sqrt p} \times k$ and $k \times k \times m$ tensors respectively. When the data is sparse, $\ten{X}$ is stored in compressed sparse row(CSR) format. Despite the sparsity of $\ten{X}$, the factors $\mat{A}$ and $\ten{R}$ are dense.  For the diagonal ranks in the 2D virtual grid, $\mat[i][]{A}=(\mat[j][]{A})^T$.  In this data distribution scheme, the algorithm is primarily comprised of dedicated row and column sub-communications which reduces the overhead required for carrying out global communication. 

Algorithm~\ref{alg:distrescal} illustrates the proposed distributed non-negative RESCAL algorithm. The input to RESCAL is a data tensor $\ten{X}$ and decomposition rank $k$. Then, the factors $\mat[i][]{A}$, $\mat[j][]{A}$ and $\ten{R}$ are initialized randomly (line~\ref{alg:init} in Algorithm~\ref{alg:distrescal}) or with NNDSVD initialization. The update operation for factors $\mat{A}$ and $\ten{R}$ is performed in alternating fashion as shown in Algorithm~\ref{alg:distrescal} where the update of $\ten{R}$ is followed by the update of $\mat{A}$. As distributed matrix multiplication along a sub-communicator is performed repeatedly (see lines~ \ref{alg:ATA},\ref{alg:XA},\ref{alg:ATXA},\ref{alg:XTAR} in Algorithm~\ref{alg:distrescal}), the code is designed so that the distributed function calls can be reused at multiple instances for each update iteration. Algorithm~\ref{alg:distMMt} provides insight into the distributed matrix multiplication between two matrices $\mat{A}$ and $\mat{B}$ along a sub-communicator $comm$. Based on the distribution of $\mat{A}$ and $\mat{B}$, first the multiplication of the local blocks $\mat{A}^i$ and $\mat{B}^j$ is performed (as shown in line~\ref{alg:distmm_mul} from Algorithm~\ref{alg:distMMt}) and then, depending upon the type of sub-communicator(row/column), the all-reduce along the sub-communicator(as shown line~\ref{alg:distmm_row} for row-comm and line~\ref{alg:distmm_col} for col-comm from Algorithm~\ref{alg:distMMt}) is performed. For sparse data cases where either $\mat{A}$ or $\mat{B}$ is sparse, a sparse matrix multiplication is utilized instead of the regular dot operation to minimize the computation cost and memory requirements.

The updates of the factors $\mat{A}$ and $\ten{R}$ are performed with respect to a slice of data $\ten{X}$ along the relation dimension whose dimension is $m$. Algorithm~\ref{alg:distrescal} involves two loops, one loop corresponding to update iterations (see line~\ref{alg:outer_itr} from Algorithm~\ref{alg:distrescal}) and other loop corresponding to this slice-based update (see line~\ref{alg:inner_itr} from Algorithm~\ref{alg:distrescal}). Here, rather than directly performing tensor-based multiplication operations, we instead slice the tensor into matrices and then perform matrix operations to minimize the hardware computation cost and avoid communication bottlenecks while communicating the intermediate results which are larger for larger $m$. Therefore, to minimize the computation and communication burden for performing the factor update, we first distribute the data along the $\ten{X}$ entities (i.e $n\times n$ dimensions) with the 2D virtual grid and further locally partition the data along the $\ten{X}$ relations($m$ dimension).
One of the advantage for slicing $\ten{X}$ along $m$ dimension locally is to be able to utilize multithreading for updating the slices of $\ten{R}$ parallely. To illustrate from Algorithm~\ref{alg:distrescal}, the update of factors can be performed along two subsequent loops, one for the update of $\ten{R}$ and other for the update of $\ten{A}$ as these are independent of each other. With this split, lines \ref{alg:inner_itr}-\ref{alg:r_update} correspond to $\ten{R}$ update whereas line \ref{alg:inner_itr},\ref{alg:XART}-\ref{alg:a_update} correspond to $\mat{A}$ update. For $\ten{R}$ update, the steps involve computing the intermediate factors can be performed independently. Due to such independent nature of the operations, a multithreading can be used to accelerate these operations. 


Sparse operations involving $\ten{X}$ utilize sparse matrix multiplication where the resultant product is dense. Therefore, the communication requirements for intermediate factors in Algorithm~\ref{alg:distrescal} remain unchanged for sparse data compared to the decomposition of dense $\ten{X}$. Similarly, the decomposed factors $\mat{A}$ and $\ten{R}$ are still dense and hence, similar complexity analysis can be applied. 

Finally, the update algorithm also requires two broadcast operations (see lines~\ref{alg:broadXTAR},\ref{alg:broadA}) to establish $\mat[i][]{A}=(\mat[j][]{A})^T$ constraint for every update iteration. The alternating  updates of factors $\mat{A}$ and $\ten{R}$ are performed until the reconstruction error $||\ten{X}-\widetilde{\mat{A}} \widetilde{\ten{R}}\widetilde{\mat{A}}^T||_{F}^{2}$ is lower than threshold $\tau$. The complexity analysis for this distributed implementation is presented in 
\ref{sec:complexity_rescal}.

\subsection{Distributed Resampling}

Resampling is the process of perturbing each element of a tensor by multiplying with a small random noise from a uniform distribution. If such perturbation of the tensor is performed on distributed data, then it is called distributed resampling. The overview of distributed resampling is shown in Algorithm~\ref{alg:distperturb}. The resampling/perturbation of the tensor $\ten{X}$, represented as $\ten{X}^q$, is performed $r$ times with different random seeds for RESCALk(see line~\ref{line:rescalk_pert} from Algorithm~\ref{alg:rescalk}) such that the mean of these perturbations is $\ten{X}$. The variance of the noise($\delta$) in Algorithm~\ref{alg:distperturb} is chosen over a range [0.005,.03] based on the noise present in the $\ten{X}$ estimated by the reconstruction error. As from line~\ref{line:perturb}, the perturbation step involves the element-wise multiplication of $\ten[i,j][]{X}$ and $\Delta$, there is no communication involved among different MPI ranks for this process. Both the local (per MPI process) tensors $\ten{X}$ and  $\ten{X}^q$ are of size $\frac{n}{\sqrt{p}} \times \frac{n}{\sqrt{p}} \times m$ as shown in Figure~\ref{fig:dist-nmf} where a different random seed is used to generate uniform random numbers for different MPI processes.
For sparse $\ten{X}$, only the elements with non-zero values are perturbed in  line~\ref{line:perturb} of Algorithm~\ref{alg:distperturb} to retain sparsity of the unperturbed $\ten{X}$.

\begin{algorithm}[htp]
	\caption{$\ten{X'}$=$\operatorname{Perturb}$($\ten{X},\delta$): Distributed resampling (perturbation) of input matrix} \label{alg:distperturb}
	\begin{algorithmic}[1]
		\Require $\ten[i,j]{X} \in \mathbb{R}^{\frac{n}{\sqrt p} \times \frac{n}{\sqrt p} \times m}_{+}$, $\delta \in \mathbb{R}_+$
		\State Initialize $\mat{\Delta} \in \operatorname{uniformRandom}(\operatorname{dim}(\ten[i,j]{X}),\delta)$ \Comment{Generate a Uniformly distributed random tensor whose elements $\Delta_{a,b,c}\in [1-\delta,1+\delta]$} 
		
		\label{line:perturb_rand}
		\State $\ten[i,j]{{X'}} = \ten[i,j]{X} \odot \mat{\Delta}$ \Comment{Element-wise product} \label{line:perturb}
		\Ensure $\ten[i,j]{{X'}}\in \mathbb{R}^{\frac{n}{\sqrt p} \times \frac{n}{\sqrt p} \times m}_{+}$ resampled representative of $\ten[i,j]{X}$ 
	\end{algorithmic}
\end{algorithm}

\begin{figure}[htp]
    \centering
    \includegraphics[width=0.4\textwidth]{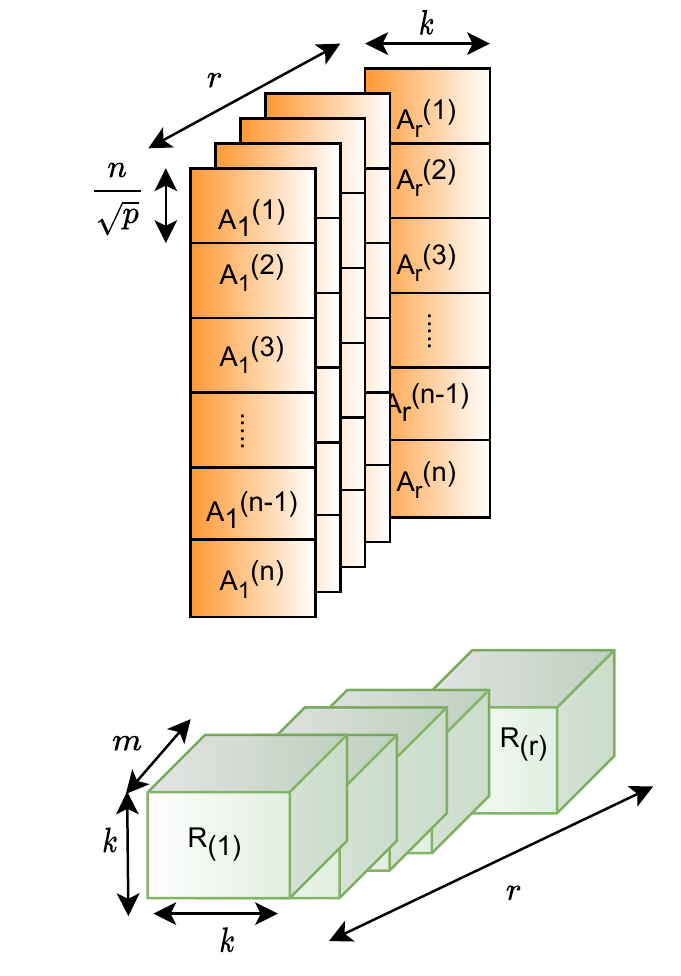}

    \caption{The distributed $\ten{A}$ and  $\ten{R^*}$ across $p$ processors for {\em r} perturbations, where $\ten[i]{A}$ is an $n/\sqrt{p} \times k \times r$ tensor and $\ten[i][:,:,q]{A}$ is an $n/\sqrt{p} \times k$ outer low-rank matrix corresponding to the $q^{th}$ resampling. Similarly, $\ten{R}$ is a tensor of  size $k \times k \times m \times  r$ and $\ten[i][:,:,:,q]{R^*}$ is a $k \times k \times m$ inner low-rank factor corresponding to the $q^{th}$ resampling.}\label{fig:wall}\vspace{-1em}
    \label{fig:dist_factors}

\end{figure}
\if 1

line 3: $k^2 n / sqrt(p)$
m times ( 
line 5: $k n^2 / p $
line 6: $k^2 n / sqrt(p)$
line 7: $k^3$
line 8: $k^3$
line 9: $k^2$
line10: $k^2 n / sqrt(p)$
line11: $n^2 k / p$
line12: $k^2 n / sqrt(p)$
line13: no computation
line14: $n k / sqrt(p)$
line15: $k^3$
line16: $k^2 n / sqrt(p)$
line17: $k^2 n / sqrt(p)$
line18: $k^3$
line19: $k^2 n / sqrt(p)$
line20: $n k / sqrt(p)$
 )
line 22: $n k / sqrt(p)$
line 23: no computation

\O{k^2 n / sqrt(p) + n k / sqrt(p) + m(k n^2 / p + k^2 n / sqrt(p) + k^3)}

\O{mkn^2}
\fi

\subsection{Distributed Clustering with Equal Cluster Size}
\label{sec:distclust}
The RESCAL decomposition is performed on the $r$ perturbations of the data $\ten{X}$ represented by $\ten{X}^q$ to  extract corresponding low-rank factors $\ten{A}_{:,:,q}$ and $\ten{R^*}_{:,:,:,q}$. These stored factor tensors $\ten{A}$ and $\ten{R}$ for a given low rank $k$
(see line~\ref{line:rescalk_rescal} in Algorithm~\ref{alg:rescalk}) are shown in Figure~\ref{fig:dist_factors}. Irrespective of the sparsity of $X$, the extracted low-rank factors $\mat{A}$ and $\ten{R}$ are dense.  The distributed outer low rank factors $\mat{A}$ are stored along a one-dimensional grid of $\sqrt p$ dimension so that each chunk is of size $n/\sqrt{p} \times k$. The inner low rank factor $\ten{R}$ is identically present throughout all MPI processes with size $n \times k \times r$.  Each of the $r$-slice of the collective tensor $\ten{A}$ and $\ten{R}$ correspond to factors $\ten{A}_{:,:,q}$  and $\ten{R^*}_{:,:,:,q}$ respectively of perturbation $\ten{X}^q$. In other words, each MPI process has an immediate access to $\ten[i]{A}$ with $n/\sqrt{p}\times k\times r$ dimensions and $\ten[i]{R^*}$ with size $k \times k \times n \times r$.

\begin{algorithm}[!ht]
the 	\caption{$\ten{A'} = \operatorname{customCluster}(\ten{A})$ -- distributed clustering to reorder the columns of factors.}\label{alg:distKmeans}
	\begin{algorithmic}[1]
		\Require $\ten[i]{A} \in \mathbb{R}^{\frac{n}{ \sqrt p}\times k\times r}_{+}$  are row-processor specific tensors
		\State $\mat[i]{M} = \ten[i][:,:,1]{A}$ \Comment{initialize centroid ($\mat{M}$) on $i^{th}$ processor} \label{line:kclust_centroid}
		\For{$iter$ \textbf{in} 1 to $n\_iter$} \label{line:kclust_loop_niter}
		
		\For{$q$ \textbf{in} 1 to $r$} \Comment{for each perturbation} \label{line:kclust5}
		
		\State $\ten[i][:,:,q]{D} = {\mat[i]{M}}^{T} \times \ten[i][:,:,q]{A}$ \Comment{
			compute partial similarity $\ten{D}$ on $i^{th}$ processor} \label{line:kclust_locdist}
		
		\EndFor \Comment{$\ten[i]{D}$ is a $k\times k\times r$ tensor per processor}
		
		\State $\ten{G} = \sum_{i=1}^{\sqrt p} \ten[i]{D}$ \Comment{ compute total similarity $\ten{G}$ on all processors with $\operatorname{all\_reduce}$}\label{line:kclust_globdist}
		
		\For{$q$ \textbf{in} 1 to $r$} \Comment{for each perturbation} \label{line:kclust_loop_perm}
		
		\State $\vect{porder}$ = $\operatorname{LSA}(\ten[][:,:,q]{G})$ \Comment{Linear Sum assignment operation to appropriately permute the columns}  \label{line:kclust_perm}
		\State $\ten[i][:,:,q]{A'} = \ten[i][:,:,q]{A}[\vect{porder}]$ \label{line:kclust_reorderW}
		\EndFor \label{line:kclust_loop_permend}
		\State $\mat[i]{\widetilde{A}}$ = $\operatorname{median}(\ten[i]{A'})$ \Comment{computes medians of clusters on $i^{th}$ processor} \label{line:medians}
		\State $\mat[i]{M} = \mat[i]{\widetilde{A}}$ \Comment{update centroid on $i^{th}$ processor} \label{line:kclust13}
		\EndFor \label{line:kclust_loopend_niter}
		\Ensure $\mat[i]{\widetilde{A}}$ $\in \mathbb{R}^{\frac{n}{\sqrt p}\times k}_{+}$, 
		distributed in $p$ row processors of each $\ten[][:,:,q]{A'}$, are clustered.
	\end{algorithmic}
\end{algorithm}

The distributed version of the custom clustering algorithm is presented in algorithm~\ref{alg:distKmeans}. This custom clustering is a variation of the k-medians clustering. The objective of this clustering is to rearrange the columns of $\mat{A}$ so that the latent communities obtained from different perturbations are aligned. With this clustering, the latent communities corresponding to both $\ten[i][:,:,q]{A}$ and $\ten[i][:,:,:,q]{R^*}$ are aligned locally for each MPI process. The clustering process starts with initializing  medoids  $\mat[i]{M}$, which are equal to the first perturbation solution (line~\ref{line:kclust_centroid}). Then a similarity tensor of size $k \times k \times r$ is constructed by computing the similarity between the centroid and each low rank factor (line~\ref{line:kclust_locdist}) locally followed by the all\_reduce operation to compute the similarity tensor $\ten{G}$ (line~\ref{line:kclust_globdist}). Once the similarity tensor $G$ is created, we use linear sum assignment (LSA) optimization to reorder the features of the $\ten{A}$ based on this similarity tensor (line~\ref{line:kclust_perm}). Given a $k \times k$ cost matrix LSA matches each row to different column in such a way that sum of corresponding entries is minimized. Alternately, this can also be generalized as an approach to select $k$ elements of cost matrix $C$ such that there exists a condition when only one element is present in each row and one in each column and, overall, the sum of corresponding cost is minimal. The unique requirement of the cost $G_{ij}$ to be non-negative is accomplished by our cost matrix $G$ as it corresponds to the distance between features. Once the features are permuted with this ordering (line~\ref{line:kclust_reorderW}), the new permuted set of features based on this order is obtained followed by the medoid computation (line~\ref{line:medians}). The median is then updated for the next iteration (line~\ref{line:kclust13}) and this reordering of the features is performed iteratively until the median doesn't change, which ensures the convergence of the clustering operation. It is important to note that the median is computed locally for each processor-specific  $\ten[i]{{A'}}$ tensor along the perturbation axis, which doesn't involve any partitioning and, hence, no communication is involved.

\subsection{Distributed Silhouette Statistics}
\label{sec:distcluststability}

The distributed implementation of the silhouette statistics is presented in Algorithm~\ref{alg:distSilhoutte} designed to quantify cluster stability. The idea is to find an optimal $k$, for which the cluster stability is maximum and the reconstruction error is minimum. The silhouette statistics computes $r$ feature data points corresponding to each perturbation and $k$ clusters. The stability of these clusters is quantified with a minimum \textit{silhouette width}, a scalar in the range [-1,1], where -1 corresponds to highly overlapping clusters/unstable clusters and 1 corresponds to disjoint stable clusters. To correctly estimate  $k$, the cluster stability analysis is performed on the outer low-rank factors ($\ten{A}$), which are the output of distributed clustering Algorithm~\ref{alg:distKmeans}.

\begin{algorithm}[htp]
	\caption{$\vect{s} = \operatorname{clusterStability}(\ten{A'})$ -- Distributed cluster stability through Silhouette analysis}\label{alg:distSilhoutte}
	\begin{algorithmic}[1]
		\Require The clustered left low-rank factors, $\ten[i]{A}$ $\in \mathbb{R}^{\frac{n}{\sqrt p} \times k\times r}_{+}$, distributed row-wise
		\For{$q$ \textbf{in} 1 to $k$} \label{line:sil_loopa} 
		\State $\mat[i]{U} = \ten[i][:,q,:]{A'}$ \Comment{gets a $\frac{n}{\sqrt p}\times r$ matrix for a given cluster}
		\State $\ten[i][:,:,q]{D} = {\mat[i]{U}}^T \times \mat[i]{U}$ \label{line:sil_locdist}
		\EndFor \Comment{$\ten[i]{D}$ is $r\times r\times k$ tensor of partial similarities on $i^{th}$ processor. } \label{line:sil_loopaend}
		\State $\ten{G} = \sum_{i=1}^{\sqrt p} \ten[i]{D}$ \Comment{ compute total similarity $\ten{G}$ on all processors with $\operatorname{all\_reduce}$} \label{line:sil_globdist}
		\State $\mat{I} = \operatorname{mean}(\ten{G},axis=1)$ \Comment{mean along one perturbation axis for $k$ clusters, $\mat{I}$ is $r \times k$}\label{line:sil_Imean}
		\For{$q$ \textbf{in} 1 to $k$} \label{line:sil_loopb} 
		\State $\mat[i]{U} = \ten[i][:,q,:]{A'}$ \Comment{gets $\frac{n}{\sqrt p} \times r$ matrix for a given cluster} \label{line:sil13}
		\For{$k_{\beta}$ \textbf{in} 1 to $k$} \label{line:sil14}
		\If{$q$ != $k_{\beta}$} \label{line:sil15}
		\State $\mat[i]{V} = \ten[i][:,k_{\beta},:]{A'}$ \Comment{gets $\frac{n}{\sqrt p} \times r$ matrix for a given cluster} \label{line:sil16}
		\State $\ten[i][:,:,k_{\beta}]{D} = {\mat[i]{U}}^T \times \mat[i]{V}$ \Comment{$r\times r\times k$} \label{line:sil_locDb}
		\EndIf
		\EndFor
		\State $\ten[][]{Z} = \sum_{i=1}^{\sqrt p} \ten[i][]{D}$ using $\operatorname{all\_reduce}$ \Comment{$i^{th}$ processor has $r\times r\times k$ tensor} \label{line:sil_globDb}
		\State $\mat{Y} = \operatorname{mean}(\ten[][]{Z},axis=1)$; \Comment{$r\times k$} \label{line:sil_Jmean}
		\State $\mat[][:,q]{J}$ = $\operatorname{min}(\mat{Y},axis=1)$;  \label{line:sil_Jmin}
		\EndFor \Comment{$i^{th}$ processor owns a copy of $\mat{J}$ of size $r\times k$}\label{line:sil_loopbend}
		\State \begin{math} \vect{s}_k = \operatorname{min}\biggl(\operatorname{mean}\biggl( \dfrac{\mat{J}-\mat{I}}{\operatorname{max}(\mat{J}, \mat{I})}\biggr)\biggr)\end{math}; $\vect{s}_k$ is minimum Silhouette width and ranges in $[-1, 1]$ \label{line:sil_avg}
		\Statex Similarly we can find the average Silhouette width
		\Ensure  $\vect{s}_k \in \mathbb{R}$
	\end{algorithmic}
\end{algorithm}

The cluster stability analysis involves computation of two different parameters, which are respectively: the average similarity of the features within a cluster ($\mat{I}$, see lines~\ref{line:sil_loopa}--\ref{line:sil_Imean}) and average dissimilarity between the clusters ($\mat{J}$, see lines~\ref{line:sil_loopb}--\ref{line:sil_Jmin}). In other words, metrics $\mat{I}$ and $\mat{J}$  correspond to, respectively, the average distance between all data points within a cluster and the smallest average distance between a data point belonging to a cluster and all other points corresponding to different clusters. The distance metric incorporated for this analysis is cosine distance for the computation of both $\mat{I}$ and $\mat{J}$. After $\mat{I}$ and $\mat{J}$ are computed, the final average silhouette value is computed for a given $k$(see line~\ref{line:sil_avg}).

The computation of $\mat{I}$ involves three major steps. First, we compute the local cosine similarity ($\ten[i]{D}$) between the data points for all the clusters (lines~\ref{line:sil_loopa}--\ref{line:sil_locdist}). Second, we compute the global similarity ($\ten[i]{G}$) from all processes (line~\ref{line:sil_globdist}) by performing MPI $\operatorname{all\_reduce}$ on $\ten[i]{D}$. This yields  a tensor of size $r\times r\times k$ for each MPI process. Finally, the mean of the tensor $\ten[i]{G}$ is computed to obtain $\mat{I}$ of size $r\times k$. The obtained $\mat{I}$ is computed to estimate cohesion between data points (line~\ref{line:sil_Imean}).

The computation of $\mat{I}$ is followed by the  the computation of $\mat{J}$. Similar to the compuation of $\mat{I}$, computing $\mat{J}$ also involves three steps. First, we compute the local cosine similarity  ($\ten[i]{D}$) between the current data point corresponding to a cluster and the rest of the data points corresponding to different clusters (lines~\ref{line:sil_loopb}--\ref{line:sil_locDb}). Second, we compute  the global similarity ($\ten{Z}$) from all processes (line~\ref{line:sil_globDb}) by performing MPI $\operatorname{all\_reduce}$ on $\ten[i]{D}$. This operation yields $\ten{Z}$ which is of the size $r\times r\times k$ for each MPI process. Finally, we compute the mean  of  $\ten{Z}$ to obtain $\mat{Y}$(line~\ref{line:sil_Jmean}).We then find the minimum of $\mat{Y}$ to estimate $\mat{J}$ (line~\ref{line:sil_Jmin}). The computed $\mat{J}$ is of size $r\times k$. Finally, the silhouette width is computed based on the metrics $\mat{I}$ and $\mat{J}$ (line~\ref{line:sil_avg}), which ranges [-1,1]. As observed from line~\ref{line:sil_avg}, the computation of minimum silhouette statistics requires a minimum operation on the average of ratio of difference of $\mat{J}$ and  $\mat{I}$ i.e.($\mat{J}$-$\mat{I}$) and the maximum i.e $\operatorname{max}(\mat{J}, \mat{I})$. Similarly, the average silhouette statistic can be computed by replacing the minimum operator in line~\ref{line:sil_avg} by the mean operator.  
In addition to this silhouette metric, the reconstruction error for the median outer low rank factor $\mat{A}$ and equivalent inner low rank factor $\ten{R}$ is used to determine overall optimal latent feature count $k_{opt}$.

\section{Cost Analysis of distributed RESCALk}	
\label{sec:complexity}

We break up the analysis of distributed RESCALk and individually evaluate the two main components, RESCAL, described in Algorithm~\ref{alg:distrescal}, and the clustering and silhouette computation, which are described in Algorithm~\ref{alg:distKmeans} and Algorithm~\ref{alg:distSilhoutte}. For both of these components, we report both the memory complexities and the time complexities for computation and communication operations.

\subsection{Analysis of distributed RESCAL}
\label{sec:complexity_rescal}
\subsubsection{Computation Complexity}
\label{subsubsec:compute_rescal}
Here, we estimate the computational complexity of Algorithm~\ref{alg:distrescal} by analyzing its individual lines. The computational complexity for calculating $\mat{A}^T\mat{A}$ on line~\ref{alg:ATA}  is $\O{\frac{n}{\sqrt{p}}k^2}$.  Line~\ref{alg:XA} computing $\ten{X}\mat{A}$ is $\O{m\frac{n^2}{p}k}$ for the dense and $\O{m\delta\frac{n^2}{p}k}$ for the sparse case, where $\delta$ is the density of $\ten{X}$. 
Similarly, the computational complexity of line~\ref{alg:ATXA}, line~\ref{alg:XART}, and line~\ref{alg:XTAR} is $\O{m\frac{n}{\sqrt{p}}k^2}$. Element-wise multiplication to update \ten{R} on line~\ref{alg:Rt} is $\O{mk^3}$. The complexity involved in calculating NumA on line~\ref{alg:NumA} and DenoA on line~\ref{alg:DenoA} is $\O{m\frac{n}{\sqrt{p}}k}$. Thus, the total computational complexity for RESCAL algorithm presented in Algorithm~\ref{alg:distrescal} per update  iteration($l$) is $\O{m\frac{n^2}{p}k}$ for dense $\ten{X}$ and $\O{m\delta\frac{n^2}{p}k}$ for sparse $\ten{X}$ since $k \leq n$. For the total RESCAL update iterations $max\_iters$, the overall complexity is obtained  as $\O{max\_iters(m\delta\frac{n^2}{p}k)}$ where $\delta$ is the density and $\delta=1$ for dense $\ten{X}$. 

\subsubsection{Communication Complexity}
\label{subsubsec:rescal_comm}
In this section, we estimate the communication complexities for the 2D distributed scheme of multiplicative updates used in our pyDRESCALk implementation. As the communication factors for both sparse and dense $\ten{X}$ are dense, the equivalent complexity analysis can be applied to both cases.There are 4 occurrences of all\_reduce in Algorithm~\ref{alg:distrescal}, and two occurrences of broadcast, all of which are done between $\sqrt{p}$ processors. The all\_reduce on line~\ref{alg:ATA} and line~\ref{alg:ATXA} are done on $k \times k$ matrices, while the all\_reduce on line~\ref{alg:XA} and line~\ref{alg:XTAR} are on $\frac{n}{\sqrt{p}} \times k$ matrices. The broadcasts on line~\ref{alg:broadXTAR} and line~\ref{alg:broadA} are both on size $\frac{n}{\sqrt{p}} \times k$ matrices. Thus, the communication complexity for  Algorithm~\ref{alg:distrescal} per RESCAL update iteration is $\O{mk\frac{n}{\sqrt{p}} \log(p) }$ and overall $max\_iters$ is $\O{max\_iters(mk\frac{n}{\sqrt{p}} \log(p))}$ based on~\cite{chan2007collective}. 

\subsubsection{Memory Analysis}
\label{subsubsub:memory_rescal}
Here, we evaluate the memory requirements for the intermediate factors of the distributed RESCAL algorithm. The space bound for the local data chunk $\ten[i][]{X}$, the local factors $\mat[i][]{A}$ and $\ten[i][]{R}$ are respectively  $\frac{mn^2}{p}$,$\frac{nk}{\sqrt p}$, and $mk^2$ words. For $r$ different perturbations, the space bound for the ensemble of local factors is $\frac{rnk}{\sqrt p}$ and $rmk^2$ words. Since, for each decomposition, we need to store the outer low rank factor $\mat{A}$, the space bound corresponding to this factor would be then $rk\frac{n}{\sqrt p}$ words. Finally, while the overall memory requirement for the tensor is $\O{\frac{mn^2}{p}}$, the memory requirement for the factors is linear in both $m$ and $n$ at $\O{\frac{mkrn}{\sqrt p}}$ words.

\subsection{Clustering}
\label{sec:cluster_complex}
\subsubsection{Computation Complexity}
\label{subsubsec:clustering_comp}
Clustering, in line~\ref{line:nmfk_cluster} of Algorithm~\ref{alg:rescalk}, is done with a call to Algorithm~\ref{alg:distKmeans}. 
The custom clustering algorithm involves computing products of pairs of matrices sized $k\times \frac{n}{\sqrt p}$ and $\frac{n}{\sqrt p} \times k$ in line~\ref{line:kclust_locdist} leading to a computational time complexity of  $\O{\frac{k^2nr}{\sqrt p}}$ for $r$ perturbations. On the other hand, the computational complexity of the LSA in line~\ref{line:kclust_perm} is $\O{k^3}$~\cite{burkard2012assignment}. Based on the ordering index achieved by LSA, the time taken to permute the columns of $\ten[i]{A}$ (line~\ref{line:kclust_reorderW}) 
is $\O{\frac{n}{\sqrt p}k}$. Further, the computation of median in line~\ref{line:medians} takes $\O{\frac{n}{\sqrt p} k r \log{r}}$ time. So with all these complexity components, the overall computational complexity of distributed custom clustering algorithm is $\O{k^2r \frac{n}{\sqrt p} \log{r}}$ per iteration of the clustering. 

Further, we analyze the complexity associated with Algorithm~\ref{alg:distSilhoutte}, which is the distributed silhouette analysis. The algorithm starts by computing $k$ similarity matrices of size $r\times r$ requiring $\O{\frac{r^2nk}{\sqrt p}}$ time with the execution of lines~\ref{line:sil_loopa}--\ref{line:sil_loopaend}. This is then followed by the computation of the mean (line~\ref{line:sil_Imean}), which requires $\O{r^2k}$ time. Then, another set of similarity matrices are computed, which includes computing $k$ similarity matrices of size $r\times r$ requiring $\O{\frac{r^2nk^2}{\sqrt p}}$ time with the execution of lines~\ref{line:sil_loopb}--\ref{line:sil_loopbend}. This is then followed by the computation of a second mean (line~\ref{line:sil_Jmean}), which requires $\O{r^2k^2}$ time and a minimum operation (line~\ref{line:sil_Jmin}), which takes $\O{k^2r}$ time. Hence, the overall time complexity of the cluster stability algorithm is $\O{r^2k^2\frac{n}{\sqrt p}}$. 
\subsubsection{Communication Complexity}\label{subsubsec:clustering_comm}
The time taken for $\operatorname{all\_reduce}$ (line~\ref{line:kclust_globdist}) reduction operation of Algorithm~\ref{alg:distKmeans} across $\sqrt p$ processors on $k\times k\times r$ tensor is $\O{k^2 r \log p}$ per iteration. 
The total communication complexity for Algorithm $\operatorname{customCluster}$ would be  $\O{k^2 r \log p}$ as the algorithm comprises of only one $\operatorname{all\_reduce}$.

Algorithm~\ref{alg:distSilhoutte} ($\operatorname{clusterStability}$) requires two  $\operatorname{all\_reduce}$ operations (line~\ref{line:sil_globdist} and line~\ref{line:sil_globDb}). However, the second all\_reduce operation dominates due to its $k$ times execution.
Hence, the total communication complexity of   Algorithm~\ref{alg:distSilhoutte} is $\O{k^3 r^2 \log p}$

Adding these communication bounds, the total communication complexity for  clustering and stability analysis is $\O{k^3 r^2 \log p}.$
\subsubsection{Memory Analysis}
\label{subsubsub:memory_clust}

The memory analysis of the factors for RESCAL is presented in subsection~\ref{subsubsub:memory_rescal}. The clustering algorithm requires memory to store the temporary matrices   $\ten{G}$ and $\ten{Z}$ (in Algorithm~\ref{alg:distKmeans}), which requires $\O{r^2k}$ words, and  $\mat{I}$ and $\mat{J}$ (in Algorithm~\ref{alg:distSilhoutte}) require $\O{rk}$ words. As these temporary matrices are of same dimensions throughout all the MPI processes, there is no requirement to normalize by the size of the grid. i.e., $\sqrt p$. Adding these memory requirement components, we obtain an overall clustering space complexity of 
$\O{r^2k}.$

\subsection{Compexity of distributed pyDRESCALk}
\subsubsection{Computation Complexity}
\label{subsubsec:compute_rescalk}



The function $\operatorname{Perturb}$ in Algorithm~\ref{alg:distperturb} (line~\ref{line:rescalk_pert} in Algorithm~\ref{alg:rescalk}) takes $\O{\frac{mn}{p}}$ time per call and $\O{\frac{mnr}{p}}$ time for $r$ perturbations. 
Adding together time bounds for Algorithms~\ref{alg:distrescal},\ref{alg:distperturb},\ref{alg:distKmeans} and \ref{alg:distSilhoutte} from sections~\ref{subsubsec:compute_rescal} and \ref{subsubsec:clustering_comp} and taking into account that $k \leq n$, with a fixed number of convergence iterations, we get a bound on the time for one $k$-iteration of Algorithm~\ref{alg:rescalk} as
$$\OO{\frac{mn^2r^2k}{p}}.$$

\subsubsection{Communication cost}
Adding the communication cost associated with  RESCAL (\ref{subsubsec:rescal_comm}) and clustering (\ref{subsubsec:clustering_comm}) operations, we get an overall communication cost bound of 
\begin{align*}
\centering
\begin{split}
    \OO{\frac{mnr^2k^2}{\sqrt p} \log p}.
\end{split}
\end{align*}



\subsubsection{Memory analysis}
Adding the memory requirements for  RESCAL and clustering steps from sections~\ref{subsubsub:memory_rescal} and \ref{subsubsub:memory_clust}, we get a memory bound per processor of $$\OO{\frac{mknr^2}{\sqrt p}}.$$ 

\input{Isoefficiency}

\section{Experiments}
\label{sec:experiments}

In this section, we present details of the experimental setup,  results,  and analysis. We demonstrate the efficacy and efficiency of the proposed distributed pyDRESCALk  via correctness and  scalability tests. To demonstrate the correctness, we synthetically generate 100 tensors with predetermined values for the latent dimension $k$ and show that we can correctly determine these values with the proposed framework. In addition to this, we also validate the result on a well-known real world relational dataset. Next, to demonstrate the scalability of this framework, we perform strong, weak, and k-scaling experiments with both CPU and GPU hardware using both dense and sparse synthetic datasets.

\subsection{Runtime Specifications}
\subsubsection{Hardware Specifications}

We perform benchmark tests on two different HPC clusters to illustrate the portability and scalability of \textit{pyDRESCALk}. We use two Los Alamos National Laboratory (LANL) HPC clusters, Grizzly and Kodiak, to perform scaling experiments.
For scaling experiments on CPUs only, we utilize the Grizzly cluster, while, for scaling experiments utilizing GPU accelerators, we utilize the Kodiak cluster. 

Grizzly has 1490 compute nodes each with a Intel Xeon Broadwell (E5-2695v4) processor, which is a $18$-core dual socket Ivy Bridge processor. Each processor within a node has $2.1$ GHz clock speed with caches $L1$ and $L2$ with memory $64$KB and $256$KB, respectively. Each node comprises of $128$GB memory and an Intel OmniPath interconnect with fat-tree topology. Grizzly is an institutional computing  cluster ideal for performing benchmarking experiments with its 53640 total CPU cores, 190.7TB total cluster memory and 1.8 Pflops of peak operating speed. This provides a platform to perform record benchmark experiments.

Kodiak has 133 compute nodes with dual Xeon E5-2695 v4 CPUs and four NVIDIA Pascal P100 GPGPUs each. Each NVIDIA Pascal P100 GPGPU has 16GB VRAM and uses PCIE 16X Gen3 Links. The cluster peaks at  1850TF/s and uses an nfiniBand band interconnect. 

\subsubsection{Software Specifications}
The software implementation of the proposed distributed framework is written in Python and labelled pyDRESCALk, which is built upon the state of the art pyDNMFk library\footnote{\hyperlink{https://github.com/lanl/pyDNMFk}{https://github.com/lanl/pyDNMFk}}~\cite{pyDNMFk}. pyDRESCALk depends on MPI4PY~\cite{MPI4PY} (a Python MPI library), Numpy, CuPy and sklearn.pyDRESCALk supports both dense and sparse matrices for both CPU and GPU platforms. The CPU version operates with  Numpy backend for dense and Scipy Sparse for sparse tensors whereas the GPU version operates with CuPy backend for both dense and sparse tensors.   We use the Python 3.8 and OpenMPI (v2.1.2) library available on the cluster.


\begin{figure*}[htb]
  \begin{subfigure}{0.5\textwidth}
    \centering
    \includegraphics[width=\textwidth]{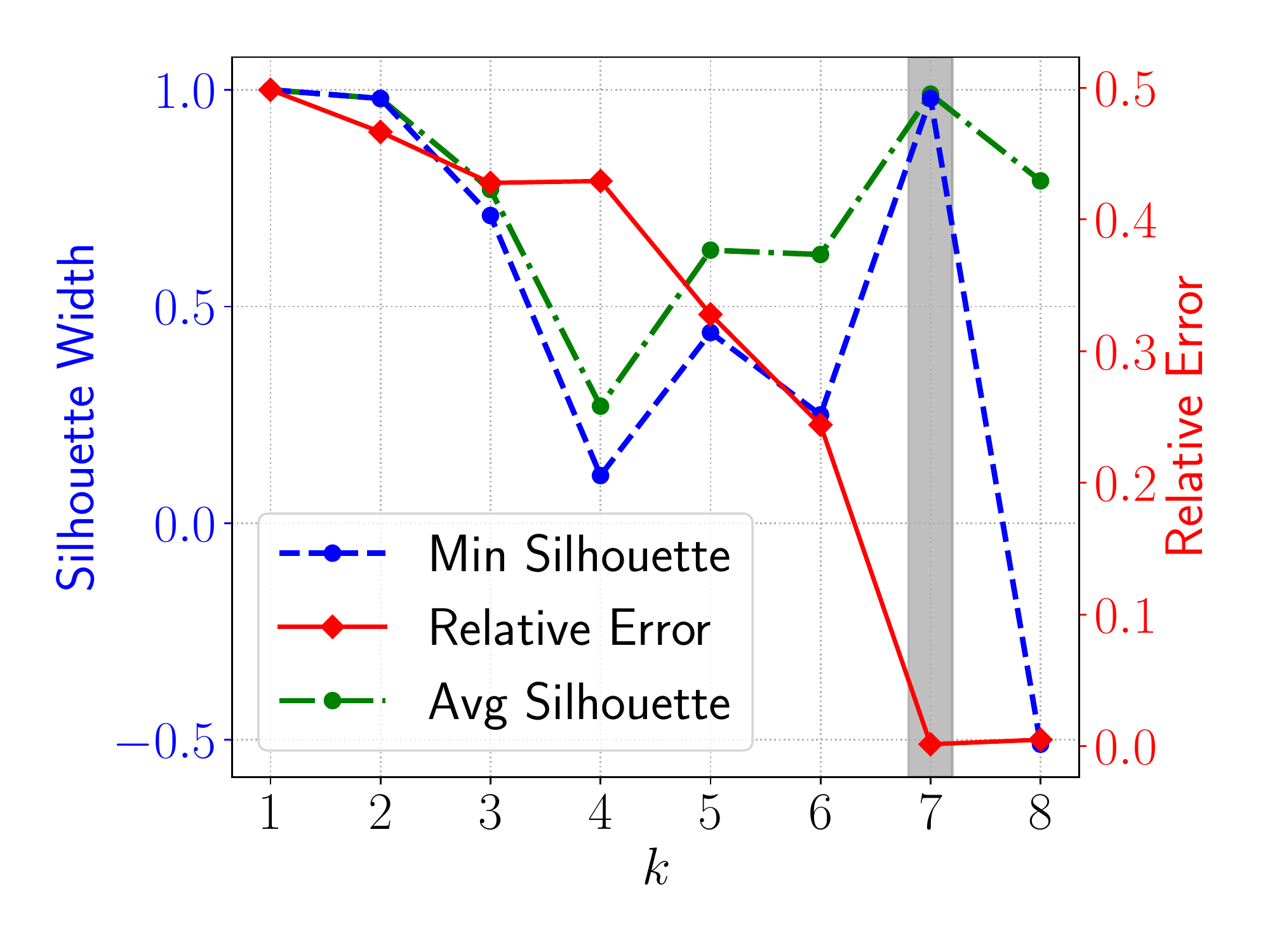}
    \caption{Feature identification on synthetic Data~1 ($k=7$). }
    \label{fig:sill_data1}
  \end{subfigure}%
  \begin{subfigure}{0.5\textwidth}
    \centering
    \includegraphics[width=\textwidth]{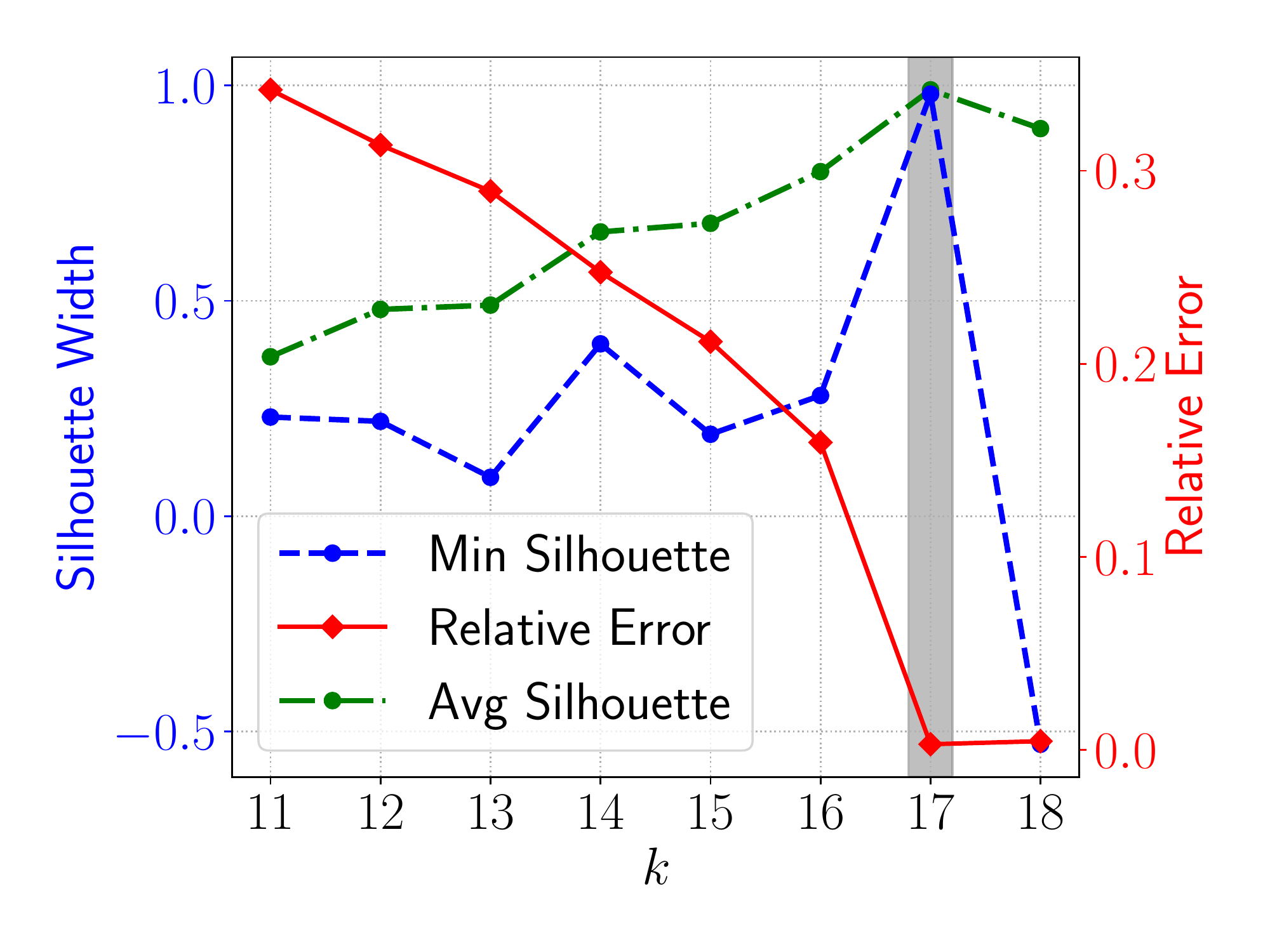}
    \caption{Feature identification on synthetic Data~2 ($k=17$)}
    \label{fig:sill_data2}
  \end{subfigure}
  

   \begin{subfigure}{0.5\textwidth}
    \centering
    \includegraphics[width=\textwidth]{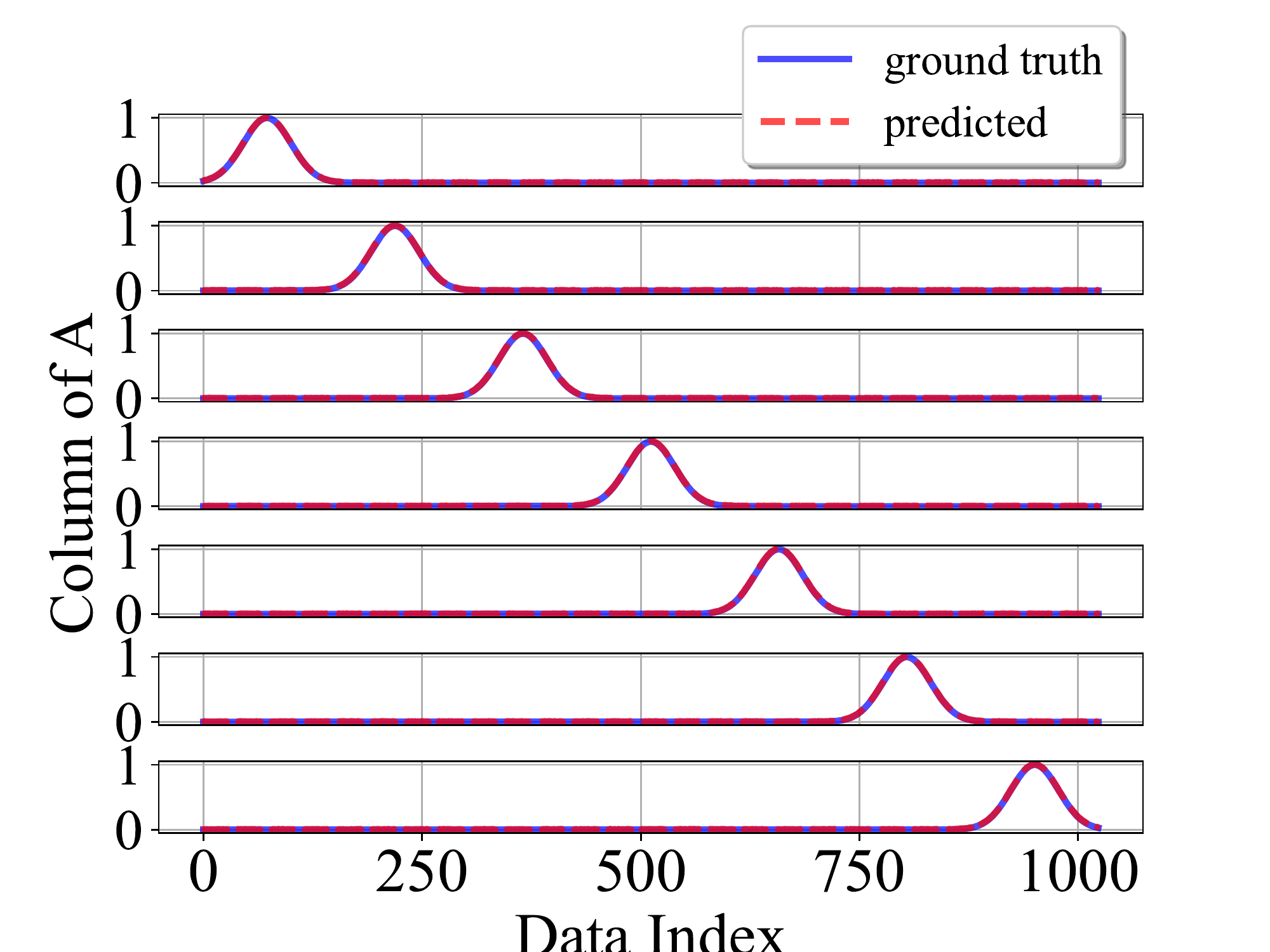}
    \caption{Feature reconstruction for Data~1. }
    \label{fig:feat_data1}
  \end{subfigure}%
  \begin{subfigure}{0.5\textwidth}
    \centering
     \includegraphics[width=\textwidth]{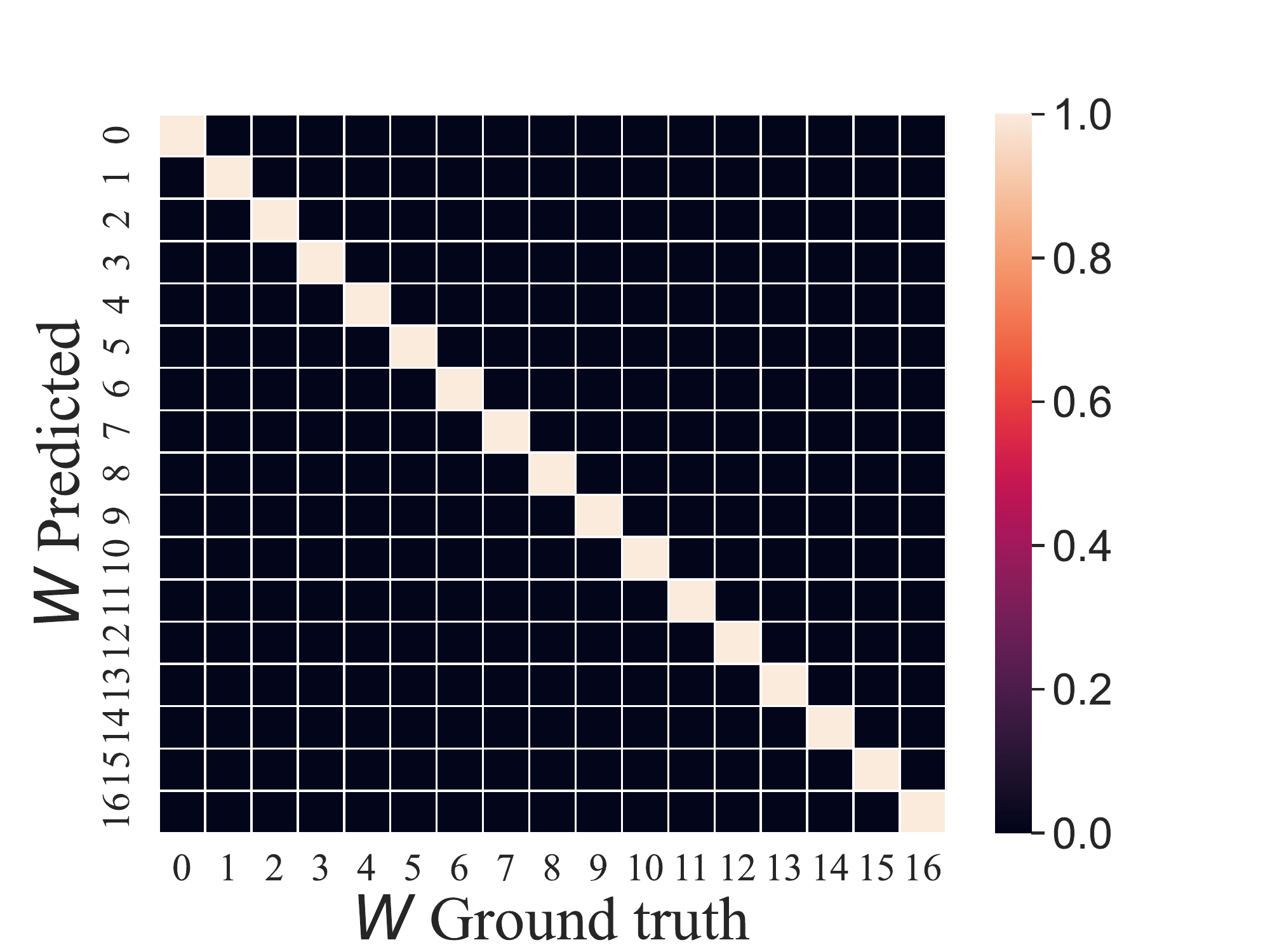}
    \caption{Pearson correlation matrix for features of Data~2.}
    \label{fig:feat_data2}
  \end{subfigure}

 \caption{Results on Synthetic dataset}
 \label{fig:synthetic}
\end{figure*}

\subsubsection{Parameter Settings}

The algorithmic  parameters are set as follows. We use Frobenius norm based multiplicative updates for RESCAL decomposition in all the experiments. Also, we utilize 1) a random initialization of both $\mat{A}$ and $\ten{R}$ using a different seed at each perturbation, or 2) a custom NNDSVD-based initialization, where we perform NNDSVD-based decomposition of concatenated unfoldings of $\ten{X}$ along axis 1 and 2 to obtain matrix $\mat{A}$ and then utilize $\ten{R}$ update steps (lines~ \ref{alg:outer_itr}-\ref{alg:r_update}) from Algorithm~\ref{alg:distrescal} to obtain the corresponding $\ten{R}$. During the resampling stage, each MPI process generates a unique seed, which is a function of its MPI rank to produce a random matrix for the randomly initialized factors and also to generate the noise for perturbing the input data. Here, because of the design constraints, we ensure $p_r = p_c$ so that the input data is distributed symmetrically along the $2D$ grid. To the best of our knowledge, {\em{pyDRESCALk}} is the first large-scale non-negative implementation of RESCAL tensor factorization with ability to estimate the number of latent features in both dense and sparse extra-large data on CPU/GPU heterogeneous architectures.

\subsection{Model selection}
\label{sec:modeldet}
\subsubsection{Synthetic Data}
\label{sec:datagen}

To demonstrate the correctness of the proposed distributed framework, we generate synthetic relational data with known latent features and prove the efficacy of the framework by comparing the determined features with the ground-truth ones. To achieve this, we generate 100 different data tensors $\ten{X}$ with dimension ${n \times n \times m}$ and different values for $n$, $m$,  and the latent dimension $k$. Before generating a data tensor $\ten{X}$, we begin with generating its corresponding latent feature matrix $\mat{A}$, whose columns correspond to random vectors with Gaussian distributions with a given mean and variance. To study the robustness of latent feature extraction with the RESCALk framework, we synthesize these features with variable inter-feature correlation by manipulating the mean and variance of the Gaussian features. Furthermore, once these features are generated, they are then multiplied with a tensor $\ten{R}$ generated with an exponential distribution with scale 1 to produce a tensor $\ten{X}^0=\mat{A}\ten{R}\mat{A}^T$. Finally, a noise tensor $\ten{D}$  with uniformly distributed elements over interval $[-.01\ten{X},.01\ten{X}]$ i.e with zero mean and 10\% variance is added to $\ten{X}^0$.
The final tensor is computed as $\ten{X}=\ten{X}^0+\ten{D}$. Each test tensor $\ten{X}$ generated in this way has dimensions either $64 \times 64 \times 128$, $128 \times 128 \times 32$, $512 \times 512 \times 10$, $1024 \times 1024 \times 20$, $2056 \times 2056 \times 25$, or  $128 \times 128 \times 128$. The 2D virtual processor grid utilized for running these experiments was chosen over one of the grid configurations, $2\times 2, 3\times 3,4\times 4$ or $8\times 8$. 

The values for $k$ used to generate matrices $\mat{A}$ were randomly sampled from $2$ to $32$. Then, to estimate the number of latent components from the these datasets, we performed pyDRESCALk for different $k$ ranges with 30 perturbations(r)  and 1000 RESCAL iterations with randomly initialized factors $\mat{A}$ and $\ten{X}$. Once the RESCALk was performed with pyDRESCALk to compute a decomposition $\ten{X}\approx\tilde{\mat{A}}\tilde{\ten{R}}\tilde{\mat{A}}^T$, the resultant $\tilde{\mat{A}}$ corresponding to $k_{opt}$ was compared against $\mat{A}$, using Pearson correlation coefficient \cite{benesty2009pearson} to estimate the accuracy. 

\begin{figure*}[htp]
     \begin{subfigure}{0.5\textwidth}
    \centering
    \includegraphics[width=\linewidth]{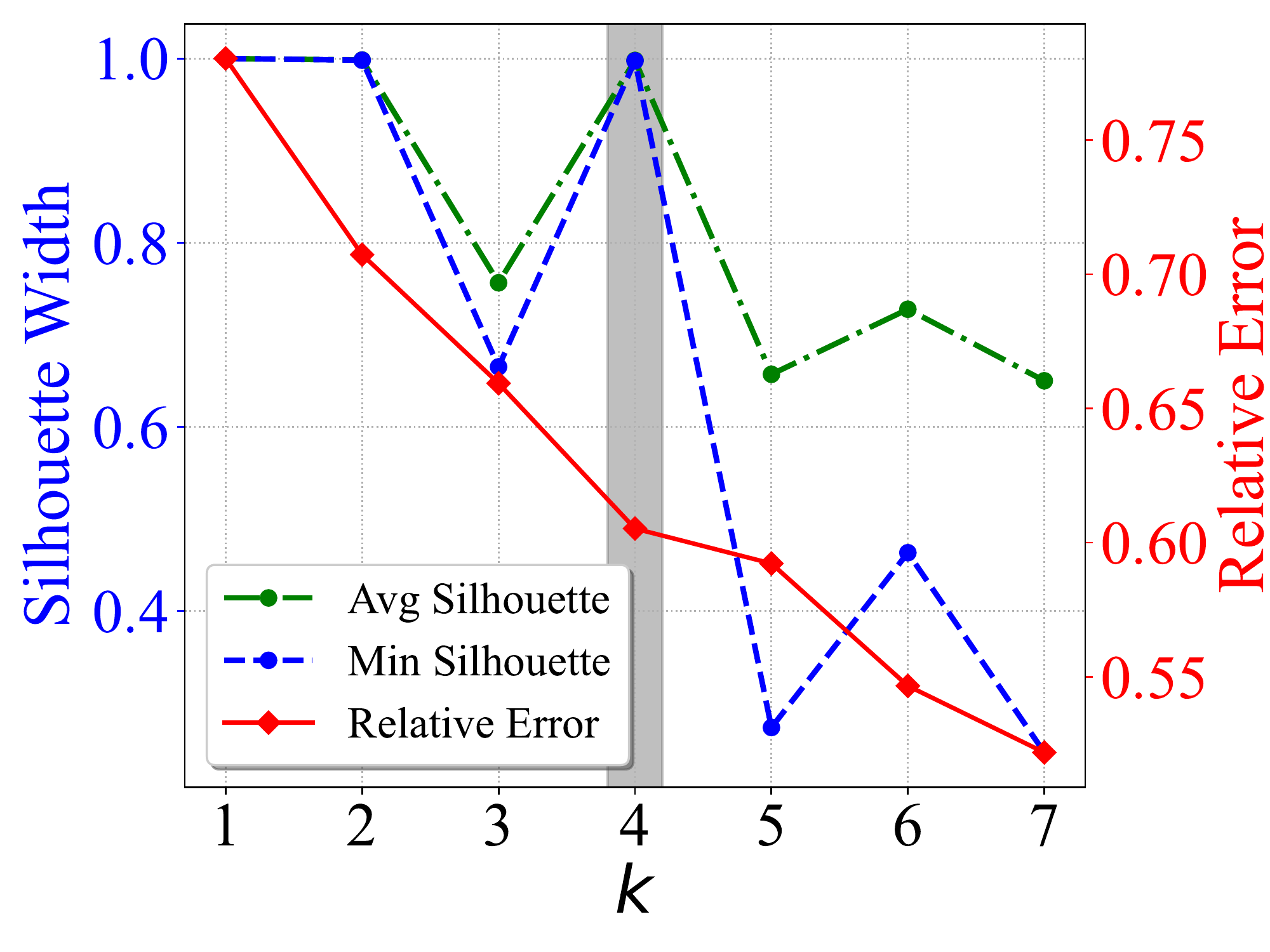}
    \caption{Feature identification on \emph{Nations} dataset ($k=4$). }
    \label{fig:sill_nations}
  \end{subfigure}%
  \begin{subfigure}{0.5\textwidth}
    \centering
    \includegraphics[width=\linewidth]{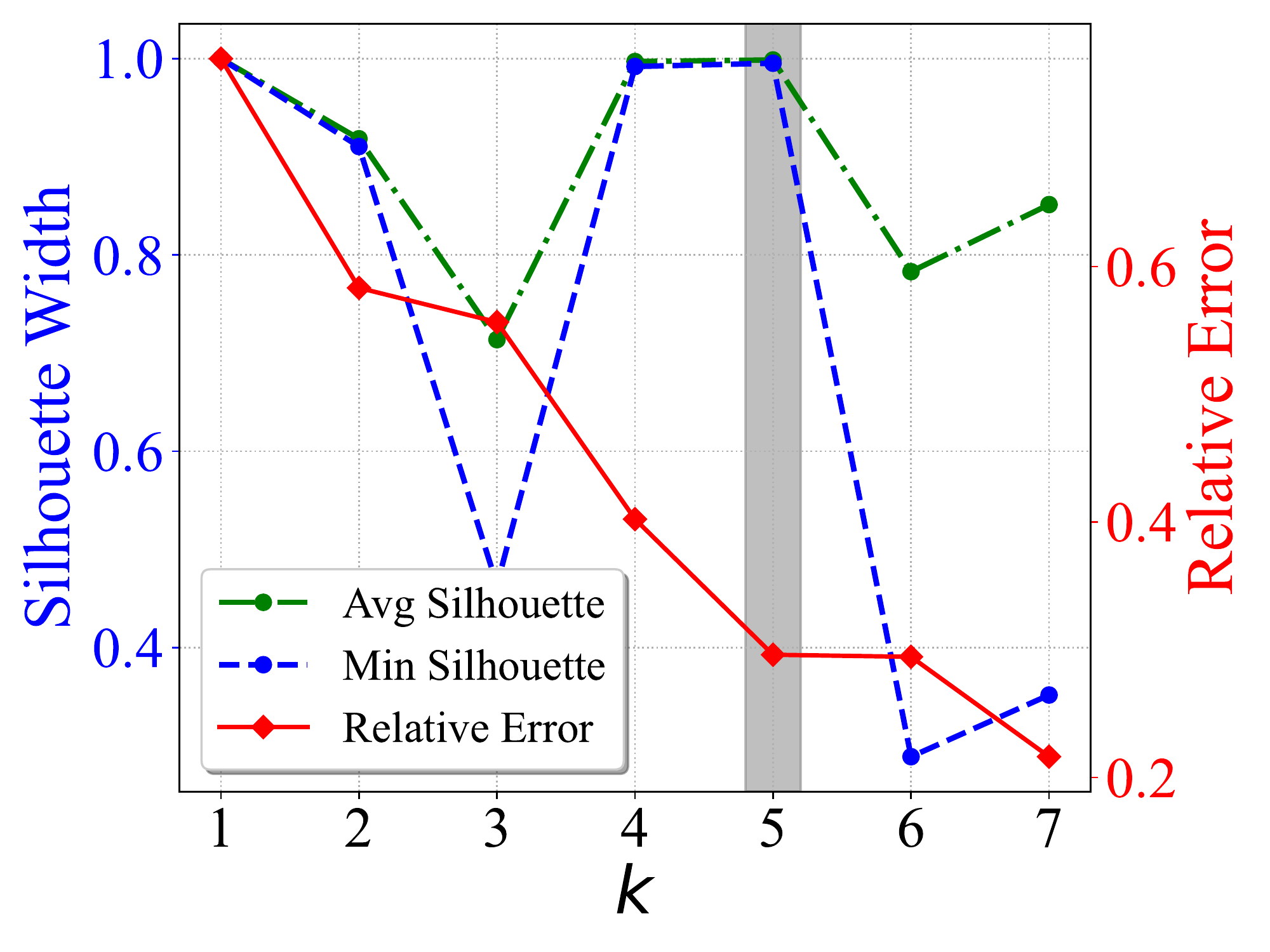}
    \caption{Feature identification on \emph{Trade} dataset ($k=5$). }
    \label{fig:sill_trade}
  \end{subfigure}%

  \begin{subfigure}{0.5\textwidth}
    \centering
    \includegraphics[width=\textwidth]{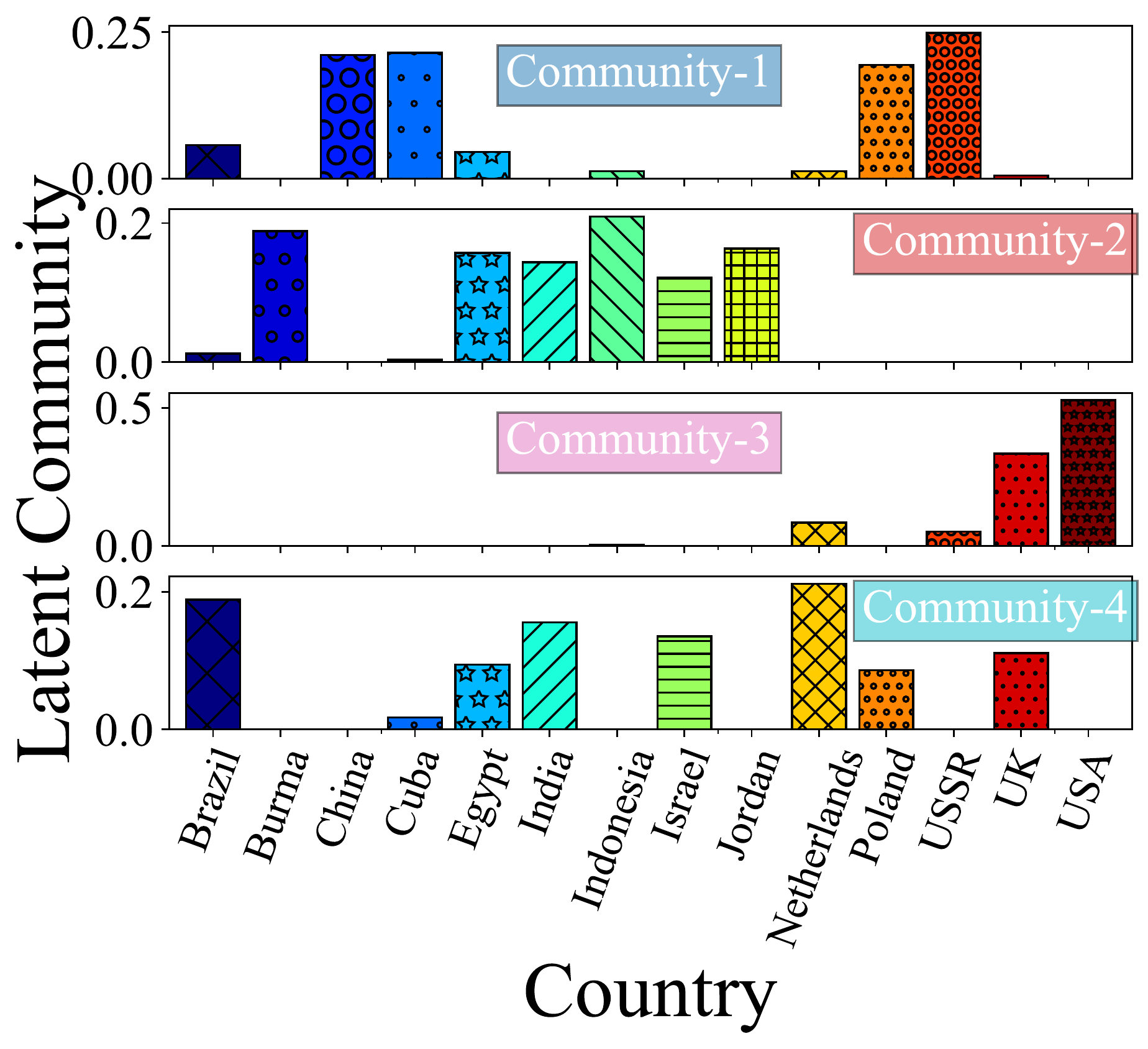}
    \caption{Extracted latent communities in \emph{Nations} dataset}
    \label{fig:feat_nations}
  \end{subfigure} 
  \begin{subfigure}{0.5\textwidth}
    \centering
    \includegraphics[width=\textwidth]{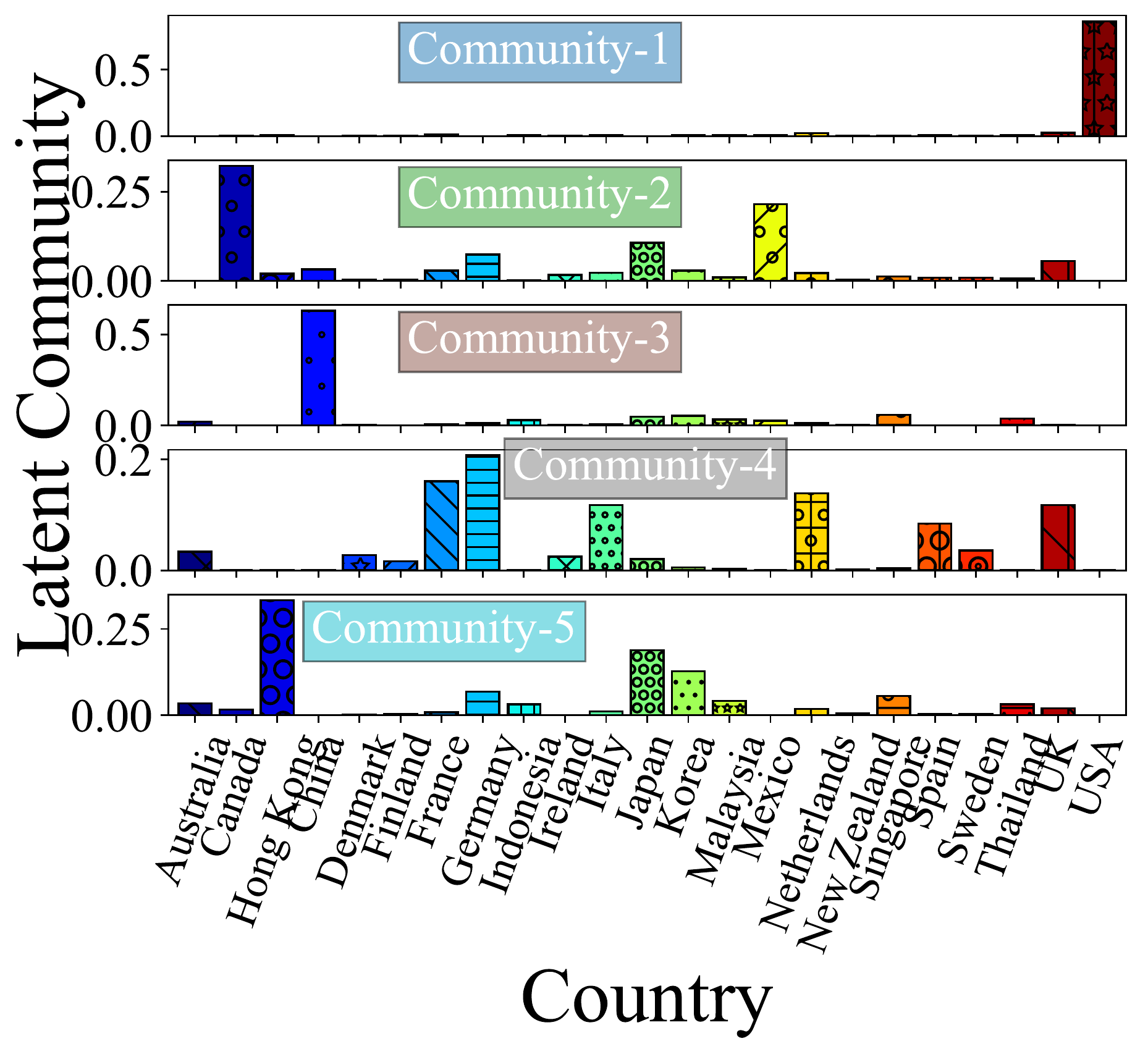}
    \caption{Extracted latent communities in \emph{Trade} dataset}
    \label{fig:feat_trade}
  \end{subfigure}

       \begin{subfigure}{0.5\textwidth}
    \centering
    \includegraphics[width=\linewidth]{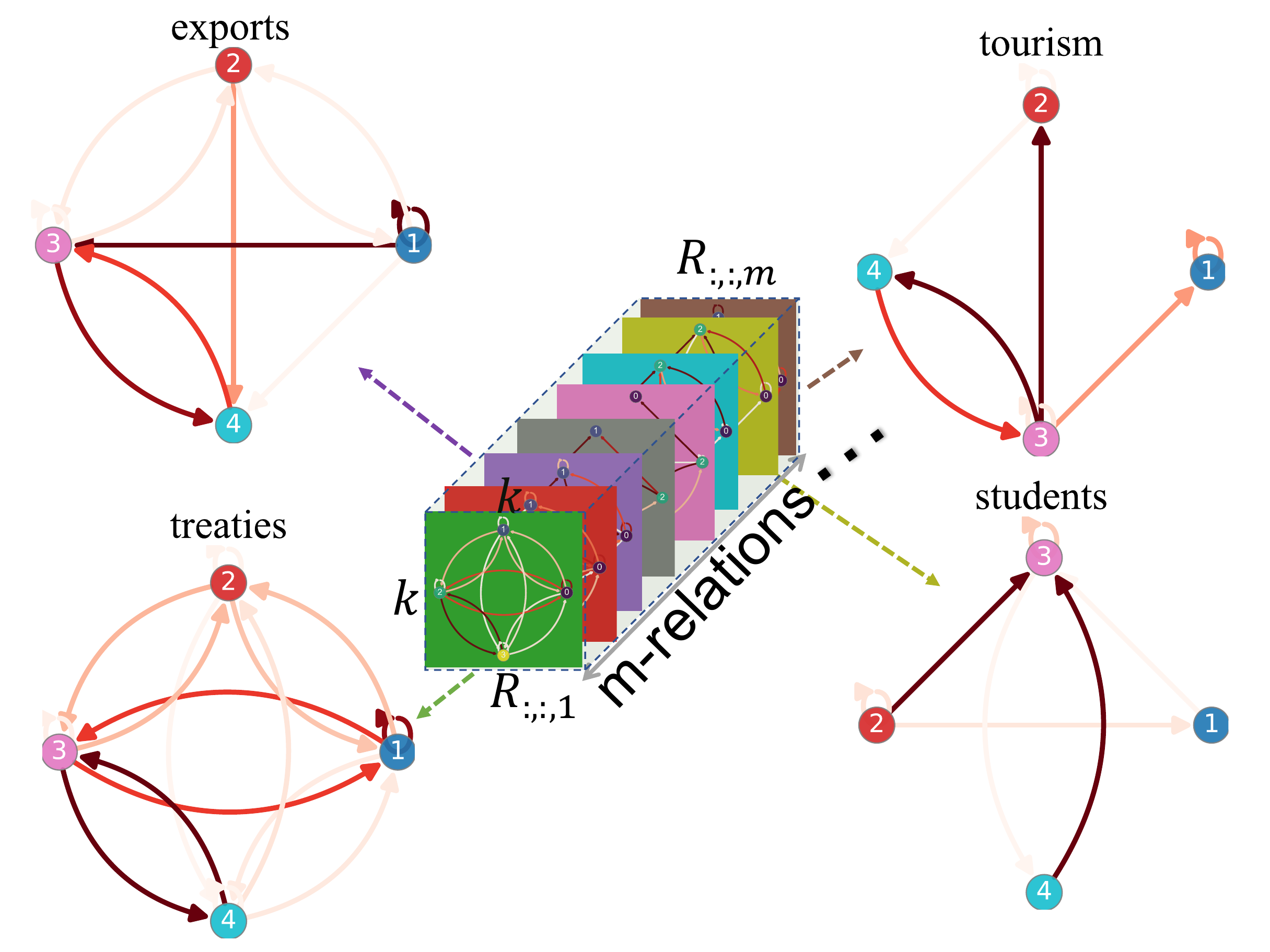}
    \caption{Graphical representation of Interaction between latent \\ communities in  \emph{Nations} dataset for relations \textit{exports}\\ \textit{tourism, treaties} and \textit{students}. }
    \label{fig:nations_graph}
  \end{subfigure}%
      \begin{subfigure}{0.5\textwidth}
    \centering
    \includegraphics[height=6.5cm]{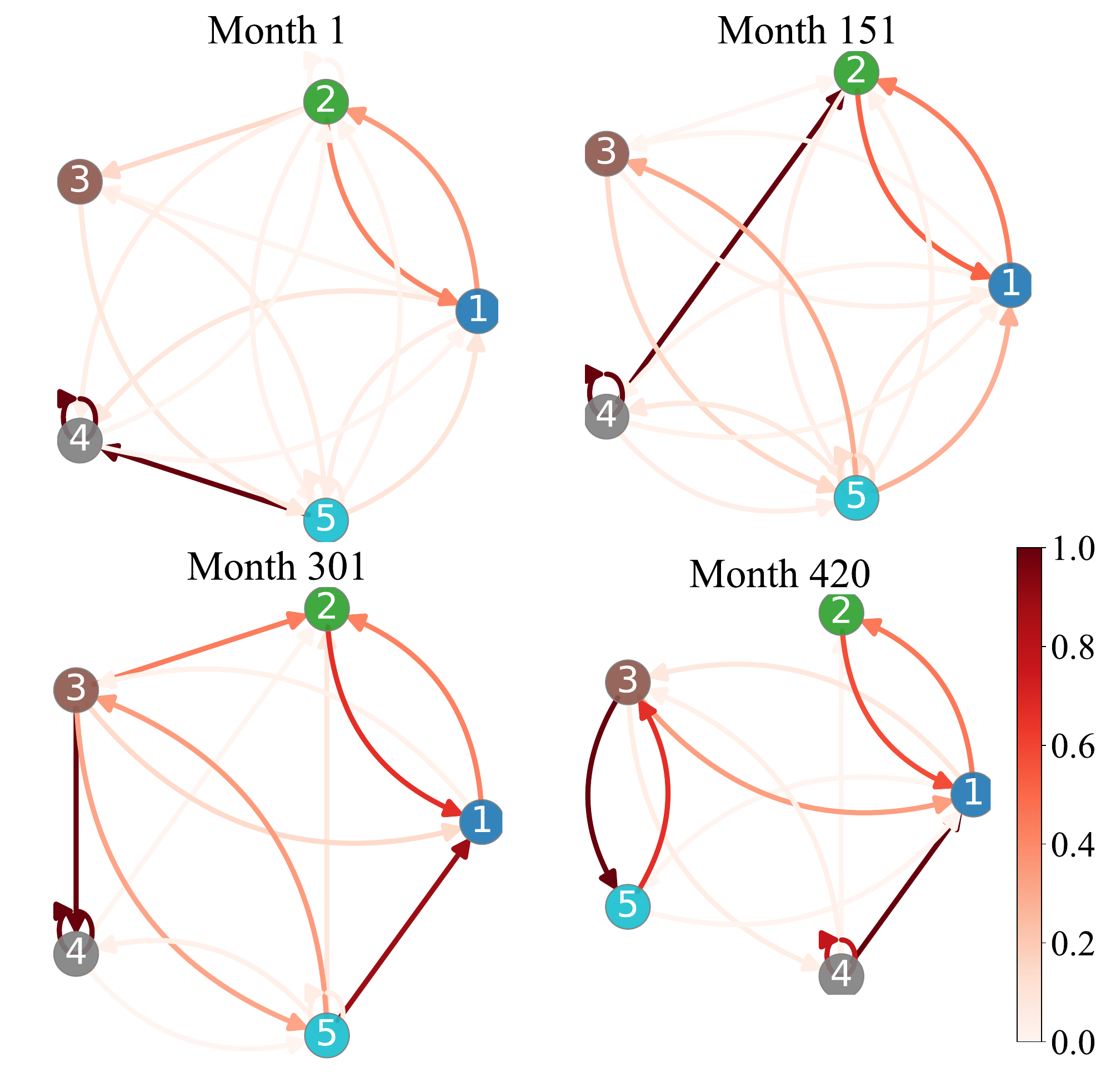}
    \caption{Graphical representation of Interaction between latent communities in  \emph{Trade} dataset for different time periods.}
    \label{fig:trade_graph}
  \end{subfigure}%


 \caption{Results on real-world datasets.} 
\end{figure*}
We found that the pyDRESCALk extracted feature components $\tilde{\mat{A}}$  corresponding to $k_{opt}$ were consistent with respect to the ground truth factors $\mat{A}$ with a correlation factor as high as .98 for weakly correlated factors and a correlation of 0.84 for highly correlated factors. 
A demonstration of feature identification for two different synthetic tensors of size  $1024 \times 1024 \times 10$ with $k=7$ and $2160 \times 2160 \times 20$ with $k=17$  is presented in Figure~\ref{fig:synthetic}. As observed from Figure~\ref{fig:sill_data1} and Figure~\ref{fig:sill_data2}, we utilize the reconstruction error and minimum silhouette width to correctly estimate the number of latent figures, which are respectively $7$ and $17$. We utilized the criteria of lower reconstruction error ($e_k$), higher minimum silhouette width ($s_k$) and largest separability between  $s_k$ and $e_k$ data points  for the estimation of latent feature $k$\cite{vangara2021finding}. In both Figures~\ref{fig:sill_data1} and \ref{fig:sill_data2}, we can see that the reconstruction error is lowest and minimum silhouette width is close to $1$ for $k=7$ and $k=17$ respectively. The silhouette value suddenly drops past the correct $k$ as the clustering tends to overfit the noise leading into weakly stable clusters.

Once these factors are extracted from the synthetic relational tensors, we can find precisely the underlying physical process for the demonstration of explainability as shown in Figure~\ref{fig:synthetic}. 
Furthermore, the visualization of the decomposed features promotes explainability of the data. Each column of the extracted latent features for data~1 is shown in Figure~\ref{fig:feat_data1} where each row represent one of the underlying processes, which is a Gaussian. Similarly, for data~2, we visualize the correlation between the original features and the reconstructed ones in Figure~\ref{fig:feat_data2}. We can see that there exists a significant correlation between the  original and the reconstructed features.

\subsubsection{Latent Feature Identification in \emph{Trade} and \emph{Nations} Data}
\label{sec:swimmer}
In this section, we showcase the efficacy of the proposed framework on real-world relational datasets. We have chosen  \emph{Trade} data and \emph{Nations} data to estimate the number of latent features. The details about these datasets are presented as follow:
\begin{itemize}
\item \emph{Trade:} This dataset is based on direction of trade statistics, IMF \cite{marini2018new}, and encodes  information about monthly imports/exports between 23 countries over 420 months into a relational tensor, $\ten{X}$, of dimensions $23\times 23\times 420$. In this tensor, each frontal slice $\ten[][:,:,k]{X}$ corresponds to trade flow statistics between countries in a given month, each horizontal slice $\ten[][i,:,:]{X}$ corresponds to trade statistics between the $i^{th}$ nation and all the  nations over the 420 months, and each lateral slice $\ten[][:,j,:]{X}$ corresponds to trade statistics between the all the nations and $j^{th}$ nation. The nations are Australia, Canada, China Mainland, Denmark, Finland, France, Germany, Hong Kong, Indonesia, Ireland, Italy, Japan, Korea, Malaysia, Mexico, Netherlands, New Zealand, Singapore, Spain, Sweden, Thailand, United Kingdom, and the United States.

\item \emph{Nations:} This dataset comprises 14 countries (Brazil, Burma, China, Cuba, Egypt, India, Indonesia, Israel, Jordan, Nederlands, Poland, USSR, UK, USA) across different continents and their relationship characterized by various 56 entities representing social, cultural, political, and other interactions. The details are provided in  \cite{nations}. The data tensor is of dimension $14\times 14 \times 56$. In contrast to \emph{Trade} dataset, where each frontal slice exhibits time evolution, each frontal slice of the \emph{Nations} dataset exhibits the interaction between nations over a given entity. This dataset comprises only binary interactions between the nations, whereas the \emph{Trade} dataset comprises a continuous valued relationship between the nations. 
\end{itemize}
To showcase the efficacy of distributed RESCAL on these datasets, we correctly estimate the number of latent features for both datasets, which are  4 and 5 for \emph{Nations} and \emph{Trade} dataset, respectively. 
For estimating the number of latent components from the \emph{Nations} and \emph{Trade} dataset, we performed pyDRESCALk for $k\in\{1,2,..,7\}$ on a processor grid of size $2\times 2$ with  $r=50$ perturbations and 10,000 RESCAL iterations  with random initialization of the factors $\mat{A}$ and $\ten{R}$. To address the issue for the division of data size by processor count for \emph{Trade} dataset (i.e 23/2), we padded the  rows and columns for all the slices with zeros such that the modified dataset size (i.e $24\times 24\times 420)$ is divisible by processor size.   The minimum silhouette, the average silhouette, and the reconstruction error are shown in Figure~\ref{fig:sill_nations} and Figure~\ref{fig:sill_trade} for \emph{Nations} and \emph{Trade} dataset respectively, based on these statistics, the estimated number of latent feature is 4 and 5 respectively as highlighted in the figures. 

Also, the columns of $\mat{A}$, which are of size $14\times 4$ and $23\times 5$ for \emph{Nations} and \emph{Trade}, respectively, represent feature groups/latent communities. For the \emph{Nations} dataset, the extracted latent groups corresponding to the columns of extracted $\mat{A}$ are shown along the rows in Figure~\ref{fig:feat_nations}. From Figure~\ref{fig:feat_nations}, the four major latent communities from $\mat{A}$ consist of the following groups of nations: i) \textbf{community-1}: China, Cuba, Poland, and USSR, ii) \textbf{community-2}: Burma, Egypt, India, Indonesia, Israel and Jordan, iii) \textbf{community-3}: US and UK, iv) \textbf{community-4}: Brazil, Egypt, India, Israel, Netherlands, Poland and UK.  

Similarly, for the \emph{Trade} dataset, the extracted five groups correspond to five economic regions as displayed along the five rows in Figure~\ref{fig:feat_trade}, and include i) \textbf{community-1}: USA, \textbf{community-2}: NAFTA (Canada, Mexico, and the USA), iii) \textbf{community-3}: China, iv) \textbf{community-4}: Europe, and v) \textbf{community-5}: Asia and Pacific (without China), which are major game players in international trade. 

Furthermore, to show interpretability of these results, we also evaluate the interactions between the extracted latent communities  of nations in \emph{Nations} and \emph{Trade} datasets through a probabilistic graphical visualization of the extracted $\ten{R}$ tensor. For the \emph{Nations} dataset, each slice $i$, i.e., $\ten[][:,:,i]{R}$, provides information about interactions within the group for relation $i$, which would be one of the $m=56$ relations representing social, cultural, political, and other interactions, whereas $\ten[][i,j,:]{R}$ correspond to interactions between $i^{th}$ and $j^{th}$ groups over all the $m$ relations. If $i=j$, the relations correspond to interactions between countries within a group $i$. Figure~\ref{fig:nations_graph} shows the interactions between the groups over relations \textit{exports}, \textit{tourism}, \textit{treaties} and \textit{students}. Each slice of the $m$-relation $\ten{R}$ can be expressed as a directed graph of $k$ nodes as shown in Figure~\ref{fig:nations_graph}, where nodes represent  groups and  edges connecting these nodes reflect the interactions for the corresponding relations. These interactions can be characterized by weights, where  value of 0 corresponds to no-interaction and  value of 1 corresponds to the strongest interaction. For example, for the \textit{exports}, \textbf{community-1} strongly relies on \textbf{community-3} whereas \textbf{community-4} also marginally relies on \textbf{community-3}. On the other hand, \textbf{community-3} strongly relies on \textbf{community-4} for \textit{export}. For  \textit{tourism}, \textbf{community-3} has more tourists visiting \textbf{community-2}, and \textbf{community-4} countries mostly and relatively lower visits to \textbf{community-1} countries. For treaties, a strong relationship is observed between i) \textbf{community-1} and \textbf{community-3} nations and ii) \textbf{community-3} and \textbf{community-4} nations.  Regarding the studies, students travel from \textbf{community-4} and \textbf{community-2} to \textbf{community-3} nations. 

A similar analysis can be made for $\ten{R}$ from the decomposition of the \emph{Trade} dataset. As shown in Figure~\ref{fig:trade_graph}, we have sampled the slices of $\ten{R}$ for four significant months, which are 1, 151, 301, and 420, which corresponds to slices 1,151,301, and 420 respectively. The trade analysis is then performed for the extracted 5 groups over these periods by constructing four directed graphs. From the sub-figures in Figure~\ref{fig:trade_graph}, the graph edges clearly show an active import-export pattern between the groups of nations over the periods. These graphs also show the evolution of trade relations between the groups over time. A minimal trade interaction is seen for month 1, which grows over time and is maximum for month 420. For the most recent result, which corresponds to month 420, \textbf{community-3} and \textbf{community-5} mostly rely on each other for import and export. Similarly, \textbf{community-1} and \textbf{community-2} rely on each other for  trades. In addition, there is uni-directional trade from \textbf{community-4} to \textbf{community-1} and from \textbf{community-3} to \textbf{community-1}. 

This analysis demonstrates the explainability of latent factors obtained by the pyDRESCALk decomposition on relational data tensors. Such analysis, when applied to large real-world datasets, will enable us to achieve a knowledge graph that provides a reasonable representation of the big data, which could be a very useful tool for data mining.

\subsection{Scalability}
\begin{figure*}
\centering
\begin{subfigure}{.45\textwidth}
  \centering
   \includegraphics[width=\linewidth]{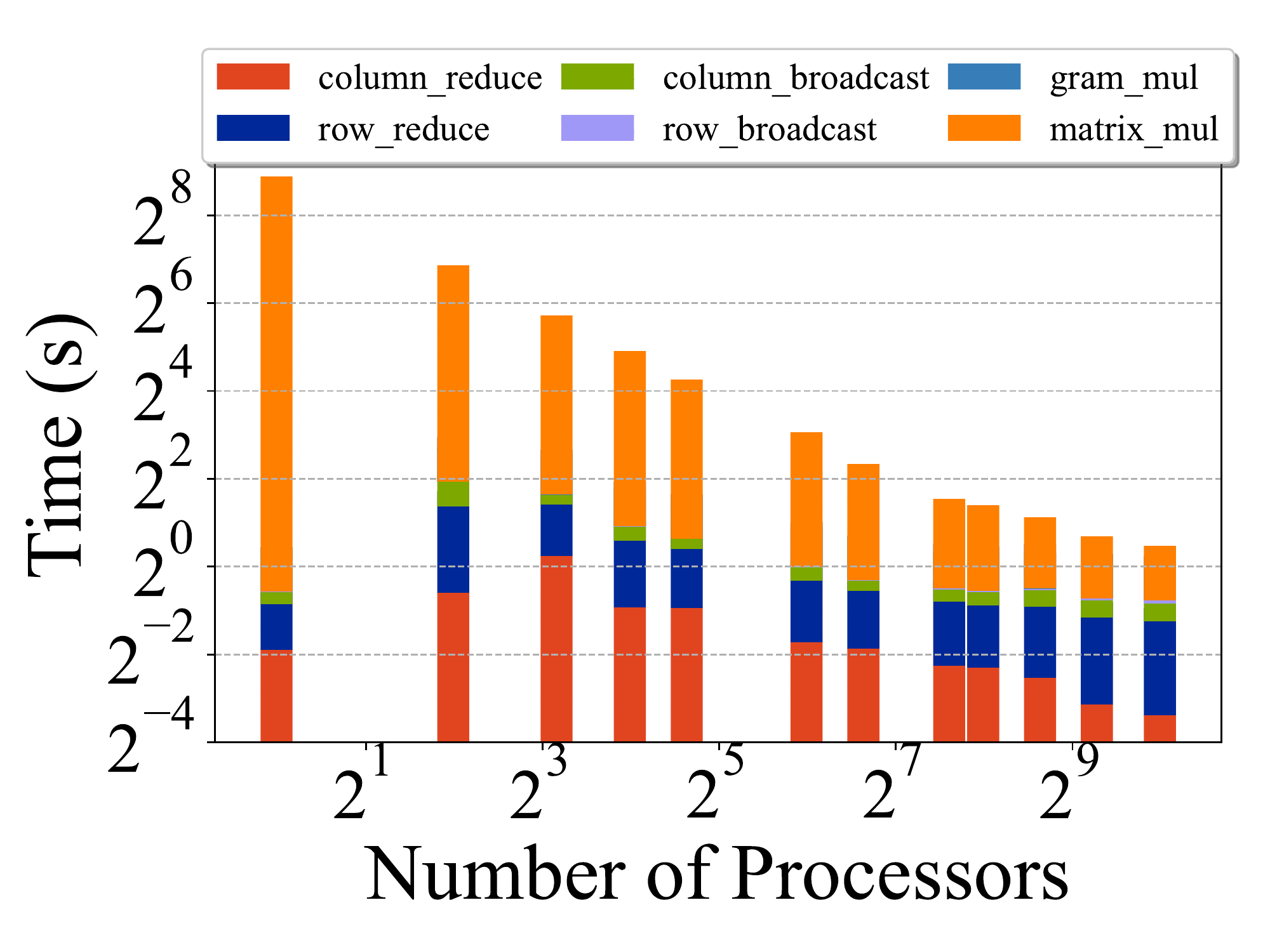}
  \caption{Strong scaling (Overall)}
  \label{fig:rescal_strong_a}
\end{subfigure}%
\begin{subfigure}{.45\textwidth}
  \centering
\includegraphics[width=\linewidth]{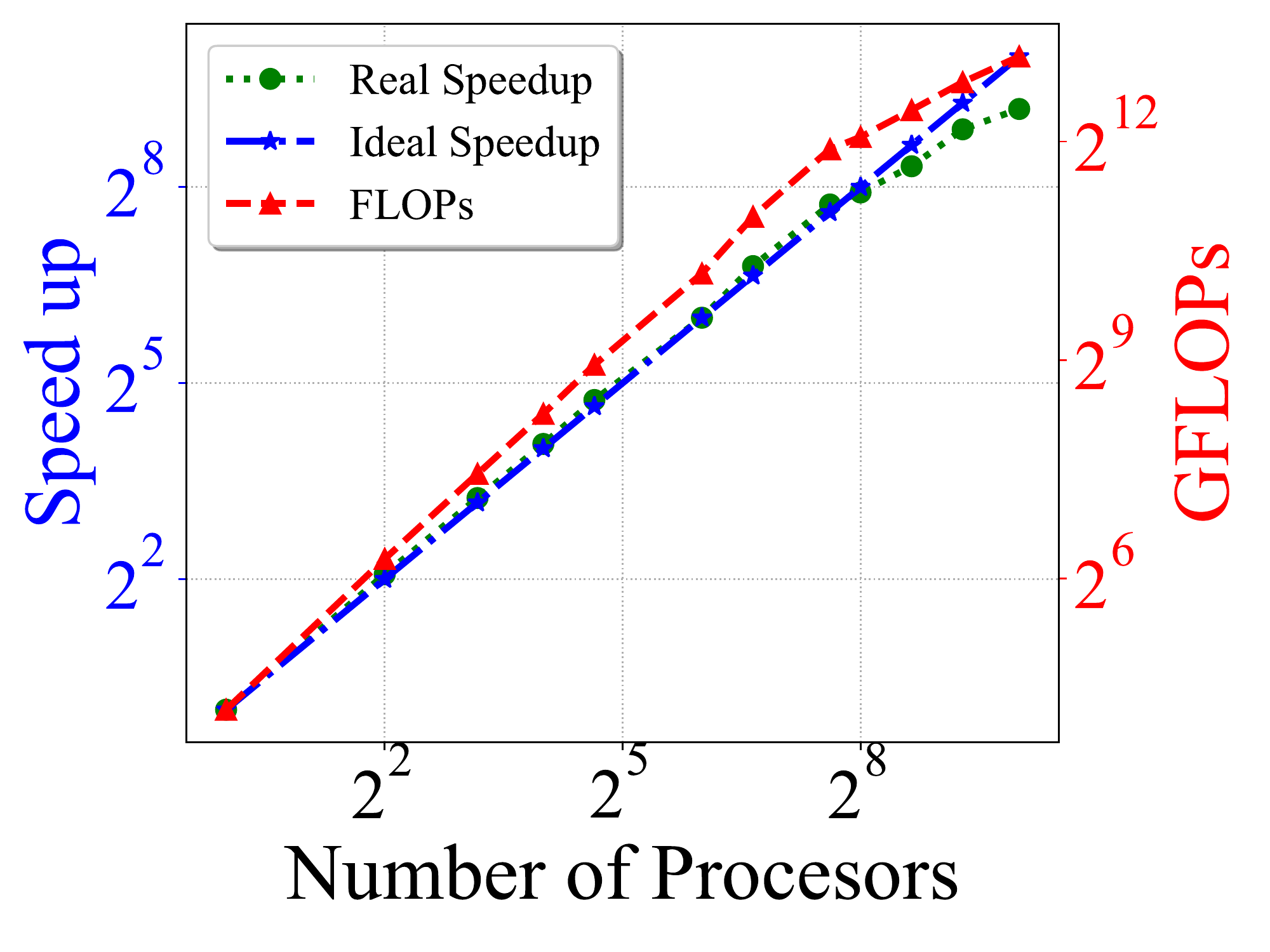}
  \caption{GFlOPS and Speedup}
  \label{fig:rescal_strong_b}
\end{subfigure}
\caption{Strong scaling experiments (RESCAL) }
\label{fig:strong}
\end{figure*}

In this section, we assess the scalability of pyDRESCALk. We estimate the strong scaling, the weak scaling, and the scaling with respect to the parameter $k$. We use  processors organized in a virtual  $p_r\times p_c$ square grid such that  $p_r= p_c$. For both strong and weak scaling for dense and sparse datasets, we choose the number of MPI processes to be in \{1, 4, 9, 16, 25, 64, 100, 196, 256, 400, 625 and 1024\}. From $1$ to $44$ Grizzly nodes were used to run these MPI Processes, where $1$ node can handle up to $25$ MPI Processes and $44$ nodes were used to run 1024 MPI processes.  Similarly, for performing scaling experiments on  GPUs, we select the GPU count to be in \{1, 4, 9, 16, 25, 64, 81\}. As the communications on GPUs within a node and across multiple nodes is handled by the CUDA aware MPI communicator, we call each MPI process as a CUDA aware MPI Process. For the Kodiak cluster, $1$ node can handle up to $4$ CUDA aware MPI Processes and $21$ nodes were used to run the $81$ CUDA aware MPI process. To ensure consistency of these calculations, we perform exactly 10 RESCAL update iterations and for RESCALk, we fix the number of perturbations to be 10 and vary the latent factor $k$ from 1 to 10. 

To analyze the computation vs communication-based scaling of the distributed framework, we utilize computational metrics such as Gram operation ($\operatorname{gram\_mul}$), matrix multiplication ($\operatorname{matrix\_mul}$), and sparse matrix multiplication ($\operatorname{matrix\_mul\_sparse}$), and MPI primitives such as reduce and broadcast. As pyDRESCALk only involves the row and column subcommunicator-based MPI operations reduce and broadcast, we quantify the scaling performance of the framework by these operations. The communication operations are row\_reduce, $\operatorname{column\_reduce}$, $\operatorname{row\_broadcast}$, and $\operatorname{column\_broadcast}$. Here, the Gram operation of a matrix $\mat{Y}$ of dimensions $p\times k$ represented by $Y^T Y$ involves the inner product of a matrix with itself such that the resultant product is of dimension $k \times k$. On the other hand, matrix multiplication is for local rank-based dense-dense matrix multiplication, whereas sparse matrix multiplication is for the local product of dense-sparse or sparse-sparse matrices. Both gram and matrix multiplication on the CPU are performed with Numpy with the OpenBLAS library as thebackend, whereas, GPU compute operations are performed with CuPY with the cuBLAS backend. As cuBLAS can accelerate GPU matrix operations, the scaling results for the GPU implementation are expected to be better than the CPU only implementation. 

The communication overhead is managed by the mpi4py library operating with OpenMPI backend for CPU-based pyDRESCALk whereas the CUDA-aware MPI takes care of the communication overhead for the GPU-based pyDRESCALk. For both  (CPU and GPU) implementations, the communication cost includes $\operatorname{row\_reduce}$, $\operatorname{column\_reduce}$, $\operatorname{row\_broadcast}$, and $\operatorname{column\_broadcast}$. $\operatorname{Row\_reduce}$ and $\operatorname{column\_reduce}$ correspond to the costs for dense data reduction along row and column sub-communicator, respectively. Similarly, $\operatorname{column\_reduce}$ and $\operatorname{row\_broadcast}$ correspond to the costs associated with broadcasting a matrix along a row and column sub-communicator, respectively. All scaling benchmarks were performed using single-precision arithmetic. 

To quantify the scaling performance, we use the metrics runtime in seconds, speedup, and Giga floating-point operations per second(GFLOPS). Here, the runtime is the actual time taken by a certain operation while performing the RESCAL decomposition. To compute the overall runtime, the runtime for each MPI process is computed and then an average is computed across these individual runtimes. Furthermore, an average of ten different independent runs is calculated to ensure consistent recordings of the timings. We also use these timing values to compute the speedup metric, which corresponds to the ratio between the p MPI process runtime and the 1 process runtime. This evaluates the merit of utilizing multiple cores over a single core for solving equivalent tasks. We also utilize the GFLOPS metric to analyze the computational efficiency of the algorithmic implementation. 
\par
In the prior published works of distributed RESCAL, the authors do not demonstrate strong and weak scaling performance. This limits us to having a one-to-one comparison of the scalability performance analysis of our approach with the existing distributed implementations. In addition to these, the inefficient design of the existing distributed implementations severely constrains their scalability for large-sized datasets utilized in this paper. As a result, we are limited in our ability to compare against these approaches.   

\subsubsection{Strong Scaling}

We perform strong scaling experiments with a dense random tensor of size $20 \times 2^{14} \times 2^{14}$ and $20 \times 2^{13} \times 2^{13}$  for RESCAL and pyDRESCALk respectively. The chosen data was the largest size that could fit in the memory of the node for single processor  operation. For strong scaling on the sparse dataset, we utilize sparse random tensor of size $20 \times 2^{17} \times 2^{17}$ and $20 \times 2^{16} \times 2^{16}$ for sparse RESCAL and pyDRESCALk respectively. In all the generated datasets, we use $k=10$ and evaluate the RESCAL performance for the same $k$. For strong scaling, we fix the data and vary the processor count such that each process executes on the corresponding chunk of the data whose size is reduced by the factor of processor count. Ideally, the execution time would be the largest when run on a single core and would be divided by the number of MPI processes during parallel execution. We have set the number of MPI processes to be in \{1, 4, 9, 16, 25, 64, 100, 196, 256, 400, 625, and 1024\} for all strong scaling experiments. For a fixed problem size where $m,n$ and $k$ are fixed, the computation complexity estimated in Sections~\ref{subsubsec:compute_rescal} and \ref{subsubsec:compute_rescalk} for RESCAL and  RESCALk respectively is reduced to $\O{\frac{1}{p}}$ as a function of $p$. This trend can be seen in the strong scaling plots (Figure~\ref{fig:rescal_strong_a}) as run times are decreased by a factor approximately equal to the number of MPI processes. 
The run times are dominated by matrix multiplication with little contribution from communication cost. In the speedup plots, 
we see sub-linear trends, indicating that the efficiency of the algorithm decreases as more and more MPI ranks are used to solve a problem of fixed size (Figure~\ref{fig:rescal_strong_b}). From Figure~\ref{fig:rescal_strong_b}. Additionally, the GFLOPS follow a similar trend (overlapping) as the speedup as increasing the processor count leads to an increase in total floating-point operations performed per second. For a large number of processors, the communication bottleneck constrains the speedup/GFLOPs performance as the communication operations are dominant compared to computational operations. As a result, the speedup peaks at 590 for 1000 cores with approximate linear scaling. 

\begin{figure*}
\centering
\begin{subfigure}{.45\textwidth}
  \centering
    \includegraphics[width=\linewidth]{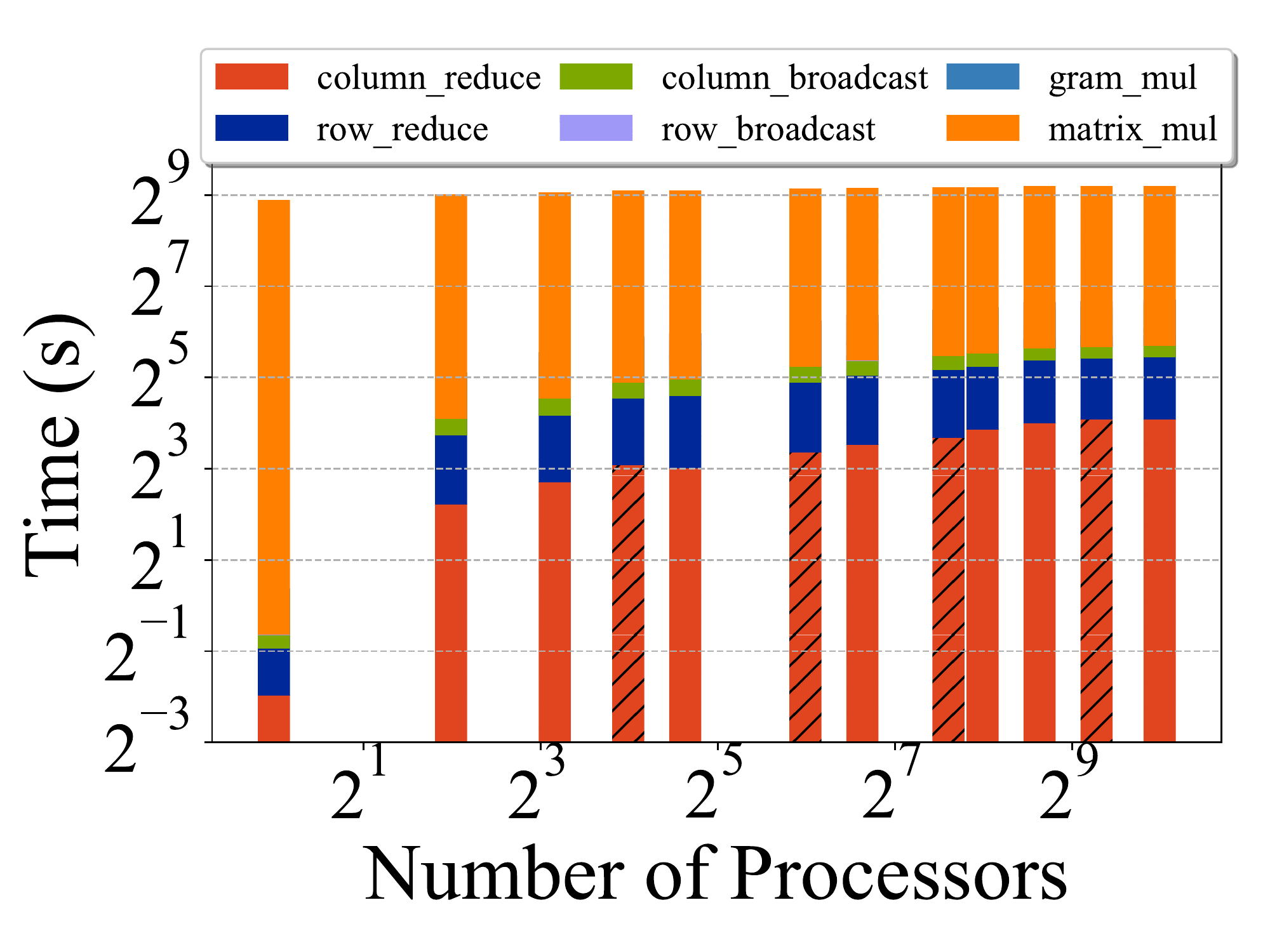}
  \caption{Weak scaling (Overall)}
   \label{fig:rescal_weak_a}
\end{subfigure}%
\begin{subfigure}{.45\textwidth}
  \centering
\includegraphics[width=\linewidth]{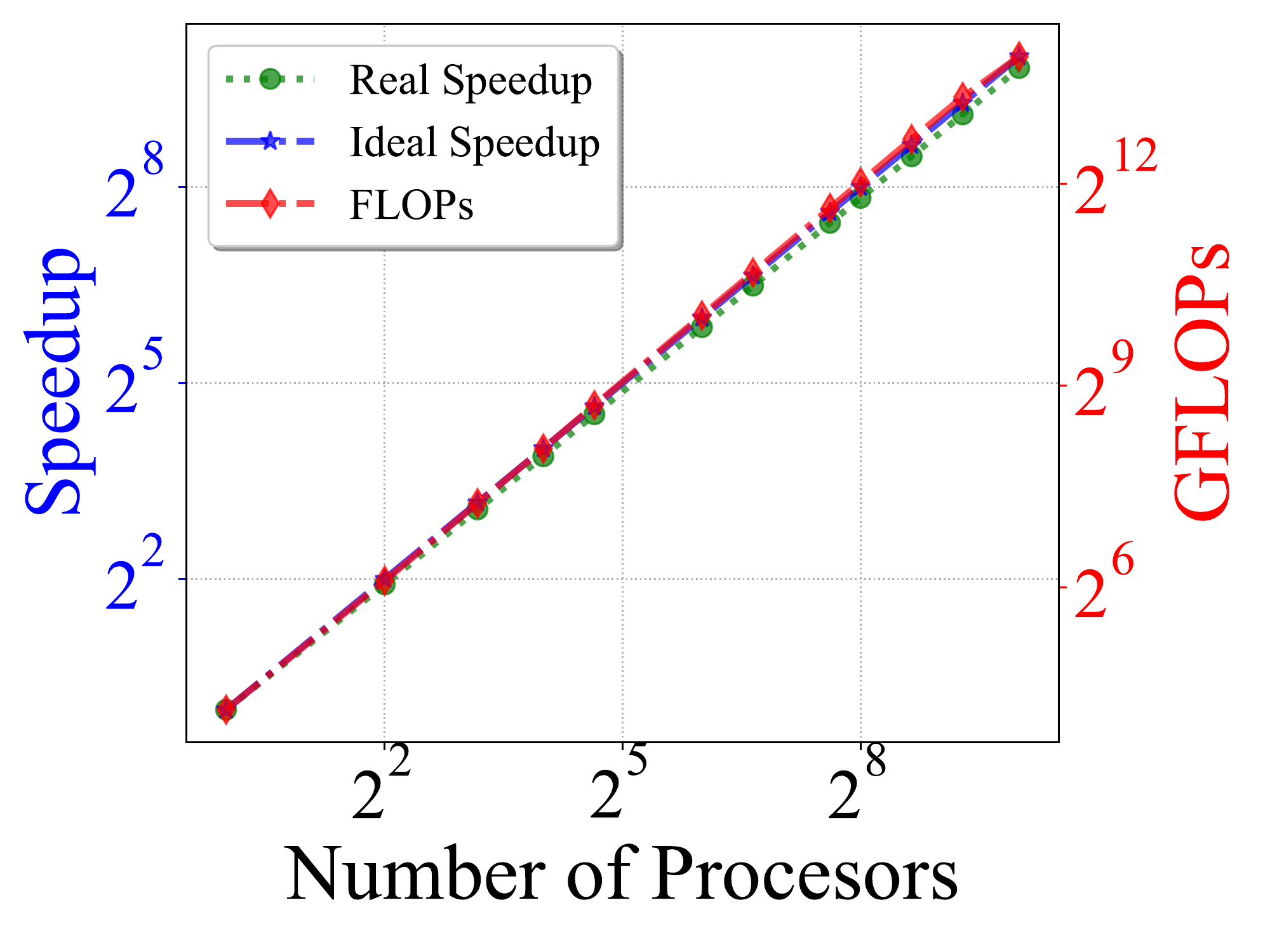}

  \caption{GFlOPS and Speedup}
   \label{fig:rescal_weak_b}
\end{subfigure}
\caption{Weak scaling experiments(RESCAL) }
\label{fig:weak}
\end{figure*}


\subsubsection{Weak Scaling}

\begin{figure*}
\centering
\begin{subfigure}{.45\textwidth}
  \centering
  \includegraphics[width=\linewidth]{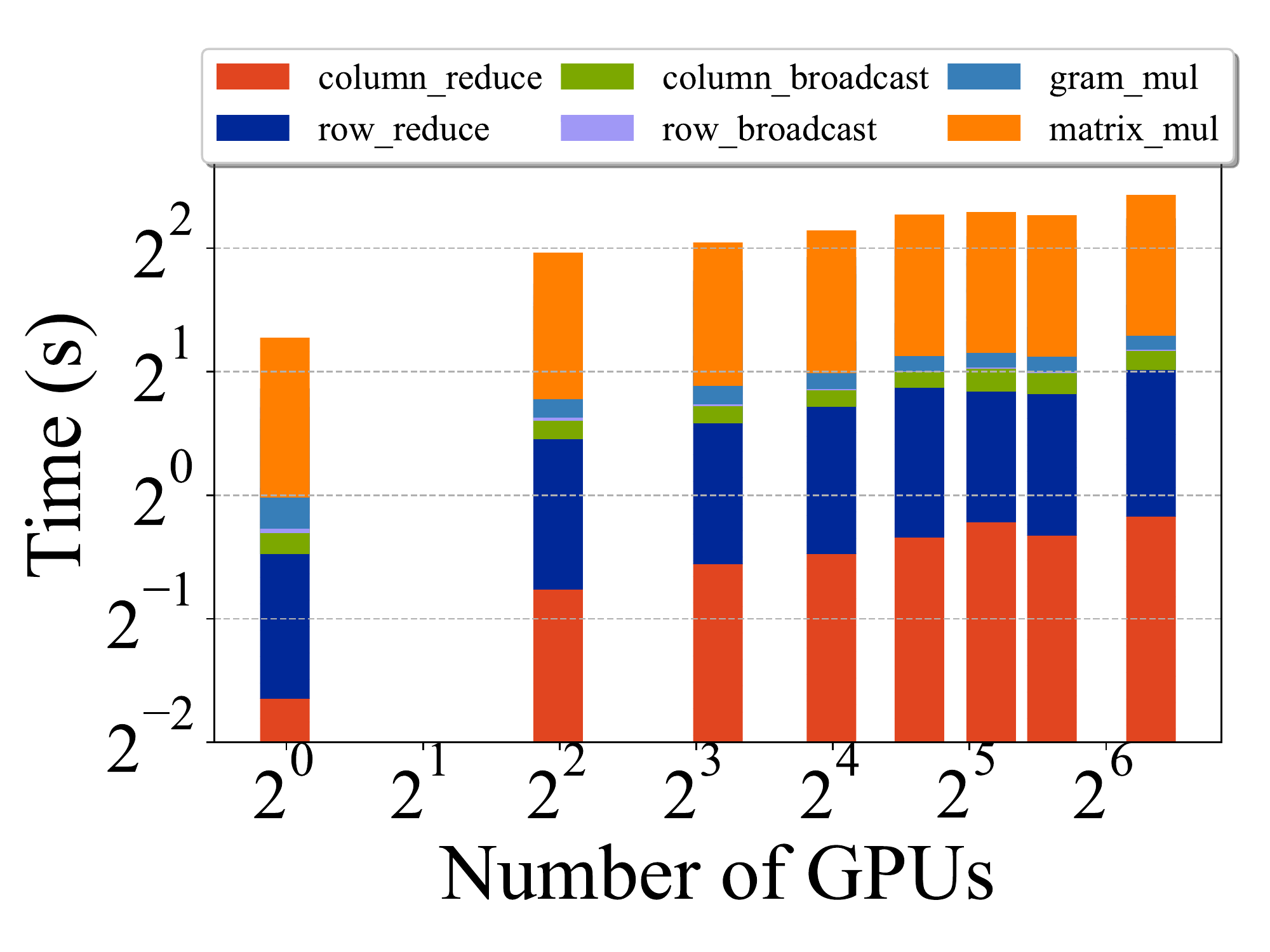}
  \caption{Weak scaling (Overall)}
  \label{fig:weak_gpua}
\end{subfigure}%
\begin{subfigure}{.45\textwidth}
  \centering
  \includegraphics[width=\linewidth]{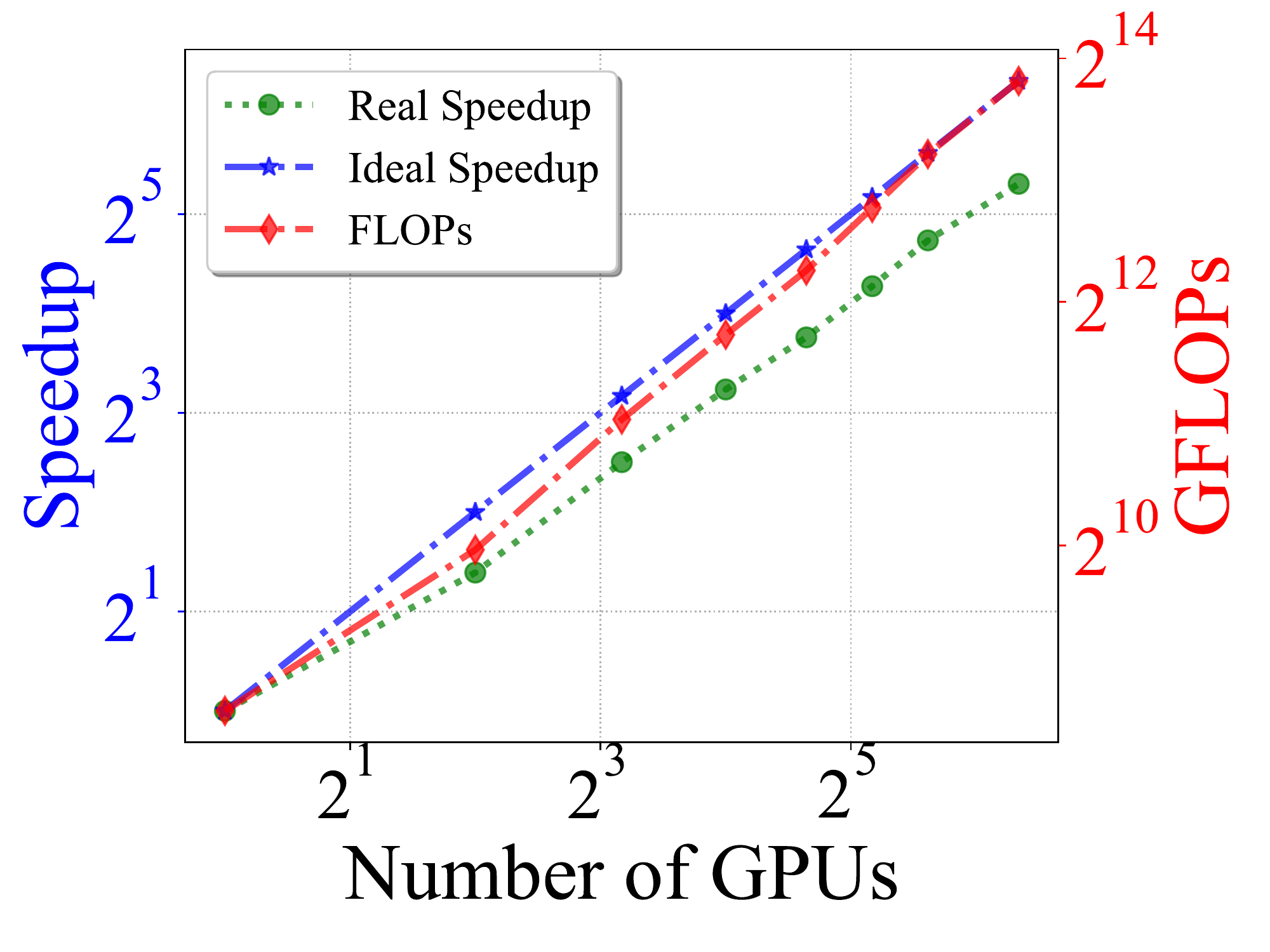}
  \caption{GFlOPS and Speedup}
  \label{fig:weak_gpub}
\end{subfigure}
\caption{Weak scaling experiments(RESCAL-GPU) }

\end{figure*}

\begin{figure*}
\centering
\begin{subfigure}{.45\textwidth}
  \centering
  \includegraphics[width=\linewidth]{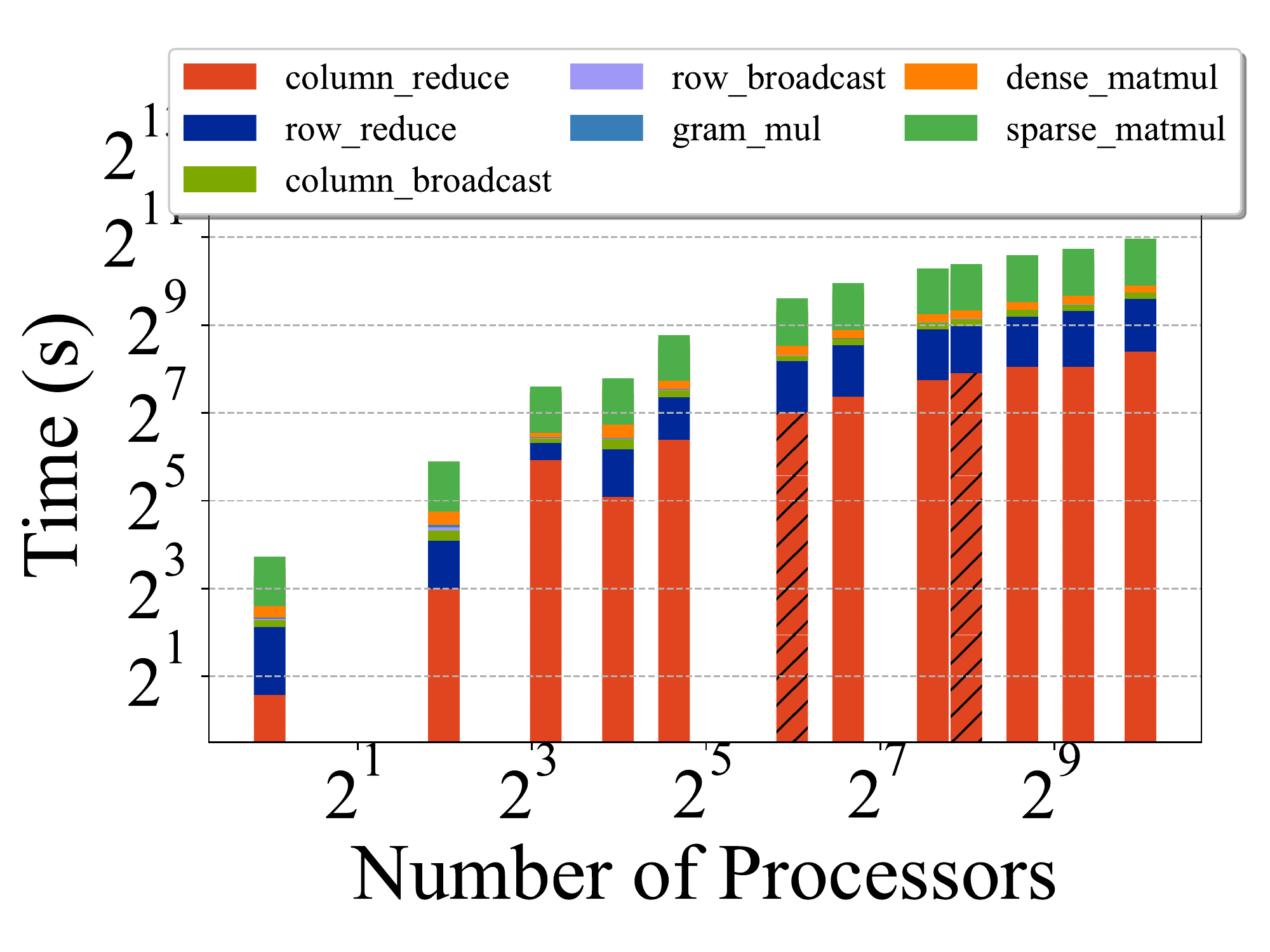}
  \caption{Weak scaling plot(RESCAL-sparse)}\label{fig:efficient_a}
\end{subfigure}
\begin{subfigure}{.45\textwidth}
  \centering
  \includegraphics[width=\linewidth]{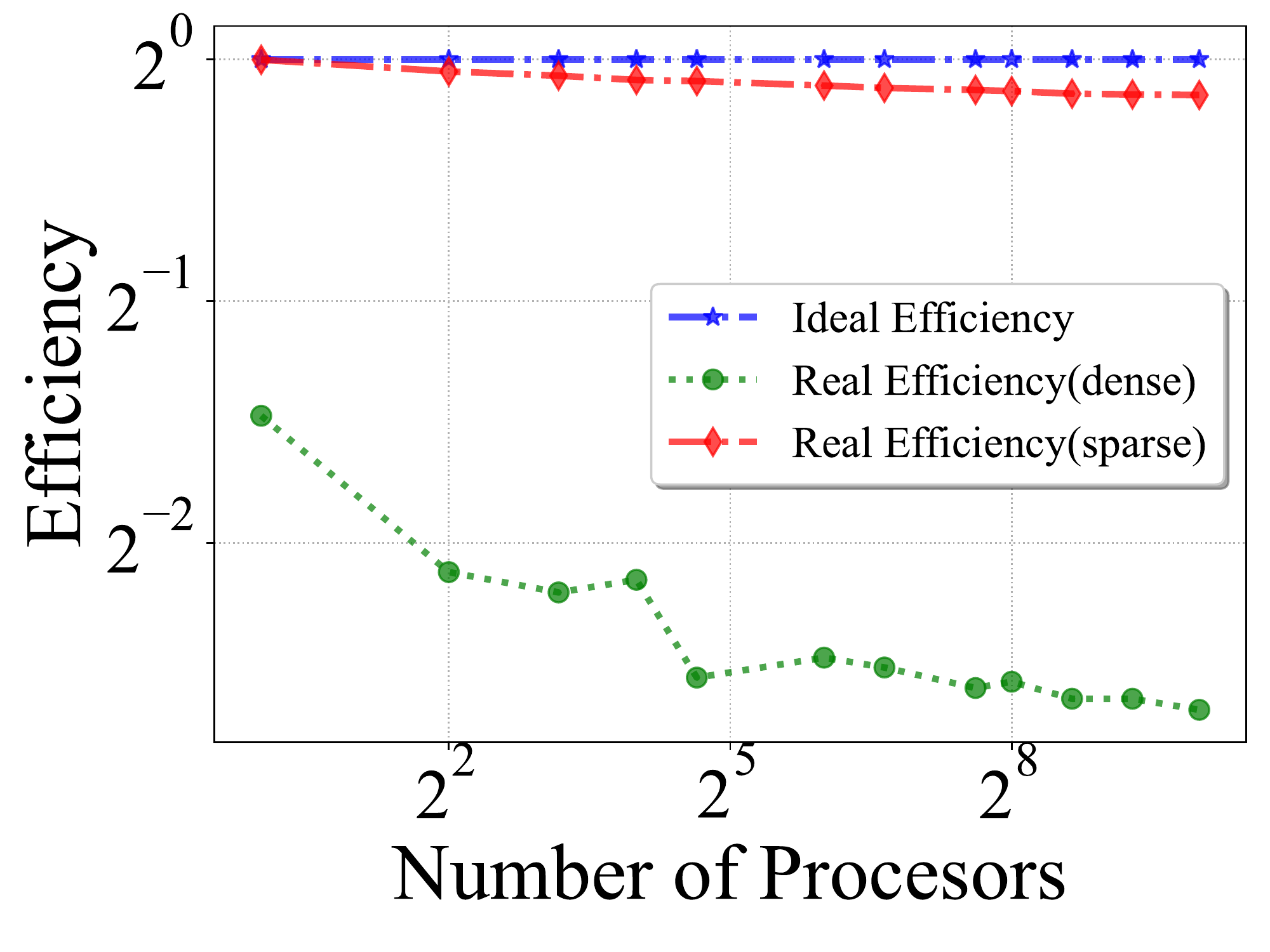}
  \caption{Efficiency plot for RESCAL(dense and sparse)}
  \label{fig:efficient_b}
\end{subfigure}%
\caption{RESCAl sparse experiments }
\label{fig:weak5}
\end{figure*}

For weak scaling, the experiments are set such that each MPI process has same amount of computational burden, meaning the increase in data size matches the  increase in the number of cores. Similar to the strong scaling experiment, we set the number of cores to be in \{1, 4, 9, 16, 25, 64, 100, 196, 256, 400, 625 and 1024\}. The data size was chosen to maximize the memory usage within a node for a given maximum core count. The tensor size  is varied as a function of p such that data size for given processor count is {20 $\times$ $2^{13}\sqrt p$ $\times$ $2^{13} \sqrt p$}. Based on this, the data size is varied from  {20 $\times$ $2^{13}$ $\times$ $2^{13}$} on one core up to {20 $\times$ $2^{18}$ $\times$ $2^{18}$} on 1024 cores with a fixed $k = 10$ for dense data. Similarly, for dense RESCALk, the data size as a function of MPI process count is {20 $\times 6200\sqrt p \times 6200\sqrt p$} such that the data sizes for 1024 cores is {20 $\times$ 198,400 $\times$ 198,400}. For sparse RESCAL, the tensor size as a function of MPI process count is {$20 \times 98304\sqrt p \times 98304\sqrt p$} with a fixed $k = 10$. The local problem size in all cases was fixed to  $20 \times 8192 \times 8192$ for dense RESCAL, $20 \times 6200 \times 6200$  for dense RESCALk and  $20 \times 98304 \times 98304$ for sparse RESCAL. 
 From (Figure~\ref{fig:rescal_weak_a}), we can observe that the scaling performance approximately flattens for MPI process counts greater than 9. For MPI process counts less then 64, the communication cost is minimum as  internode communication only exists. As The runtime for the scaling experiment is 
 is mostly dominated by matrix multiplication, the RESCAL decomposition is computation bound for the dense CPU implementation. When we substitute $n=\sqrt p \log p$ into the computation complexities derived in Section~\ref{subsubsec:compute_rescal}, we expect the run times for weak scaling to follow $\O{\log^2 p}$. This trend is approximately observed in (Figure~\ref{fig:rescal_weak_a}).
In the speedup plots from Figure~\ref{fig:rescal_weak_b},  we see an almost perfect linear correlation between the speedup and the number of CPUs, indicating a constant efficiency. A similar pattern is observed for FLOPS as a function of processor counts.
A comparable weak scaling profile can be observed in GPUs as shown in Figure~\ref{fig:weak_gpua} and Figure~\ref{fig:weak_gpub}. However, compared to CPUs' computational bottleneck, GPUs' computational advantage causes the communication operations to become the bottleneck as shown in Figure~\ref{fig:weak_gpua}. Still, with such a bottleneck, the GPU-based implementation performs at least 10 times faster compared to the CPU implementation. To illustrate this, we can observe from Figures~\ref{fig:rescal_weak_a} and \ref{fig:weak_gpua}, the execution time corresponding to GPU counts in \{1,4,...64\} is at least 10 times lower then that of equivalent processor counts. On the other hand, the speedup for GPU is limited for larger GPU counts due to the communication bottleneck compared to CPUs. Still, the superior computational abilities of GPUs enable the GPU scaling to have the same GFLOPS achieved with 1000 cores with just 81 GPUs as seen from Figure~\ref{fig:weak_gpub}.

A weak-scaling performance of sparse-RESCAL on sparse tensors for CPU is shown in Figure~\ref{fig:efficient_a}. From Figure~\ref{fig:efficient_a}, while the efficiency of the weak scaling for dense implementation is close to 90\%  , the sparse implementation has efficiencies less than 20\% . This is likely due to the communication costs being more significant in the sparse cases. Also note that, unlike the dense implementation, the sparse implementation only performs local sparse computations, which are significantly faster than dense operations; however, the communication cost is still the same as that of dense as observed in Figure~\ref{fig:efficient_b}. This significantly limits the efficiency of sparse implementation as this implementation is heavily constrained by communication bottlenecks. 

\begin{figure*}[ht!]
\centering
\begin{subfigure}{.45\textwidth}
  \centering
\includegraphics[width=\linewidth]{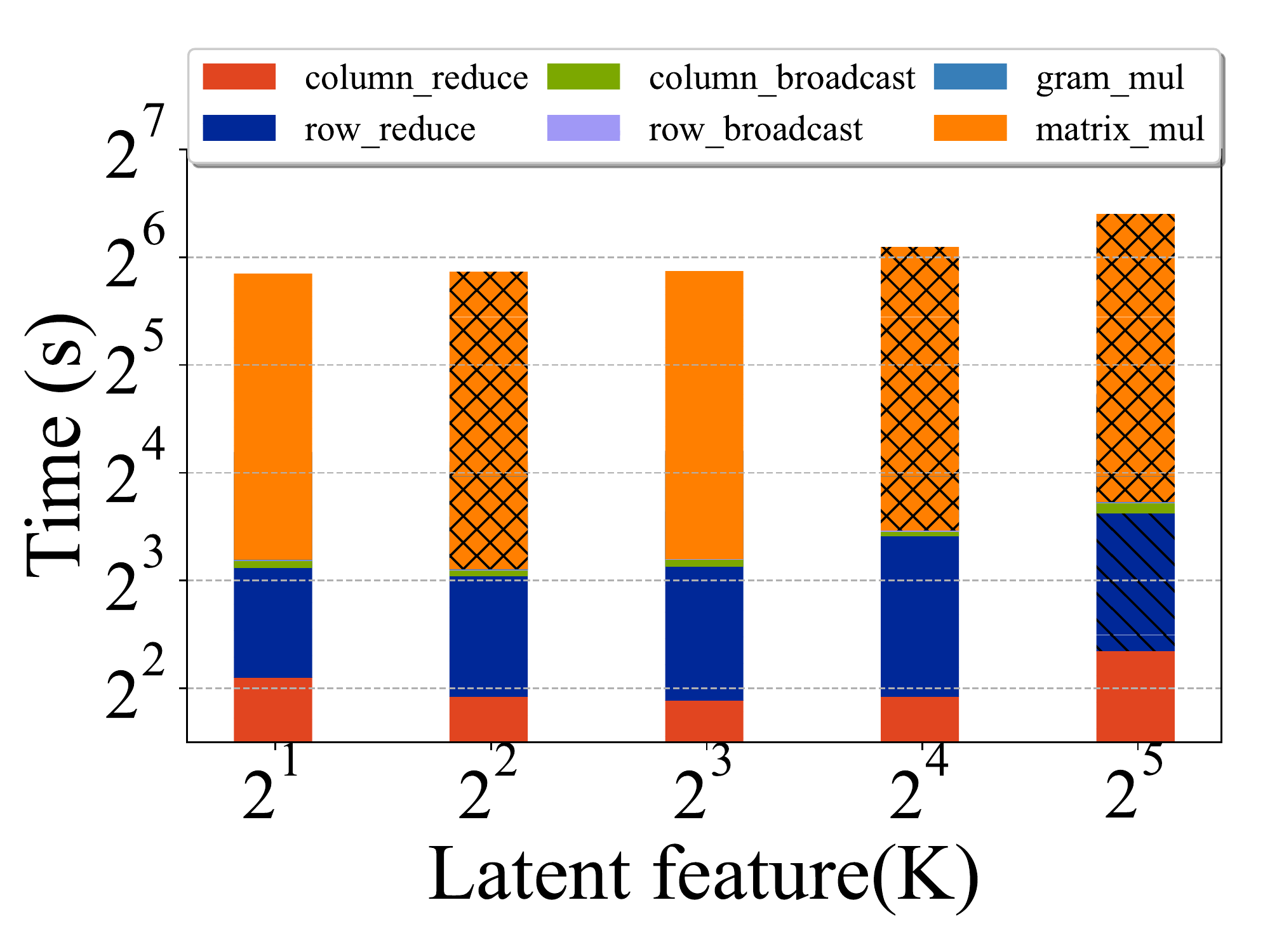}
\caption{K scaling-CPU (Overall)}
   \label{fig:rescal_k_cpu}
\end{subfigure}%
\begin{subfigure}{.45\textwidth}
  \centering
  \includegraphics[width=\linewidth]{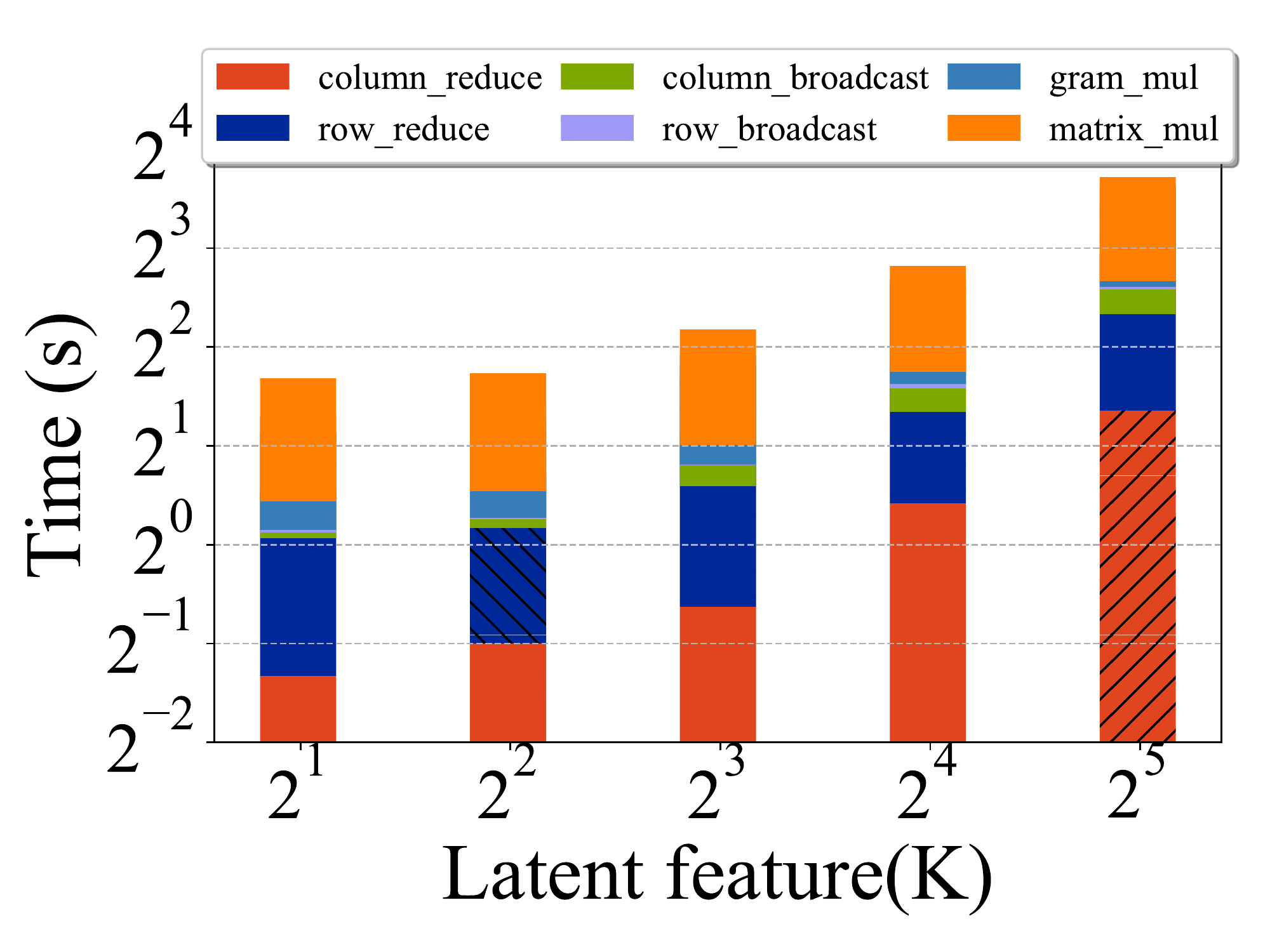}
  \caption{K scaling-GPU (Overall)}
   \label{fig:rescal_k_gpu}
\end{subfigure}
\caption{K scaling experiments(RESCAL) }
\label{fig:k-scale}
\end{figure*}

\subsubsection{Scaling at k}

For the $k$-scaling experiment, we set the tensor size and grid size to be fixed and vary the number of latent feature $k$. We choose the largest size of the data and core count from weak scaling i.e {20 $\times$ $2^{18}$ $\times$ $2^{18}$} for 1024 cores and with $k$ in \{2, 4, 8, 16, 32, 64, 128, 256\}. We performed the $k$-scaling on both the CPU and GPU hardware. 

The complexity analysis informs us of an $\O{k^2}$ trend, which we observe in Figure~\ref{fig:k-scale}. Even though the GPU based k-scaling results (Figure~\ref{fig:rescal_k_gpu}) show significantly better performance over the CPU version (Figure~\ref{fig:rescal_k_cpu}), the CPU results exhibit close to ideal k-scaling  performance. Since the GPU-based implementation is much faster, the communication costs become a significant fraction of run time for higher $k$ values.


For the GPU-based implementation, we observed that the scaling performance is heavily impacted by the communication operations. Even though we can utilize the computation ability of the GPU for accelerated performance, communication bottlenecks severely impact the performance compared to CPUs. One reason would be an inefficient communication backend, i.e., CUDA-aware MPI 
and limited communication bandwidth of the cluster that is causing a significant bottleneck for the implementation. We aim to address these issues in the future with the utilization of an efficient NCCL-based efficient communicator design and utilization of high throughput communication servers.

\begin{figure*}[ht!]
\centering
\begin{subfigure}{.45\textwidth}
  \centering
\includegraphics[width=\linewidth]{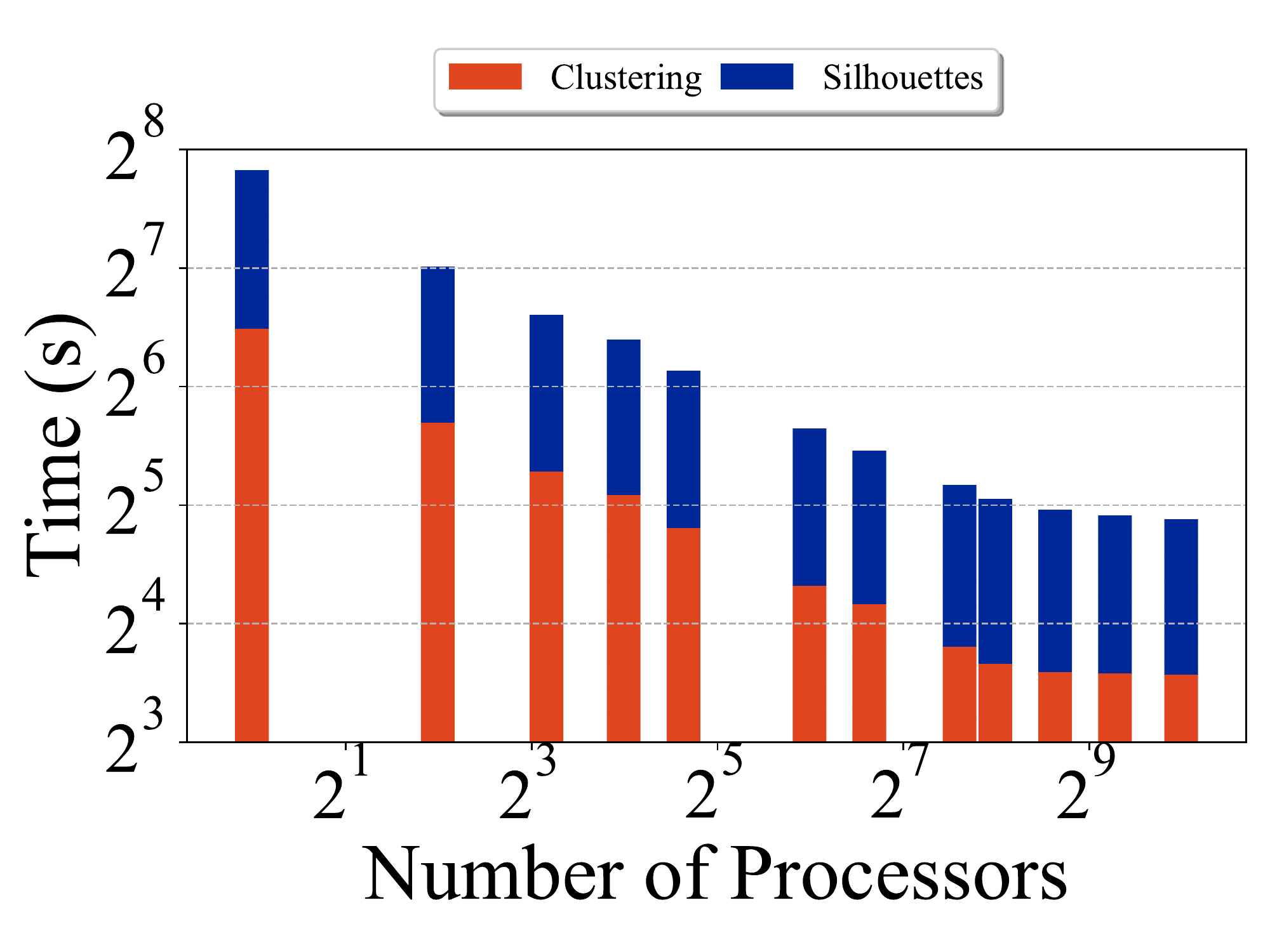}
  \caption{Strong scaling for clustering/Silhouette}
  \label{fig:strong_cls}
\end{subfigure}%
\begin{subfigure}{.45\textwidth}
  \centering
  \includegraphics[width=\linewidth]{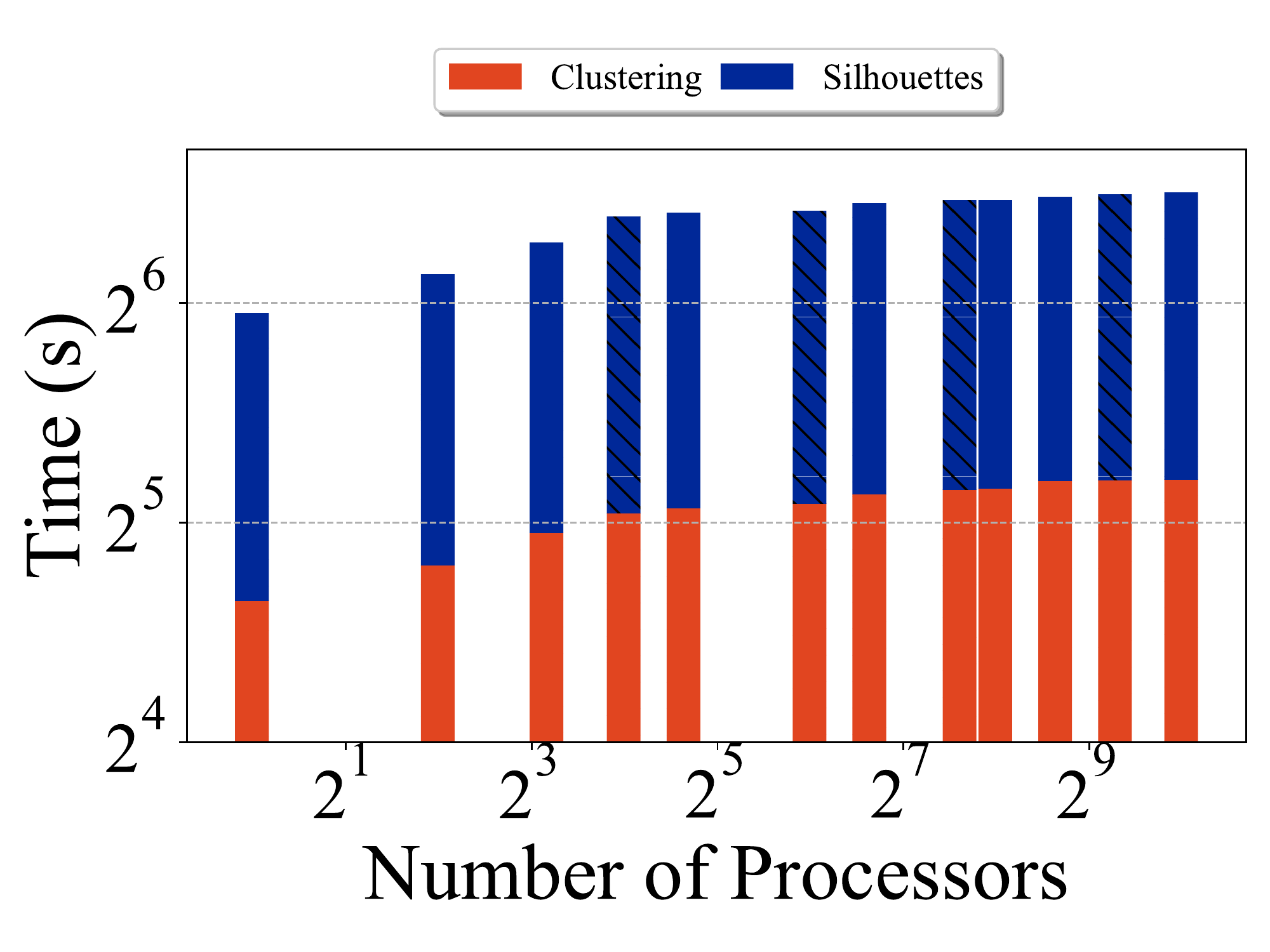}
  \caption{Weak scaling for clustering/Silhouette}\label{fig:weak_cls}
\end{subfigure}
\caption{Scaling results for clustering and silhouettes.}
\label{fig:scaling_clstr}
\end{figure*}

\subsection{Performance of Distributed Clustering and Silhouette}
In this section, we demonstrate the scalability of the distributed clustering and distributed silhouette algorithms. We evaluated the performance for processors of counts 1,4,9,....1000 similar to that of RESCAL. We perform both strong and weak scaling experiments for 10 perturbations for each $k$ in range 1 to 10. The data size chosen for the strong and weak scaling tests is equivalent to those chosen for the corresponding scaling in RESCAL. With strong scaling, we measure the speedup associated with increased MPI process counts compared to the performance of 1 core. As per the complexity analysis from \ref{subsubsec:clustering_comp}, for both clustering and silhouette operations, the complexity is $\O{ \frac{1}{\sqrt p}}$. As per this analysis, we observe a comparable speedup up until the number of MPI ranks becomes too large and performance flattenes as shown in Figure~\ref{fig:strong_cls}. This is when the communication bottlenecks overcome the computation bottleneck. Similarly, the weak scaling also demonstrates a similar performance to RESCAL as shown in Figure~\ref{fig:weak_cls}.

Notice that the strong scaling performance of the clustering and silhouette algorithms is not comparable to RESCAL. This is due to a couple of factors: First, the factors $\ten{A}$ being processed with distributed clustering are much smaller than the tensor $\ten{X}$. Also, the RESCAL decomposition performed on $\ten{X}$ employs 2D virtual grid topology, which is effective at reducing communication overhead as the communication is only involved on local sub-communicators. However, the clustering and silhouette framework is performed on a 1D grid where global communication is required for performing major operations. Unlike decomposition of large TB scale datasets, where the entire factors can't fit into node memory, the decomposed factors corresponding to data used for benchmarking the algorithm in this paper are of a lower size, which can easily fit in node memory. In such situations, there is an advantage of performing clustering on a node as the communication bottleneck can easily dominate for larger core/node counts. Because of this, we don't see a significant gain in performance with an increasing number of processor counts. However, the situation is different for large-sized factors and better scalability is observed. In a nutshell, the scalability of the clustering and silhouette is limited by the size of factors decomposed by RESCAL. Even though the RESCAL performance might scale well for a specific tensor size, the clustering and silhouette algorithms will only scale similarly if  the factors are large. 


\begin{figure*}
\centering
\begin{subfigure}{.45\textwidth}
  \centering
 \includegraphics[width=\linewidth]{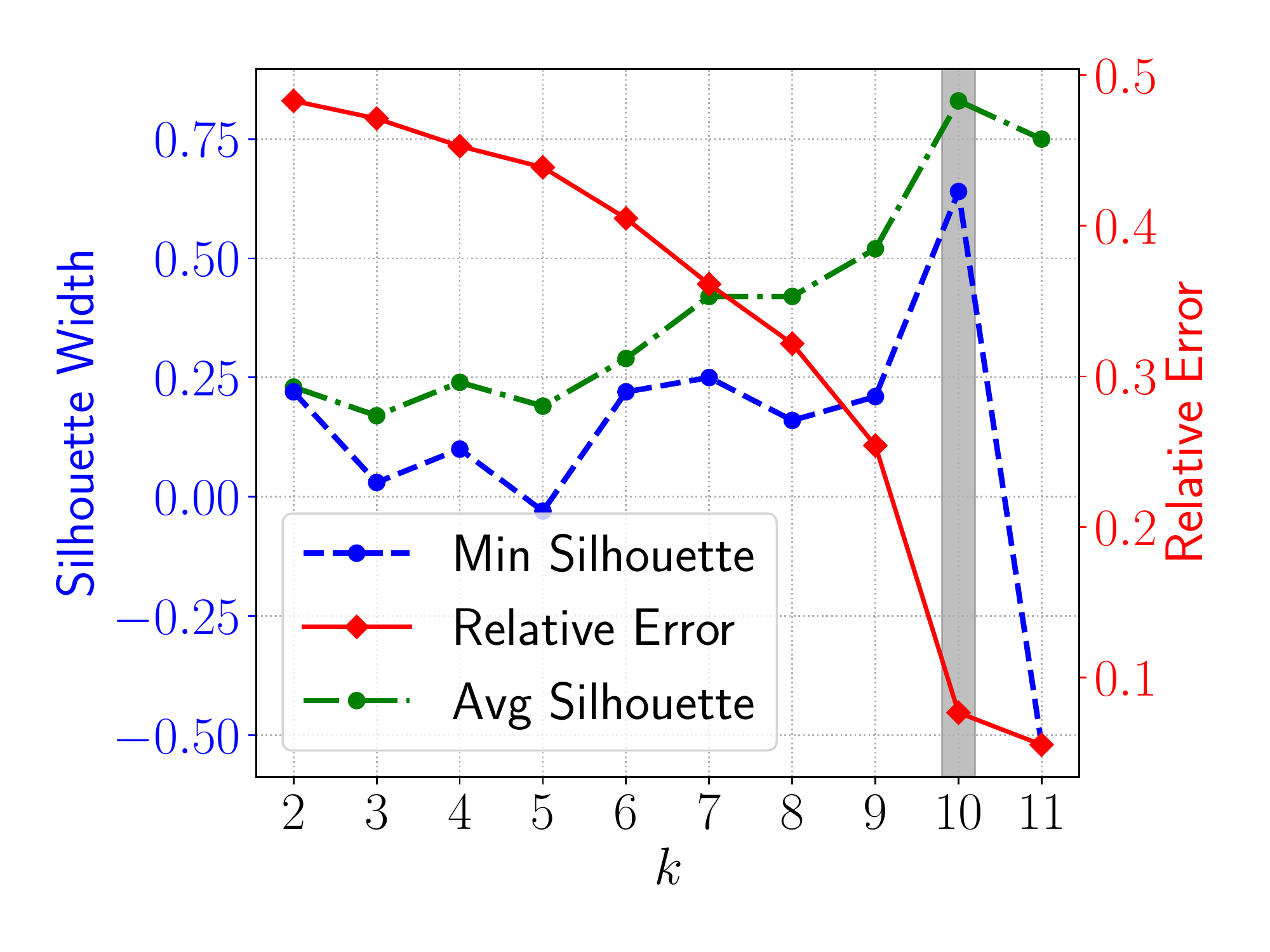}
 	\captionof{figure}{Estimating the number of hidden features in $11.5 TB$ dense data, {\em DRESCALk} finds $k$ as $10$, which is the same as the ground truth value.}
	\label{fig:TB}
\end{subfigure}%
\hfill
\begin{subfigure}{.45\textwidth}
  \centering
  \includegraphics[width=\linewidth]{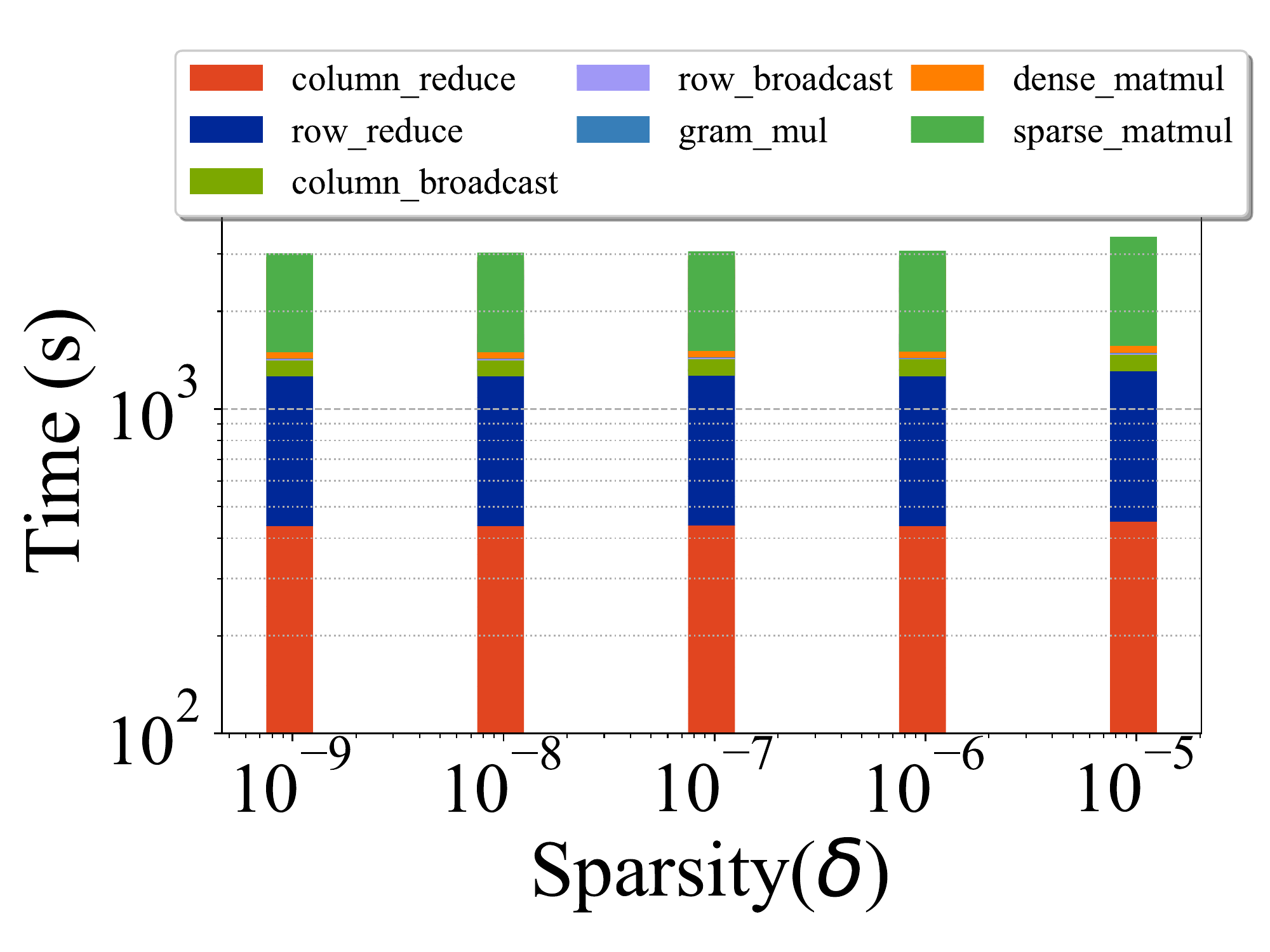}
  \captionof{figure}{Timing breakdown for exabyte data for various sparsity.}
  \label{fig:exa}
\end{subfigure}
\caption{Large scale data results}
\label{fig:large_res}
\end{figure*}

\subsection{Model Determination in Large Data}
To the best of our knowledge, we are the first to perform the pyDRESCALk on a large synthetic dense dataset of size $20\times 396800\times 396800$, i.e., three hundreds trillion of numbers, or ~$11.5TB$ with float32 precision. For this decomposition, we utilize 173 nodes with 4096 cores. The core utilization per node is reduced from 36 to 23 to increase the available memory per core. The data is generated using the approach described in subsection~ \ref{sec:datagen} with $k=10$. PyDRESCALk is able to correctly estimate the value of $k$ as shown in Figure~\ref{fig:TB}. The decomposition is run for about 3 hours to identify the correct number of latent features. For this experiment, the range of $k$ for the evaluation of latent dimension with pyDRESCALk is set from 2 to 11 with 10 perturbations for each $k$ and 200 RESCAL updates for each perturbation. The relative reconstruction error at  $k=10$ as shown in Figure~\ref{fig:TB} is $6\%$ with minimum silhouette score of $.9$.

We also utilize pyDRESCALk to perform factorization on a sparse tensor $\ten{X}$ of size $20\times 373555200\times 373555200$, i.e., $\approx 9.5E$B with  varying sparsity of $1e-5,1e-6,1e-7,1e-8$ and $1e-9$. For this tensor decomposition, we utilize 23,000 CPU cores across 963 nodes on the Grizzly Linux cluster at LANL. The runtimes for 100 iterations of sparse-RESCAL decomposition on the sparse tensors with given sparsity is shown in Figure~\ref{fig:exa}.
From Figure~\ref{fig:exa}, we observe that most of the time is spent on MPI communications, which accounts for more than $90\%$ of the total execution time, and the per-core compute time is less than $10\%$ of the total execution time. Despite the level of sparsity in the dataset, the total amount of communicated data remains the same. This leads to the same overall communication time for the sparse decompositions of $\ten{X}$ with different sparsity. On the other hand, compute time decreases with the decrease in the non-zero elements i.e. increase in the sparsity of $\ten{X}$.  
However, as communication time takes a major portion of overall execution time which is the same for different sparse decomposition experiments of $\ten{X}$,  the total time remains unaffected despite the faster compute time for increasing sparsity of the data as observed in \ref{fig:exa}.

To address such bottlenecks, an implementation on  better hardware platforms with higher communication bandwidths, higher per node memory and better compute capabilities  will be able to minimize the communication cost and the computation cost leading to faster decomposition. 

\section{Conclusions}
\label{sec:conclusions}
In this paper, we introduced pyDRESCALk, a distributed framework for RESCAL tensor factorization, suitable for large knowledge graph embeddings, with the unique ability to estimate the number of latent features in a dataset. While pyDRESCALk can leverage GPU accelerators, it is also able to run on a CPU only cluster. The latent feature identification pipeline involves distributed implementations of custom clustering and stability analysis. The efficacy of our framework is demonstrated on several synthetic and real-world datasets. Furthermore, to demonstrate the scaling of our framework, we report strong and weak scaling figures as well as the scaling of our algorithm with the number of latent communities, $k$, for both sparse and dense datasets. The complexity analysis for the entire framework is provided, which includes computational, communication, and spatial complexity for RESCAL and clustering modules. We also show the agreement between the scaling experiments and the complexity analysis. Finally, we perform numerical experiments on $11$TB dense and  $9.5$EB  sparse tensors with $10^{-6}$ sparsity and, for the first time, demonstrate ability to correctly identify latent communities on such large scale datasets. In the future, we aim to demonstrate even faster performance with optimized GPU communication primitives such as NCCL on larger datasets to approximate a realistic knowledge graph. Taken in whole, the pyDRESCALk code, which implements the algorithms discussed in this paper and is freely available on GitHub, is a feature rich code with many possible applications including business, commerce, surveillance activities, social media networks, computer macro-simulations, and large-scale experiments

\section*{Acknowledgements}
This research was funded by DOE National Nuclear Security Administration (NNSA) - Office of Defense Nuclear Nonproliferation R\&D (NA-22), and
supported by LANL Laboratory Directed Research and Development (LDRD) grant 20190020DR, and the Los Alamos National Laboratory Institutional Computing Program, supported by the U.S. Department of Energy National Nuclear Security
Administration under Contract No. 89233218CNA000001. The work of Hristo Djidjev has been also partially supported by Grant No.~BG05M2OP001-1.001-0003, financed by the Science and Education for Smart Growth Operational Program (2014-2020) and co-financed by the European Union through the European Structural and Investment Funds.

\section*{Conflict of interest}
The authors declare that they have no conflict of interest.
\bibliographystyle{IEEEtran}
\bibliography{references}{} 
\vskip 0pt plus -2fil
\begin{IEEEbiography}[{\includegraphics[width=1in,height=1.25in]{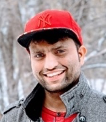}}]{Manish Bhattarai} received the M.S and the Ph.D. degree from the Department of Electrical and Computer Engineering at The University of New Mexico. He is currently a Postdoc Research Associate in the Theoretical division at the Los Alamos National Laboratory in Los Alamos, NM. At LANL, Dr. Bhattarai is part of the tensor factorizations group which specializes on large scale data factorization and improving the Lab`s high-performance processing and computing abilities.  He has extensively worked on developing HPC empowered ML algorithms for mining big data such as distributed Matrix and Tensor factorization. His current research interests include: machine learning, computer vision, deep learning, tensor factorizations and high performance computing.
\end{IEEEbiography}

\vskip 0pt plus -1fil

\begin{IEEEbiography}[{\includegraphics[width=1in,height=1.25in]{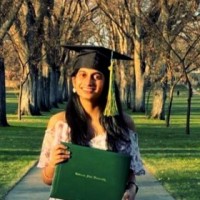}}]{Namita Kharat} is currently a Postmasters Research Associate in the Theoretical division at the Los Alamos National Laboratory in Los Alamos, NM. She received the M.S. degree from the Department of Electrical and Computer Engineering at The Colorado State University. Her research interests include: parallel programming, high performance computing, machine learning and deep learning.
\end{IEEEbiography}

\vskip 0pt plus -1fil

\begin{IEEEbiography}[{\includegraphics[width=1in,height=1.25in,clip,keepaspectratio]{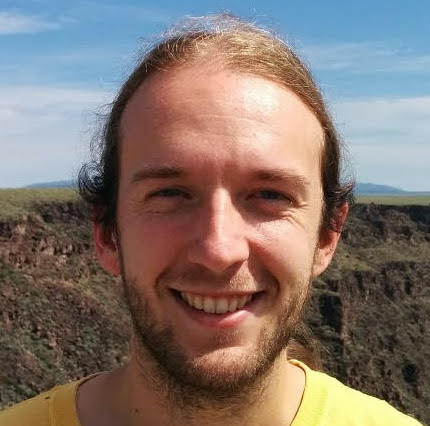}}]{Erik Skau} received the B.Sc. degree in applied mathematics and physics, and the M.Sc. and Ph.D. degrees in applied mathematics from North Carolina State University, Raleigh, NC, USA. His research expertise includes optimization techniques for matrix and tensor decompositions. Erik  is  a  scientist  in  the  Information  Sciences Group as Los Alamos National Laboratory.
 \end{IEEEbiography}
 
\vskip 0pt plus -1fil

\begin{IEEEbiography}[{\includegraphics[width=1in,height=1.25in,clip,keepaspectratio]{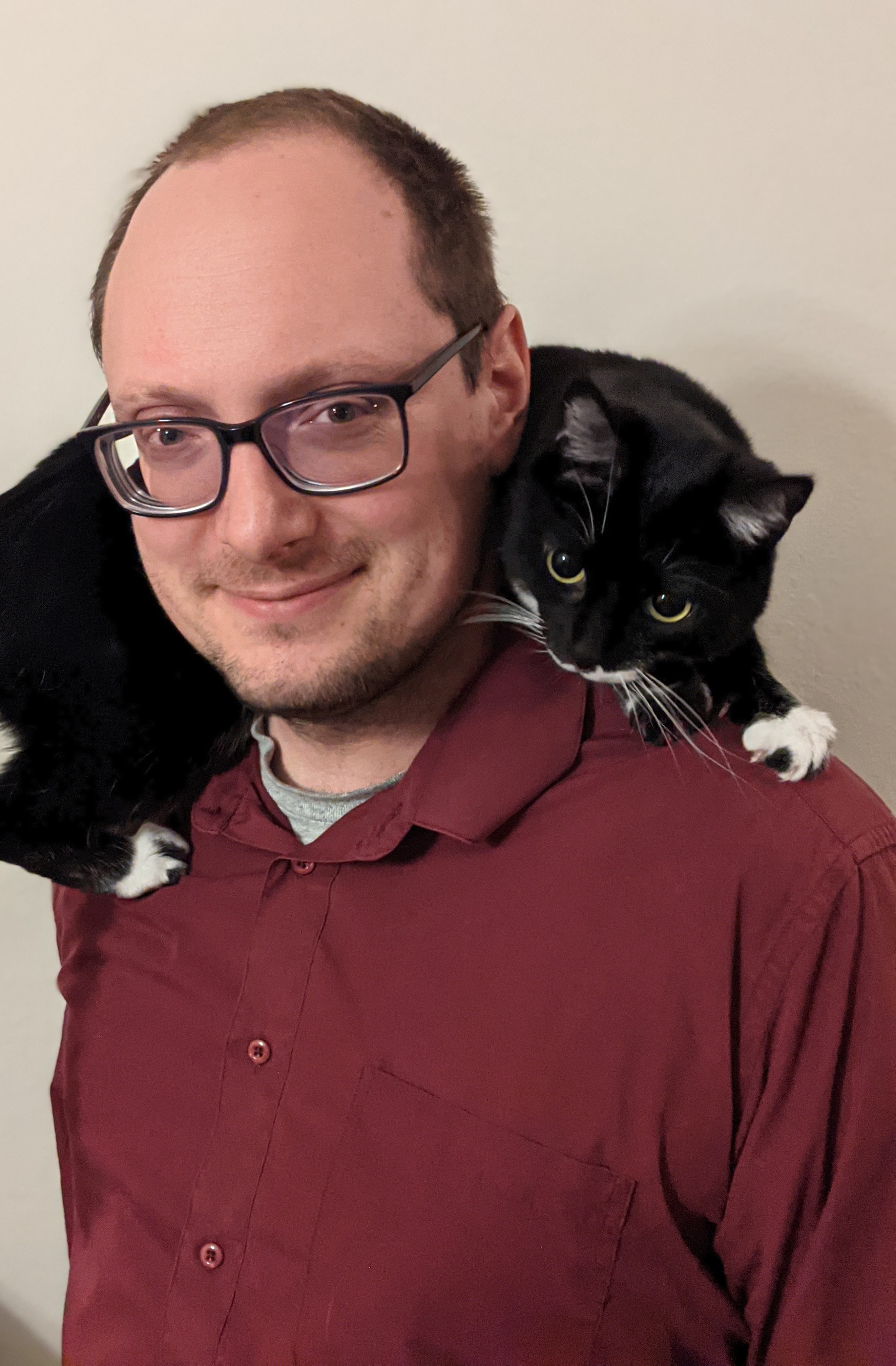}}]{Benjamin Nebgen} received the B.A. degree in Chemistry from Cornell University,Ithaca, NY, USA and  Ph.D. degrees in Chemistry from Purdue University, West Lafayette, IN, USA. He previously had Post doctoral appointments at University of Southern California: Los Angeles, CA, USA and Theoretical division at  Los Alamos National Laboratory (LANL). He is currently scientist in the Theoretical division at LANL. His research expertise includes Quantum Chemistry and optimization techniques for matrix and tensor decompositions. 
 \end{IEEEbiography}

\vskip 0pt plus -1fil

\begin{IEEEbiography}[{\includegraphics[width=1in,height=1.25in,clip,keepaspectratio]{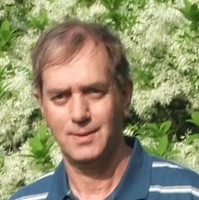}}]{Hristo Djidjev} is a computer scientist in the Information Sciences (CCS-3) group at Los Alamos National Laboratory (LANL). Before joining LANL as a scientist, Hristo worked as an Assistant Professor in Rice University, and as a Senior Lecturer in Warwick University. He is currently a Research Adjunct Professor at Carleton University, Ottawa, Canada. Hristo holds an MSc in applied mathematics and a PhD in computer science from Sofia University, Bulgaria.
 \end{IEEEbiography}

\vskip 0pt plus -1fil

\begin{IEEEbiography}[{\includegraphics[width=1in,height=1.25in]{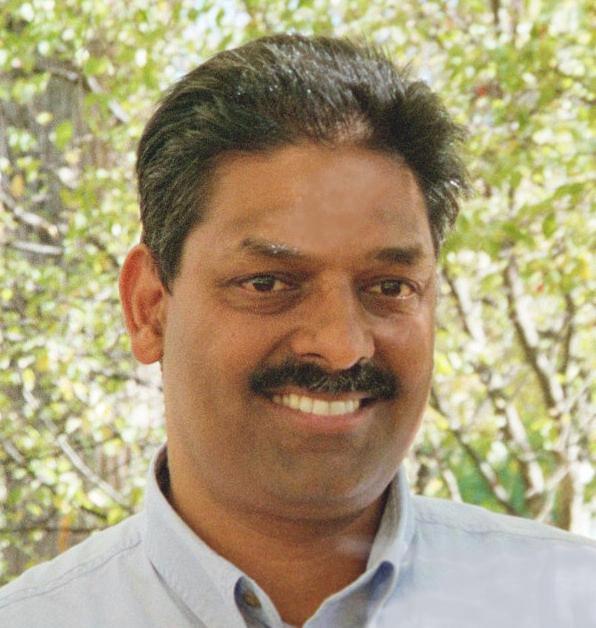}}]{Sanjay Rajopadhye} received the B.Tech. degree (Hons.) in electrical engineering from the Indian Institute of Technology Kharagpur, Kharagpur, India, in 1980, and the Ph.D. degree in computer science from the University of Utah, Salt Lake City, UT, USA, in 1986. 
He held academic positions with the University of Oregon, Eugene, OR, USA, Oregon State University, Corvallis, OR, USA, and IRISA, Rennes, France. He is currently a Professor with the Department of
Computer Science and the Department of Electrical and Computer Engineering, Colorado State University, Fort Collins, CO, USA. He is one of the inventors of the polyhedral model—a mathematical formalism for reasoning about massively parallel, regular, and compute- and data-intensive computations. It was developed to address the design of early era hardware accelerators called systolic arrays. His current research interests include very-large-scale integration, architecture, embedded systems, languages, algorithms, and compilation.
\end{IEEEbiography}

\vskip 0pt plus -1fil

\begin{IEEEbiography}[{\includegraphics[width=1in,height=1.25in]{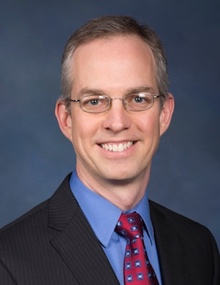}}]{Jim Smith} has a Ph.D. in Physics and two decades in leadership roles at LANL, including Group Leader, Division Senior Scientist, and Technical Director for the Principal Associate Directorate of Global Security.  He has led many large programs, including over 100 technical staff on the DHS National Infrastructure Simulation Analysis Center delivered analytic products and represented the US as Science Delegate to Quad Working Groups.  
\end{IEEEbiography}

\vskip 0pt plus -1fil

\begin{IEEEbiography}[{\includegraphics[width=1in,height=1.25in]{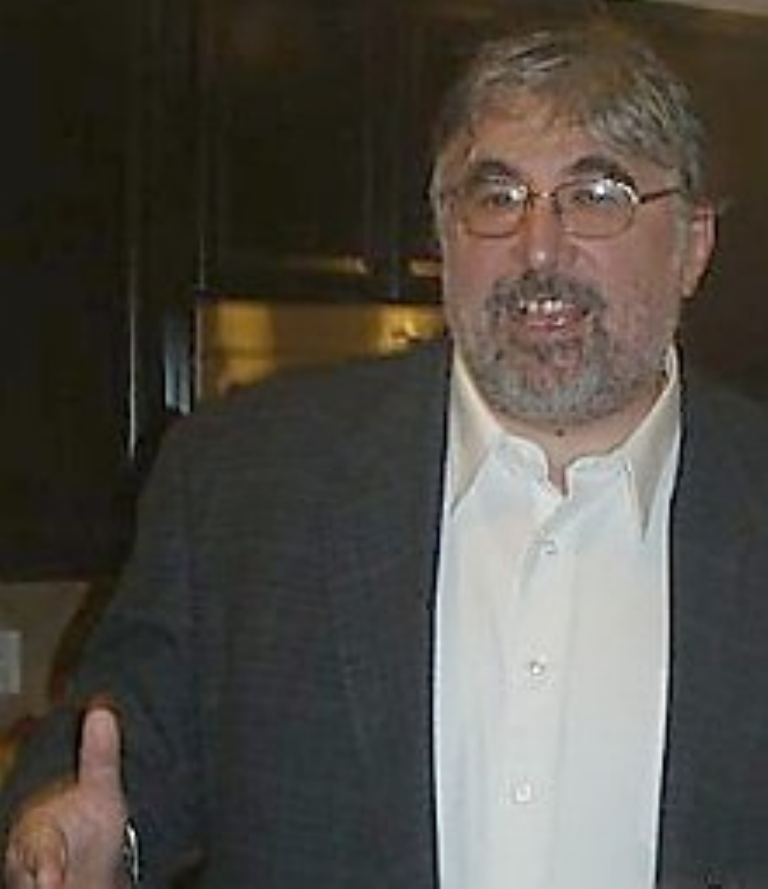}}]{Boian Alexandrov} is a senior scientist at the Theoretical Division in Los Alamos National Laboratory. He has MS in Theoretical Physics, a PhD in Nuclear Engineering and second PhD in Computational Biophysics. Alexandrov is specialized in Big Data analytics, non-negative Matrix and Tensor Factorization, Unsupervised Learning,and Latent Feature Extraction.
\end{IEEEbiography}


\end{document}

%% file: Isoefficiency.tex
\subsection{The Isoefficiency Metric of Scalability}

We observe the following characteristics about the scalability of parallel programs:
\begin{itemize}
    \item If the number of processing elements are increased for a given problem size, the overall efficiency of the parallel system decreases.
    \item Whereas, if we keep the number of processing elements constant and increase the problem size, the efficiency of the parallel system increases in many cases.
\end{itemize}

From the above two observations, we can define a scalable parallel system as one in which the efficiency can be kept constant when both the number of processing elements and the problem size is increased. Thus, it is useful to determine the rate at which the problem size must be increased with respect to the number of processors in order to keep the efficiency constant. The degree of scalability of the parallel system is determined by this rate. The problem size must increase at different rates for different parallel systems with the the number of processors to keep the efficiency fixed. A lower rate in increase of problem size  is more desirable then the higher rate increase in problem size. 

Here, in this section, the problem size is defined as the number of basic computation steps taken in the best sequential algorithm on a single processor. The problem size is a function of the size of the input since it is defined in terms of a sequential algorithm. In other words, problem size $W$ is equal to the serial runtime $T_{s}$ of the fastest known algorithm in order to solve the problem on a single processor.

\subsubsection{The Isoefficiency Function}

The parallel runtime can be defined as 
\begin{align}\label{eqn:T_p}
\centering
\begin{split}
    T_{p} = \frac{W + T_{o}(W,p)}{p}
\end{split}
\end{align}
 where $T_o$ is the total overhead time, $W$ is the problem size, and $p$ is the number of processors. The resulting expression for speedup from equation~\ref{eqn:T_p} is
 \begin{align}\label{eqn:S}
\centering
\begin{split}
    S = \frac{W}{T_p}= \frac{Wp}{W + T_o(W,p)}
\end{split}
\end{align}
Finally, utilizing equation~\ref{eqn:S}, the efficiency is expressed as
 \begin{align}\label{eqn:E}
\centering
\begin{split}
   E = \frac{S}{p} = \frac{W}{W + T_o(W,p)} = \frac{1}{1 + \frac{T_o(W,p)}{W}}
\end{split}
\end{align}
Thus, from equation~\ref{eqn:E}, the efficiency as defined in \cite{grama1993isoefficiency} can be kept constant if the ratio $\frac{T_o}{W}$ is maintained at a fixed value. If equation~\ref{eqn:E} is solved further,
 \begin{align}\label{eqn:E_2}
\centering
\begin{split}
   E = \frac{1}{1 + \frac{T_o(W,p)}{W}}\Leftrightarrow \frac{T_o(W,p)}{W} = \frac{1 - E}{E}\\ \Leftrightarrow W = \frac{E}{1 - E} T_o(W,p)
\end{split}
\end{align}

Assume, $K=\frac{E}{1 - E}$ to be constant based on the efficiency to be maintained. Thus, equation~\ref{eqn:E_2} becomes,
 \begin{align}\label{eqn:W}
\centering
\begin{split}
  W = KT_o(W,p)
\end{split}
\end{align}
This function dictates the rate of growth for the problem size $W$ required to maintain the efficiency constant as $p$ increases. This function is called the isoefficiency function of the parallel system~\cite{grama1993isoefficiency}. Thus, it determines the ease with which a parallel system can keep a fixed efficiency and get speedup with respect to the number of processors.
\begin{itemize}
    \item If the isoefficiency function is small, it means that small increase in the problem size is enough to utilize increasing number of processors efficiently.
    \item If the isoefficiency function is large, it indicates that the parallel system is poorly scalable.
\end{itemize}

Using the big O complexity of the RESCAL  algorithm derived in the previous sections, the above equation can be written
 \begin{align}\label{eqn:mn}
\centering
\begin{split}
 \frac{mn^2k}{p} = \O{ m k \frac{n}{\sqrt p} \log p }
 \end{split}
\end{align}
where $W$ is replaced by the total computation cost for a dense tensor $\ten{X}$ on a single processor and $T_o(W,p)$ is replaced by the total communication overhead calculated in the complexity section. Simplifying the above equation, we get
$n = \O{ \sqrt{p}\log p}$ as the derived isoefficiency function for dense RESCAL Algorithm. In the sparse case, we introduce a term $\delta$ which is the density of $\ten{X}$ in the computation section and hence the above sparse isoefficiency function is
$n = \O {\frac{\sqrt{p}\log(p)}{\delta}}$.
